\documentclass[aps,pra,longbibliography,10pt,twocolumn,superscriptaddress,notitlepage,floatfix,amssymb,showpacs]{revtex4-2}
\usepackage[letterpaper,hmargin={1.5cm,1.5cm},vmargin={1.5cm,2.5cm}]{geometry}
\usepackage{mathtools}
\usepackage{float}
\usepackage{amsfonts}
\usepackage{enumitem}
\usepackage{graphicx}
\usepackage{color}
\usepackage{placeins}
\usepackage{hyperref}
\hypersetup{
    colorlinks=true,    
    linkcolor=blue,     
    citecolor=green,    
    urlcolor=magenta    
}

\begin{document}
\title{Josephson traveling-wave parametric amplifier based on low-intrinsic-loss coplanar lumped-element waveguide}

\author{C. W. Sandbo Chang}
\email{chungwaisandbo.chang@riken.jp}
\affiliation{
RIKEN Center for Quantum Computing (RQC), Wako, Saitama 351\text{--}0198, Japan
}
\author{Arjan F. Van Loo}
\affiliation{
RIKEN Center for Quantum Computing (RQC), Wako, Saitama 351\text{--}0198, Japan
}
\affiliation{
The University of Tokyo, Meguro-ku, Tokyo 153\text{--}8904, Japan
}
\author{Chih-Chiao Hung}
\affiliation{
RIKEN Center for Quantum Computing (RQC), Wako, Saitama 351\text{--}0198, Japan
} 
\author{Yu Zhou}
\affiliation{
Fujitsu Limited, Nakahara-ku, Kawasaki, Kanagawa 211\text{--}8588, Japan
}
\affiliation{
RIKEN Center for Quantum Computing (RQC), Wako, Saitama 351\text{--}0198, Japan
}
\author{Christian Gnandt}
\affiliation{
RIKEN Center for Quantum Computing (RQC), Wako, Saitama 351\text{--}0198, Japan
}
\affiliation{
The University of Tokyo, Meguro-ku, Tokyo 153\text{--}8904, Japan
}
\author{Shuhei Tamate}
\affiliation{
RIKEN Center for Quantum Computing (RQC), Wako, Saitama 351\text{--}0198, Japan
}
\author{Yasunobu Nakamura}
\affiliation{
RIKEN Center for Quantum Computing (RQC), Wako, Saitama 351\text{--}0198, Japan
}
\affiliation{
The University of Tokyo, Meguro-ku, Tokyo 153\text{--}8904, Japan
}

\begin{abstract}
We present a Josephson traveling-wave parametric amplifier (JTWPA) based on a low-loss coplanar lumped-element waveguide architecture. By employing open-stub capacitors and Manhattan-pattern junctions, our device achieves an insertion loss below 1~dB up to 12~GHz. We introduce windowed sinusoidal modulation for phase matching, demonstrating that a smooth transition in the impedance-modulation strength effectively suppresses intrinsic gain ripples. Using Tukey-windowed modulation with 8\% impedance variation, we achieve 20\text{--}23-dB~gain over 5-GHz bandwidth under ideal matching conditions. In a more practical circuit having impedance mismatches, the device maintains 17\text{--}20-dB gain over 4.8-GHz bandwidth with an added noise of 0.18~quanta above standard quantum limit at 20-dB gain and $-99$-dBm saturation power, while featuring zero to negative backward gain below the band-gap frequency.
\end{abstract}

\date{\today}


\maketitle

\section{\label{sec:Introduction} Introduction}
Quantum-limited amplification plays a crucial role in superconducting quantum computing, particularly for achieving high-fidelity single-shot qubit readout without perturbing the quantum state~\cite{sankMeasurementInducedStateTransitions2016,dumasMeasurementInducedTransmonIonization2024}. Resonator-based Josephson parametric amplifiers (JPAs) have long served as the primary solution, offering high gain with near-quantum-limited noise performance~\cite{castellanos-beltranAmplificationSqueezingQuantum2008,yamamotoFluxdrivenJosephsonParametric2008,bergealPhasepreservingAmplificationQuantum2010}. However, despite extensive development efforts in extending their gain--bandwidth product and saturation power~\cite{royBroadbandParametricAmplification2015,grebelFluxpumpedImpedanceengineeredBroadband2021,whiteReadoutQuantumProcessor2023}, these JPAs still face fundamental limitations in instantaneous bandwidth from their resonant nature. Over the last decade, Josephson traveling-wave parametric amplifiers (JTWPAs) have emerged as a promising alternative, overcoming these constraints through their transmission-line-based architecture~\cite{macklinQuantumLimitedJosephson2015,whiteTravelingWaveParametric2015, planatPhotonicCrystalJosephsonTravelingWave2020,fengDesignMeasurementJosephson2020, perelshteinBroadbandContinuousVariableEntanglement2022, ranadiveKerrReversalJosephson2022}. By delivering high gain over a wide bandwidth while maintaining high saturation power, JTWPAs have enabled significant advances in quantum information processing and applications in microwave quantum optics~\cite{konoMechanicallyInducedCorrelated2024, swiadekEnhancingDispersiveReadout2024,nieParametricallyControlledMicrowavephotonic2024, meesalaQuantumEntanglementOptical2024,ahnExtensiveSearchAxion2024,fraudetDirectDetectionDownConverted2025}, particularly in simultaneous single-shot readout of frequency-multiplexed qubits~\cite{heinsooRapidHighfidelityMultiplexed2018b,remmIntermodulationDistortionJosephson2022}.

Despite their advantages, JTWPAs have not been widely adopted due to significant implementation challenges~\cite{kisslingVulnerabilityParameterSpread2023,peatainSimulatingEffectsFabrication2023}. The primary obstacle lies in fabricating hundreds to thousands of Josephson junctions with highly uniform parameters across a single device to maintain uniform line impedance. The most straightforward implementation of a JTWPA, proposed more than three decades ago, was to build a nonlinear transmission-line based on series junction or superconducting quantum interference device (SQUID) arrays, which naturally exhibits parametric gain through three-wave-mixing~(3WM) or four-wave mixing~(4WM)~\cite{sweenyTravellingwaveParametricAmplifier1985a, mohebbiAnalysisSeriesConnectedDiscrete2009}. However, achieving gain efficiently in such designs requires maintaining specific phase relationships between propagating waves, a condition that is not naturally satisfied due to nonlinear phase shifts induced by the strong pump wave required for amplification. Without phase-matched propagation among pump, signal, and idler waves, efficient parametric gain cannot be achieved~\cite{yaakobiParametricAmplificationJosephson2013}. Addressing this limitation requires additional engineering in the transmission-line architecture beyond simply implementing a junction array, typically in the directions of compensating for or reducing these nonlinear phase shifts.

Progress in JTWPA development has led to various design approaches for achieving phase-matched gain. Many successful implementations exist, with dispersion-engineered transmission lines using resonant structures to compensate for the shifts serving as one notable example~\cite{obrienResonantPhaseMatching2014,whiteTravelingWaveParametric2015, macklinQuantumLimitedJosephson2015,fengDesignMeasurementJosephson2020}. While effective in achieving phase matching and providing high gain, this method requires stringent fabrication conditions to precisely define resonator frequencies and integrate high-density resonant structures, complicating device reproducibility~\cite{kisslingVulnerabilityParameterSpread2023}. Meanwhile, nonlinearity-engineered transmission lines using SQUIDs or superconducting nonlinear asymmetric inductive elements (SNAIL)~\cite{frattini3waveMixingJosephson2017a} to reduce or change the sign of the phase shifts represent another approach~\cite{ranadiveKerrReversalJosephson2022,perelshteinBroadbandContinuousVariableEntanglement2022,fadaviroudsariThreewaveMixingTravelingwave2023}, offering more relaxed fabrication requirements but necessitating additional flux-bias control for operation. This added complexity poses challenges for integrating these JTWPAs into large-scale superconducting computing systems. A more traditional approach that was often adopted in kinetic-inductance based TWPAs (kTWPAs), the periodic modulation of transmission line parameters, offers a simpler implementation of phase-matched amplification~\cite{chaudhuriBroadbandParametricAmplifiers2017,planatPhotonicCrystalJosephsonTravelingWave2020, malnouThreeWaveMixingKinetic2021,faramarzi48GHzKinetic2024}. It requires neither precise resonant structures nor flux-bias control. However, implementations of this approach in JTWPAs have shown significant intrinsic gain ripples in frequency~\cite{planatPhotonicCrystalJosephsonTravelingWave2020}, making it less than ideal for realizing JTWPAs where a flat gain profile is often desired. 

Beyond these phase-matching considerations, another challenge common to most existing JTWPA designs lies in their significant insertion loss~\cite{macklinQuantumLimitedJosephson2015,planatPhotonicCrystalJosephsonTravelingWave2020,ranadiveKerrReversalJosephson2022}, mainly due to the choice of dielectric material used in creating the shunt capacitance or the use of normal-conductor grounding material. This intrinsic loss lowers the effective gain and limits the minimum added noise achievable~\cite{fadaviroudsariThreewaveMixingTravelingwave2023}.

\begin{figure*}[!htb]
    \includegraphics[width=17.59cm]{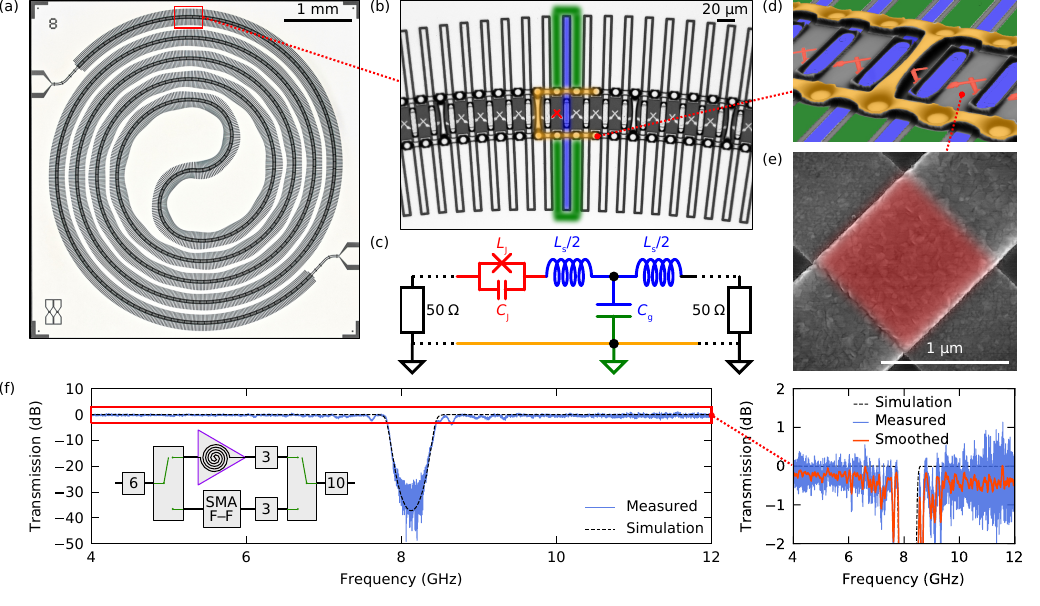}
    \caption{Overview of the CP-JTWPA architecture and its transmission characteristics. (a) Double-spiral layout of the CP-JTWPA chip fabricated on a silicon substrate (darker regions) with aluminum thin films (brighter regions), implementing a nonlinear transmission line over a 5$\times$5-mm$^2$ area. (b) Unit cell structure of length 20~\textmu$\mathrm{m}$ showing the Josephson junction (red), open-stub capacitors (blue: centerline, green: ground), and airbridges (yellow) constructed across every single open stub and across the ground planes every five unit cells along the transmission line. (c) Lumped-element circuit model of a unit cell including junction inductance and capacitance ($L_\mathrm{J}$, $C_\mathrm{J}$), open-stub capacitance ($C_\mathrm{g}$), and two extra series inductors ($L_\mathrm{s}/2$) accounting for geometric inductance, with cross-stub and cross-ground airbridges (yellow) forming the ground current path. (d) 3D visualization of the unit-cell geometry with airbridge structures, where aluminum bandages connecting the junctions to the open stubs were fabricated in the same step as the airbridges. (e) SEM image of a Manhattan-type Josephson junction of approximately 1-\textmu m$^2$ area, formed by double-angled evaporations and liftoff with electrodes of 1-\textmu m width, wider than that typically used in transmon qubits to achieve sufficiently high critical current. (f)~Device~B baseline transmission at 12~mK compared to a 50-$\Omega$ SMA F--F adapter, measured at roughly $-125$-dBm power and 6~GHz using two microwave switches. The two paths between switches were verified at room temperature to have transmission difference of $<$0.1~$\mathrm{dB}$. At low temperature, besides the band gap which appears at the designed frequency~(8.13~GHz), the typical insertion loss is less than 1~dB over the 4\text{--}12~GHz range including the sample package.}
    \label{fig:devicePic}
\end{figure*}

In this work, we address these challenges by introducing a novel JTWPA design based on coplanar lumped-element architectures, hereafter referred to as CP-JTWPAs. Our approach achieves low insertion loss through a highly accessible fabrication process. We demonstrate three CP-JTWPA variants implementing window functions on top of periodic modulation for phase correction, substantially enhancing both the performance and practicality of periodic-modulation-based JTWPAs. We present our results in the following order. Section~\ref{sec:device_design} introduces the fundamentals of our low-loss CP-JTWPA architecture. Section~\ref{sec:windowed_modulation} presents theoretical analysis and experimental validation of our windowed modulation scheme for achieving phase-matched amplification. Section~\ref{sec:characterization_tukey} provides comprehensive characterization of our best-performing Tukey-windowed CP-JTWPA, including gain profile, noise performance, and saturation power measurements.

\section{\label{sec:device_design} DEVICE DESIGN}
\subsection{\label{sec:open-stub line} Open-stub transmission line structure}

Our first objective is to obtain an underlying nonlinear transmission line with low insertion loss. We achieved this by building our device based on a coplanar waveguide~(CPW), which is known for its low loss per unit length and ease of fabrication~\cite{sageStudyLossSuperconducting2011a}. We introduced nonlinearity into the CPW by inserting Josephson junctions along its centerline to form a series junction array. To compensate for the increased series inductance due to the junctions and maintain a 50-$\Omega$ line impedance, additional shunt capacitance is required. To avoid using lossy dielectric materials, we employed open-stub shunt capacitors, where the low-loss silicon substrate and vacuum are the dominating dielectrics, as depicted in Fig.~\ref{fig:devicePic}(b). The open-stub capacitors are constructed by extending part of the CPW centerline into both sides of the ground plane, added for each junction. A CP-JTWPA unit cell thus consists of a combination of an open-stub pair and a junction. These cells then chain up in series to form the underlying open-stub lines. The CP-JTWPA unit cell can be effectively modeled by an equivalent lumped-element circuit as shown in Fig.~\ref{fig:devicePic}(c), consisting of the nonlinear inductance $L_\mathrm{J}$ and capacitance $C_\mathrm{J}$ of the series junction, and the shunt capacitance $C_\mathrm{g}$ of the open-stub pair. Additionally, we included two series inductors to take into account any stray inductance $L_\mathrm{s}$ due to the geometry. We note that a similar open-stub approach for JTWPA implementation has been recently demonstrated~\cite{wangHighEfficiencyLowLossFloquetmode2025} also as an attempt to achieve low insertion loss, though with a different phase-matching scheme rather than our windowed periodic-modulation approach.

To extract the parameters from the geometry, we perform finite-element simulations over a short chain of 40 unit cells, as simulating the full device~($>$2000~cells) is computationally prohibitive. We varied the length of the open stubs and computed the S-parameters for each length, followed by fitting the results to the lumped-element model using the ABCD-matrix method~(see Appendix~\ref{sec:appendix_numerical_analysis_1dTL}). With the established scaling between stub length and shunt capacitance, we can achieve the desired impedance modulation by adjusting the stub length in each unit cell. 

In our implementation, we modulated both the junction area and stub length in a coordinated manner, in order to achieves the required 8\% impedance modulation. That is, moving along the series junction array, we reduce the junction area (increasing $L_\mathrm{J}$) when the stub length is decreased (reducing $C_\mathrm{g}$) and vice versa. This correlated approach reduces the required stub-length variation from approximately 16\% (if modulating capacitance alone) to 8\%, as detailed in Appendix~\ref{sec:appendix_device_parameters}, thus avoiding a large geometric variation in the design. We verified through Monte Carlo simulations~(see Appendix~\ref{sec:appendix_mc_dual_vs_mono}) that this correlated modulation scheme maintains resilience against junction critical-current variations comparable to capacitive-only modulation, despite the junction areas are no longer uniform by design.

While open-stub lines on silicon substrate share the low-loss property of CPW, due to their coplanar nature, they also suffer from issues commonly observed in CPW such as coupling to slot-line modes, introducing loss and unwanted features in transmission~\cite{chenFabricationCharacterizationAluminum2014}. To suppress slot-line modes, we inserted airbridges connecting the ground planes across the centerline every five unit cells, shown in Fig.~\ref{fig:devicePic}(b)~and~\ref{fig:devicePic}(d). This spacing provides sufficient suppression of slot-line modes below 12~GHz, covering our operating bandwidth. Besides coupling to slot-line modes, stray coupling between adjacent open-stub lines should also be minimized. As the open stubs extend into ground planes, ground currents of the guided waves tend to flow along the ground metal around the tips of the open-stub extensions. This effectively reduces the separation between two adjacent open-stub lines and promotes a stray inductive coupling between the lines, potentially introducing cross-talk across the device. Although increasing spacing between open-stub lines could mitigate this issue, it would reduce device density. Instead, we inserted additional airbridges across every open stub on both sides of the open-stub lines. These cross-stub airbridges guide the ground current to flow closer to the junction-loaded centerline, effectively restoring the separation between the ground-current paths of adjacent lines. Additionally, these airbridges also provide extra ground-current paths that reduce stray geometric inductance, allowing the use of higher junction inductance for the same shunt capacitance.

\subsection{\label{sec:large Manhattan junction} Expanded-area Manhattan-type junctions}

The performance of a JTWPA critically depends on the quality of its constituent Josephson junctions. Optimal junction design must ensure both high fabrication reliability and exceptional spatial homogeneity across the device. Typically, JPAs require junctions with critical currents in the microampere range, traditionally achieved using Al-AlO$_x$-Al junctions fabricated through shadow evaporation techniques such as Dolan bridges~\cite{dolanOffsetMasksLiftoff1977} or asymmetric undercut methods~\cite{lecocqJunctionFabricationShadow2011a}. However, these approaches rely on suspended structures that risk mechanical collapse and produce junction areas sensitive to resist thickness variations. Meanwhile, Manhattan-type junctions are preferred in transmon qubit fabrication for their structural stability and reproducible junction areas~\cite{osmanSimplifiedJosephsonjunctionFabrication2021}. However, their traditionally small junction sizes yield critical currents often below 1~\textmu A, insufficient for JPA and JTWPA applications. To address this limitation, we have developed a process that extends the Manhattan-type junction architecture to areas exceeding 1~\textmu$\mathrm{m}^2$, thereby achieving the high critical current required for JTWPA operation, depicted in Fig.~\ref{fig:devicePic}(e). This was enabled by using an electron-beam resist stack consisting of a thick~(2~\textmu$\mathrm{m}$) top imaging layer over a thin~(200~nm) bottom undercut layer, which allows the shadowing of a wide electrode. Using this process, we typically obtain a standard deviation of less than two percent in the normal-state junction resistance over the area of a device, on top of a near-perfect yield. 

Building upon these design principles, we fabricated and characterized three CP-JTWPA samples in double-spiral layouts, all housing roughly 2400~unit cells each in a chip area of 5$\times$5 mm$^2$, using the recipes described in Appendix~\ref{sec:appendix_fabrication}. We will refer to these as device A, B, and C throughout this text, each implementing a different window function (boxcar, Hann, and Tukey, respectively) on the sinusoidal impedance modulation~(see Appendix~\ref{sec:appendix_device_parameters} for detailed parameters).

A first achievement of our design is the low insertion loss, which we confirmed by measuring device B in a dilution refrigerator. By comparing the device transmission to a 50-$\Omega$ SubMiniature Version A~(SMA) Female-to-Female~(F--F) adapter using identical microwave cables sandwiched by microwave switches, we demonstrated insertion loss below 1~dB up to 12~GHz, including the loss from the sample package, as shown in Fig.~\ref{fig:devicePic}(f). Consistent low-loss transmissions are also observed for device A and C. This low-loss performance addresses a persistent challenge in JTWPA development and establishes a foundation for amplifiers with noise performance approaching the standard quantum limit~(SQL), as we will demonstrate in subsequent sections.


\section{\label{sec:windowed_modulation} Windowed modulation}
\subsection{\label{sec:modulation_theory} Theoretical analysis}

\begin{figure*}[!htb]
    \includegraphics[width=18.00cm]{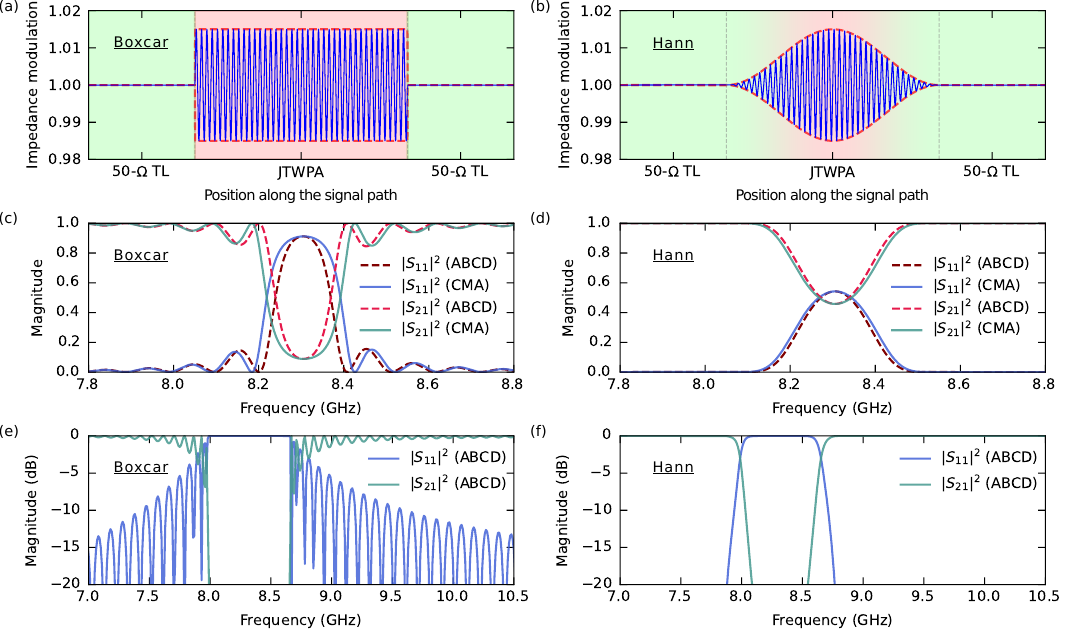}
    \caption{Comparison of impedance-modulation profiles and their transmission characteristics for boxcar (left panels) and Hann (right panels) windows. (a,b) Example impedance-modulation profiles showing the JTWPA sections embedded between 50-$\Omega$ transmission lines. Simulation parameters are provided in Appendix~\ref{sec:appendix_numerical_analysis_1dTL}. For visual clarity, the modulation patterns are shown with a reduced frequency compared to those used in the example. (c,d) Transmission ($|S_{21}|^2$) and reflection ($|S_{11}|^2$) characteristics with 1.5\% impedance-modulation amplitude, calculated using both the analytical coupled-mode analysis (CMA) and numerical ABCD-matrix method (ABCD). (e,f) ABCD-matrix simulations at 7.7\% impedance-modulation amplitude (achieved through 16\% capacitance modulation), showing that the transmission ripples in the boxcar case and their suppression in the Hann case persist similarly to the weak-modulation case.}
    \label{fig:windowedModulation_theory}
\end{figure*}

Having established the base architecture of our 50-$\Omega$ nonlinear transmission line, we now address the critical phase-matching requirements. With a series junction array and no static bias, different $4\mathrm{WM}$ processes can occur simultaneously. While we aim to utilize the $4\mathrm{WM}$ process where two pump photons are converted into signal and idler photons for parametric amplification, other mixing processes such as self-$\mathrm{phase}$ modulation~($\mathrm{SPM}$) of the pump and cross-$\mathrm{phase}$ modulation~($\mathrm{XPM}$) between the waves are also present. These processes create power-dependent phase mismatches that inhibit efficient parametric gain~\cite{yaakobiParametricAmplificationJosephson2013}. To counter these nonlinear phase shifts, we implemented a sinusoidal impedance modulation along the transmission line~(see Fig.~\ref{fig:windowedModulation_theory}(a) for example). This periodic modulation creates a frequency stopband (see Fig.~\ref{fig:windowedModulation_theory}(c)) when the wavelength of propagating waves matches twice the spatial modulation period, known as Bragg condition, and significantly modifies the dispersion relation around this stopband frequency. By applying the pump at a frequency slightly below the stopband, we obtain the necessary wave vector correction to compensate for the SPM effect. This dispersion engineering approach has been successfully demonstrated in both KTWPAs and JTWPAs to achieve phase-matched amplification~\cite{hoeomWidebandLownoiseSuperconducting2012b, chaudhuriBroadbandParametricAmplifiers2017,planatPhotonicCrystalJosephsonTravelingWave2020, malnouThreeWaveMixingKinetic2021}.

While periodic modulation provides an effective approach to phase matching, JTWPAs implementing this scheme have seen limited practical adoption. Although the presence of a stopband might initially appear problematic, it is generally acceptable when sufficient bandwidth is available in the operating range. A more fundamental challenge emerges from the observation that these devices exhibit significant gain ripples even in simulations with perfect fabrication and ideal circuit parameters~\cite{hoeomWidebandLownoiseSuperconducting2012b,planatPhotonicCrystalJosephsonTravelingWave2020}. This behavior indicates that the gain ripples are intrinsic to the periodic modulation design rather than arising from fabrication imperfections.

\begin{figure*}[!htb]
    \includegraphics[width=18.15cm]{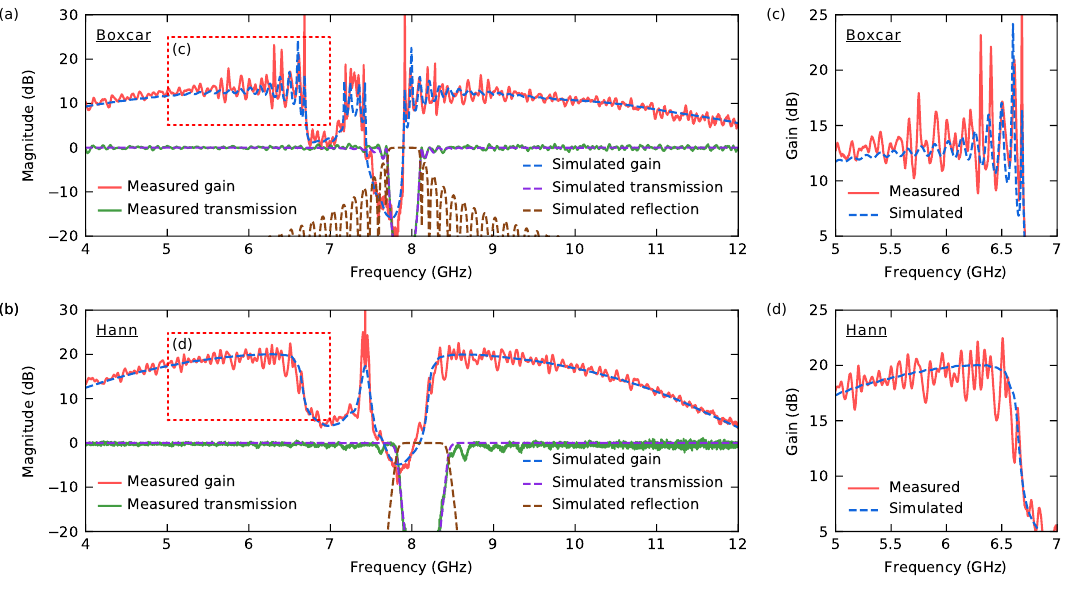}
    \caption{Comparison of transmission characteristics between device A (boxcar window, top) and device B (Hann window, bottom). Left panels~[(a)~and~(b)] show both measured and simulated data from 4\text{--}12~GHz: measured and simulated normalized forward transmission $|S_{21}|^2$ without pump (transmission), measured and simulated normalized forward transmission $|S_{21}|^2$ with pump (gain), and simulated reflection $|S_{11}|^2$ (reflection). All simulations were performed assuming lossless transmission lines. Right panels~[(c)~and~(d)] present enlarged views of the high-gain regions marked by red dotted-line boxes. Base transmission and reflection magnitudes are calculated using ABCD-matrix methods, while gain profiles are obtained through harmonic-balance simulations. The band gap appears at 7.90~GHz and 8.13~GHz for devices A and B, respectively.}
    \label{fig:windowedModulation_experiment}
\end{figure*}

Conventional JTWPA designs typically implement uniform impedance modulation along the transmission line, which mathematically corresponds to applying a rectangular (boxcar) window to the spatial impedance profile. As shown in Fig.~\ref{fig:windowedModulation_theory}, this approach creates abrupt transitions at the JTWPA boundaries, leading to pronounced ripples in both transmission and reflection characteristics near the band-gap frequency. These transmission ripples emerge through a mechanism familiar in signal processing, where a finite-duration signal produces ripples in its frequency-domain spectrum due to windowing effects. Parametric gain in JTWPAs inherently amplifies any features present in the base transmission characteristics, so small ripples in transmission can lead to significant variations in the gain profile. These gain ripples pose practical limitations as they can cause frequency-dependent variations in signal-to-noise-ratio~(SNR) improvement and complicate frequency-multiplexed readout. Moreover, the amplitude of ripples increases with the JTWPA gain, and their density increases with electrical length of the line, as they originate from the Fabry–Pérot-like interference due to impedance mismatches at the terminations between the periodic structure's Bloch impedance and the external 50-$\Omega$ environment.~\cite{takagiFrequencyDependenceBloch2001a,gaydamachenkoNumericalAnalysisThreewavemixing2022}. In some previous work on JTWPAs based on periodic modulations, gain ripples were not as strong as predicted by simulations~\cite{planatPhotonicCrystalJosephsonTravelingWave2020}, which can be attributed to the damping effect from intrinsic transmission line losses. The achievement of low insertion loss in our CP-JTWPA devices removes this inadvertent damping mechanism, necessitating the suppression of gain ripples through deliberate design choices.

Using coupled-mode theory and considering a small modulation amplitude in the shunt capacitance, we can analyze the ripple formation due to the windowing effect by examining how the periodic modulation couples forward- and backward-propagating waves near the Bragg frequency. With a weak modulation (3\% in only the shunt capacitance), the transmission coefficient t (equivalent to the scattering parameter $S_{21}$) and reflection coefficient r (equivalent to $S_{11}$) of the modulated line assuming a matched load, are approximately~(see Appendix~\ref{sec:appendix_coupled_mode})
\begin{align}
    t &\approx \frac{1}{\cosh(\kappa|F(0,l)|)}, \label{eq:transmission_coeff_main} \\ 
    r &\approx j\frac{F^*(0,l)}{|F(0,l)|}\tanh(\kappa |F(0,l)|), \label{eq:reflection_coeff_main}
\end{align}
for a modulated line of length $l$. $\kappa = \beta_0m/4$ is the coupling coefficient between the forward and backward propagating waves, determined by the wave vector of the unmodulated line $\beta_0$ and modulation amplitude $m$. The function $F$ represents an integral over the window function $W(z)$:
\begin{equation}
    F(0,l) = \int_{0}^l W(z')e^{j2\Delta_\beta z'}dz', \label{eq:sec_F_function}
\end{equation}
with $\Delta_\beta = \beta_0 - \beta_\mathrm{B}$ being the detuning from the Bragg wave vector $\beta_\mathrm{B} = \pi/\Lambda$, where $\Lambda$ is the period of the modulation.

Both $t$ and $r$ are monotonic functions of $|F|$, directly connecting transmission and reflection ripples to the behavior of the $F$ integral. Since the window function $W$ is nonzero only over the finite JTWPA length, $F$ represents a spatial Fourier transform of $W$. This mathematical framework reveals that window functions with smoother transitions in modulation amplitude contain fewer high-frequency components and consequently reduce transmission ripples. This connection between the modulation window and transmission spectrum provides insight for selecting suitable window shapes to suppress transmission and gain ripples in JTWPA lines. Fig.~\ref{fig:windowedModulation_theory} demonstrates that applying a Hann window on a spatial modulation can significantly reduces ripples while maintaining the desired band gap, as verified by both coupled-mode analysis and ABCD-matrix simulations. 

The primary tradeoff of windowed modulation is that smoother windows reduce the root-mean-square~(RMS) amplitude of the periodic modulation, decreasing the overall phase-matching correction strength. This reduction can be compensated by increasing the modulation amplitude, though at the cost of widening the band gap. Based on these theoretical considerations, we implemented and tested CP-JTWPA devices with different modulation schemes.

\subsection{\label{sec:modulation_experiment} Experimental validation}

To validate our windowed modulation approach at higher modulation amplitudes, we characterized two CP-JTWPA variants, device A with a conventional boxcar window~(5\% impedance modulation) and device B with a Hann window~(8\% impedance modulation). Both devices were designed with a band gap around $8$~$\mathrm{GHz}$ based on our open-stub-line architecture, with design parameters determined through ABCD-matrix analysis since the analytical coupled-mode theory becomes insufficient at high modulation amplitudes. We performed gain analysis using harmonic-balance simulations~\cite{obrienKpobrienJosephsonCircuitsjl2025} to account for inter-cell reflections from both impedance mismatches and periodic modulations~(see Appendix~\ref{sec:appendix_juliaPlots} for details). The devices were fabricated with similar processes and measured in dilution refrigerators over different cool-downs. In their microwave networks, they were sandwiched between cryogenic attenuators to minimize external impedance mismatches, allowing us to focus on their intrinsic characteristics. Given their low insertion loss ($<$1~$\mathrm{dB}$ up to 12~GHz, Fig.~\ref{fig:devicePic}), in the following text, we will simply present the normalized parametric gain of our CP-JTWPAs by comparing pump-on and pump-off transmissions. For gain extraction near the band gap, where pump-off transmission contains the stopband feature, we constructed the baseline by interpolating the transmission over the band-gap region from adjacent frequencies, which was then used to normalize the pump-on transmission to obtain the gain profile.

\begin{figure*}[!htb]
    \includegraphics[width=18.78cm]{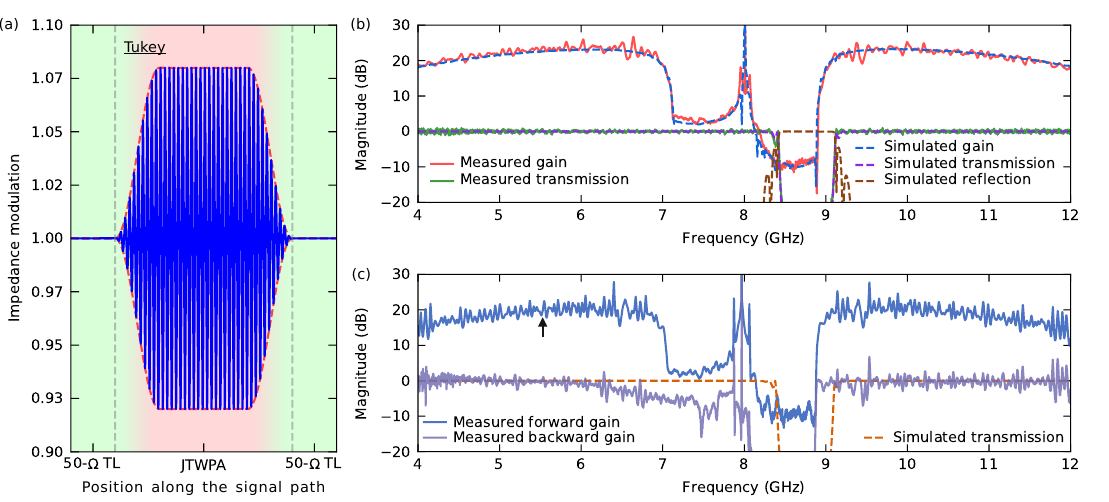}
    \caption{Characterization of device C implementing a windowed periodic impedance modulation with a Tukey window with equal length in taper and maximum-amplitude region ($\alpha = 0.5$). (a)~The impedance modulation profile along the signal path. The modulation amplitude increases smoothly from zero at the input to 8\% in the central uniform section, then decreases symmetrically to zero at the output. (b)~Measured and simulated scattering parameters with 10-dB attenuators at both ports providing 50-$\Omega$ impedance matching. (c)~Measured forward and backward scattering parameters without output attenuation (see Appendix~\ref{sec:appendix_sysAttCal} for the corresponding microwave network), representing a more realistic microwave environment with typical circuit impedance mismatches present in applications. The noise measurement was performed at 5.563~GHz~(arrow) under this configuration.}
    \label{fig:tukeyGain}
\end{figure*}

In Fig.~\ref{fig:windowedModulation_experiment}, we present the scattering characteristics of CP-JTWPAs with boxcar-windowed and Hann-windowed modulation profiles. Our initial measurements of the boxcar device used 3-dB attenuators at both ports, as the impact of impedance mismatches on JTWPA performance was not yet fully appreciated at that stage of development. We progressively increased the installed attenuation to 6~dB~(Hann) and 10~dB~(Tukey) as we observed higher gains from successive samples. While these different attenuation values might initially appear problematic, we demonstrate quantitatively in Appendix~\ref{sec:appendix_impedance_mismatch_effect} that these measurement configurations do not bias the experimental results and that the observed characteristics represent the intrinsic device behavior.

The baseline transmission of the boxcar device exhibits ripples near the band gap that match theoretical predictions for uniformly modulated transmission lines. Our simulations reveal that these ripples are accompanied by substantial reflection magnitudes exhibiting oscillatory behavior. These reflections create two critical challenges for JTWPA performance. First, enhanced reflection near the band-gap complicates optimal pump placement. Although stronger phase correction requires positioning the pump frequency closer to the band gap, increased pump reflection in this region reduces effective pump-power delivery and can generate standing waves with elevated local current amplitudes, potentially saturating certain junctions prematurely. Second, while near-band-gap regions typically provide better phase-matching conditions and the highest gain, they also exhibit the strongest gain ripples, making these desired frequency ranges unavailable for applications sensitive to gain fluctuations.

When pumping the boxcar device, we observed a gain of $<$13~$\mathrm{dB}$ with significant gain ripples near the band gap, agreeing well with the prediction by harmonic balance simulations. In particular, the frequency spacing between adjacent gain ripples closely matches that of the reflection ripples in the linear response, providing strong evidence that gain ripples originate directly from the underlying waveguide's linear characteristics. These variations in gain can be seen in both measured and simulated data, and effectively limit the usable bandwidth for applications requiring uniform gain. As indicated in the enlarged view of the 5\text{--}7-GHz region, gain variations can exceed $\pm5$~$\mathrm{dB}$ in the boxcar device. Our theoretical model in Eq.~\eqref{eq:F_boxcar} further indicates that transmission ripples and consequently gain ripples become denser with increased JTWPA cell count, suggesting that scaling up boxcar-window JTWPAs to achieve higher gain would further increase the severity of this limitation.

Next, we consider the transmission characteristics of the Hann device (lower panels of Fig.~\ref{fig:windowedModulation_experiment}) by measuring it sandwiched between two 6-dB attenuators. While its larger band-gap width reflects the stronger modulation, it lacks the periodic oscillations in its base transmission near the band~gap that were characteristic of the boxcar device. The measured transmission shows only irregular fluctuations, likely arising from fabrication imperfections. Our simulations predict significantly reduced reflections near the band gap for the Hann device with a sharp roll-off to $-20$~$\mathrm{dB}$, in contrast to the slow oscillatory decay observed in the boxcar device. This reduced reflection enables pump placement closer to the band~gap without the detrimental effect of pump reflection that the boxcar device might exhibit, allowing for stronger phase correction. The experimental results confirm this advantage through the achievement of phase-matched gain over a broad frequency range, delivering 16\text{--}18~$\mathrm{dB}$ of gain over an instantaneous bandwidth of 3~GHz. Despite having a higher gain than the boxcar device, as suggested by the simulation, the gain profile of the Hann device exhibits significantly smoother characteristics, particularly near the band~gap where the strong ripples observed in the boxcar device are absent. The gain variations in the Hann device remain within $\pm2$~$\mathrm{dB}$, validating windowed periodic modulation as an effective method to reduce gain ripples.

\section{\label{sec:characterization_tukey} Tukey-window CP-JTWPA}

The Hann window demonstrated excellent suppression of gain ripples and pump reflections through smooth modulation amplitude tapering. However, it reduces the RMS modulation amplitude by nearly half compared to a boxcar window, weakening the overall phase correction and limiting the maximum achievable gain given the same modulation strength. Increasing the modulation strength to compensate for this reduction leads to two significant drawbacks. First, it widens the band gap, reducing valuable bandwidth in the cryogenic microwave network. Second, when designing JTWPAs, we realized that a stronger modulation reduces the tolerance of the device to variations in junction inductance, due to a possible increase in internal reflection when junctions are increasingly inhomogeneous, effectively lowering the fabrication yield. We therefore sought a window function that could balance ripple suppression and phase correction strength while maintaining a good RMS modulation amplitude. 

The Tukey window, a generalized raised-cosine function, provides this balance through its adjustable parameter $\alpha$ that controls the ratio between cosine-tapered and constant-amplitude regions. When $\alpha=0.5$, the window contains equal lengths of cosine-tapered transitions and constant maximum-amplitude section. This configuration achieves an RMS modulation amplitude of 0.83 (relative to uniform modulation), significantly higher than the Hann window's 0.61, enabling stronger phase correction while maintaining effective ripple suppression through smooth transitions at the boundaries.

\subsection{Gain characteristics}
We implemented this approach in device C, with its impedance modulation profile shown in Fig.~\ref{fig:tukeyGain}(a). To study its intrinsic transmission characteristics, we measured the device between 10-dB attenuators. The result is shown in Fig.~\ref{fig:tukeyGain}(b). Simulations predicted minimal ripples in base transmission and faster reflection roll-off away from the band gap compared to the boxcar device. By applying a pump slightly below the band-gap frequency, we achieved a gain of 20\text{--}23~dB over a 5\text{-}GHz bandwidth. This gain significantly exceeded that of the Hann device overall, despite both devices shared similar designs except for their window functions. The Tukey device maintained excellent ripple suppression comparable to the Hann case, showing no strong ripples like those in the boxcar device even at higher gain.

The measurements with attenuators demonstrated the intrinsic gain characteristics of CP-JTWPAs by neglecting the effects of background impedance mismatches. To study its behavior in a more realistic environment, we moved the Tukey device to a modified setup where we reduced the attenuation between the device and input circulator from 10 dB to 7 dB, and completely removed the 10-dB attenuator at the output port, thereby exposing the JTWPA directly to potential impedance mismatches from downstream components such as the output-side circulators~(see Fig.~\ref{fig:tukey_connection}). The measured forward and backward gain are shown in Fig.~\ref{fig:tukeyGain}(c). We note that numerical fitting becomes impractical here due to insufficient knowledge of the frequency-dependent background impedance. The forward gain exhibits irregular ripples characteristic of impedance mismatches, with reduced overall amplitude compared to the impedance-matched case. This reduction stems from pump reflections generating standing waves, which can create local current amplitudes exceeding the junction critical current. This limitation forced us to reduce the input pump power, consequently lowering the achievable gain. Additionally, the gain shows faster roll-off away from the pump frequency, as the combined reflections from background impedance and band gap necessitate placing the pump further from the band-gap, reducing phase-matching effectiveness. Nevertheless, the backward gain remains zero or negative across the useful bandwidth. The small negative backward transmission below the band-gap frequency suggested coupling of backward-propagating waves either to different frequencies or to the forward direction. While the precise mechanism requires further investigation, this coupling behavior could potentially be developed into an integrated approach for backward isolation in JTWPA design.

\begin{figure}[t]    
    \includegraphics[width=8.57cm]{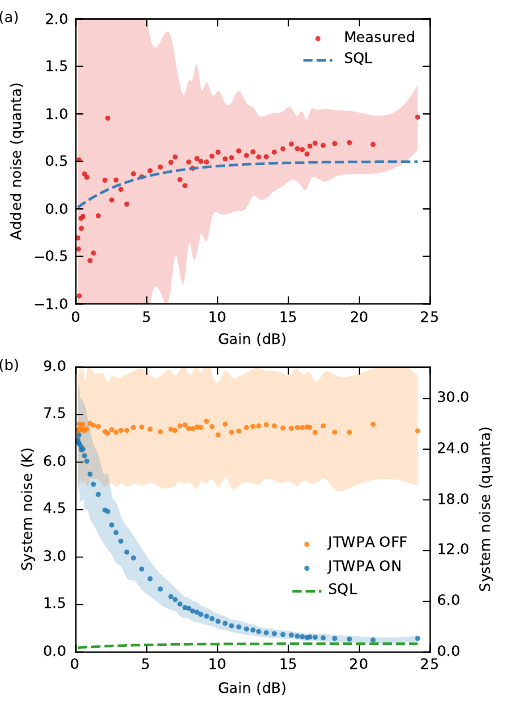}
    \caption{Noise performance of the Tukey CP-JTWPA measured at 5.563~GHz. (a)~Added noise versus gain. The added noise increases from $<$0.5~quanta at gain below 5~dB to 0.68~quanta at 21-dB gain, closely following the trend given by the SQL, before rising significantly at higher gain. (b)~System noise temperature referred to the CP-JTWPA input. With the JTWPA off (orange), the system exhibits a baseline noise of approximately 7.0~K~(26.2~quanta). When the JTWPA is activated (blue), the noise rapidly decreases to below 1.06~K~(3.96~quanta) within the first 10~dB of gain, then gradually approaches 382~mK~(1.43~quanta) at 21-dB gain, before increasing at higher gain. Shaded regions represent statistical measurement uncertainties, with the added noise varying by $\pm$0.31~quanta at high gain. The system noise uncertainty ranges from $\pm$1.72~K~($\pm$~6.42~quanta) in the off state to $\pm$90~mK~($\pm$0.34~quanta) at high gain.}
    \label{fig:jtwpaAddedNoiseVSGain}
\end{figure}

\subsection{Noise performance and saturation}

Next, we characterize the noise performance of the Tukey CP-JTWPA with the configuration in Fig.~\ref{fig:tukeyGain}(c) and microwave network described in Fig.~\ref{fig:tukey_connection}. This requires precise knowledge of power levels at the CP-JTWPA input, which demands accurate calibration of the system gain from room temperature down to the device output port. We performed this calibration at 5.563~GHz using a waveguide-QED approach with a qubit-coupled transmission line as a reference device~\cite{kannanGeneratingSpatiallyEntangled2020}. By analyzing the frequency- and power-dependent transmission of this qubit-coupled line, we established the system input attenuation $A_\mathrm{sys}$ between room temperature and the qubit. To extend this calibration to the CP-JTWPA measurements, we employed two microwave switches to alternately connect either the CP-JTWPA or the qubit-coupled line to identical measurement paths, ensuring equivalent signal paths through cables of matched lengths~(see Appendix~\ref{sec:appendix_sysAttCal} for details of the noise calibration setup and procedures). This configuration allowed us to translate the calibrated attenuation from the qubit reference to the CP-JTWPA. The system gain through the CP-JTWPA was then determined by measuring the overall transmission and compensating for the calibrated input attenuation.

\begin{figure}[t]    
    \includegraphics[width=8.98cm]{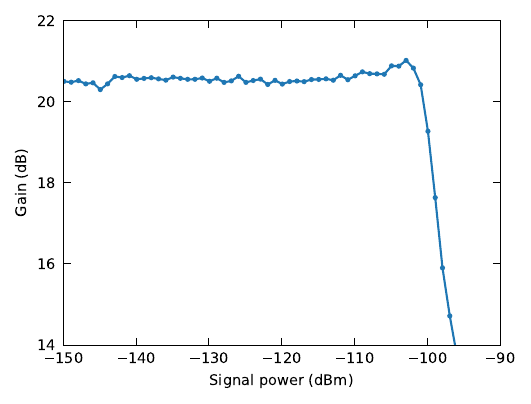}
    \caption{Saturation power of the Tukey CP-JTWPA measured at 20.5~dB gain and 5.563~GHz. The CP-JTWPA exhibits a good gain linearity from $-150$~dBm to $-104$~dBm in input signal power, and saturates at $-99$~dBm.}
    \label{fig:saturation}
\end{figure}

With the calibrated system gain, we measured noise spectra at room temperature with the CP-JTWPA pump turned on and off to determine both the added noise by the CP-JTWPA, $N_\mathrm{JTWPA}$, and the system noise $N_\mathrm{sys}$ referred to the CP-JTWPA input port. As shown in Fig.~\ref{fig:jtwpaAddedNoiseVSGain}, we achieved near-quantum-limited performance at the calibrated frequency with an added noise of 0.68~quanta for gains up to 21~dB, corresponding to a quantum efficiency at $84\%$ of an ideal quantum-noise-limited amplifier~(see Appendix~\ref{sec:appendix_juliaPlots} for details). With this gain configuration, the input-referred system noise temperature decreased to 382~mK (1.43~quanta), relatively low comparing to most previous implementations~\cite{macklinQuantumLimitedJosephson2015,planatPhotonicCrystalJosephsonTravelingWave2020,ranadiveKerrReversalJosephson2022}. 

Using the same calibration, we also extracted the saturation power $P_\mathrm{1dB}$ by performing a probe power sweep and observing at what power $G_\mathrm{JTWPA}$ compresses by 1~dB, shown in Fig.~\ref{fig:saturation}. With a reference gain of 20.5~dB, the CP-JTWPA exhibits excellent linearity in gain throughout a wide range of input signal power from $-150$~dBm to $-104$~dBm. Saturation was measured at $-99$~dBm.

\section{\label{sec:discussion} Conclusion and discussion}
In this work, we have demonstrated a CP-JTWPA architecture that successfully addresses two fundamental challenges in JTWPA implementation: device loss and gain ripple formation in periodic modulations. The achieved insertion loss below 1~dB up to 12~GHz represents a significant advance in superconducting artificial nonlinear transmission lines based on Josephson junctions. This low-loss characteristic not only enables near-quantum-limited noise performance in our amplifier but also opens possibilities for other waveguide-QED experiments where loss is critical, such as wide-band quantum-non-demolition detection of single microwave photons and traveling-wave single-mode squeezers.

Our windowed modulation approach provides a practical solution to phase matching without requiring flux-bias control or additional phase-matching resonators. Our coupled-mode analysis demonstrates the connections between the choice of spatial modulation windows and the resultant transmission spectra, providing insight for designing window functions with desired ripple suppression. While coupled-mode analysis becomes insufficient at high modulation amplitudes, we found that ripple characteristics remain qualitatively similar between weak and strong modulation regimes. This observation suggests that coupled-mode theory, despite its limitations, serves as a computationally efficient tool for preliminary evaluation of window functions before detailed numerical analysis.

Through systematic comparison of window functions, we established that the Tukey window achieves a good balance between phase-matching strength and ripple suppression. The resulting performance, with over 20-dB gain across a 5-GHz bandwidth with ideal impedance matching while maintaining intrinsic gain variations within $\pm 2$~dB, demonstrated the effectiveness of this approach. In a more realistic impedance-mismatched environment, our Tukey device still demonstrated 17-20~dB gain over 4.8~GHz, with added noise below 0.63~quanta and a saturation power of $-99$~dBm.

Several technical aspects warrant further investigation for improving device performance. While the Tukey window significantly reduces ripples, due to its relatively short taper sections, residual transmission variations persist in the linear response. Our studies across multiple devices reveal that these variations often stem from regions of strong reflection near the band gap, which constrain optimal pump placement for phase matching. These reflection features, which can merge and intensify under fabrication variations and nonlinear phase shifts, necessitate operating at frequencies further from the band gap in some devices, limiting both achievable gain and fabrication yield. Future work could explore alternative window functions that combine high RMS values with smooth reflection roll-off, such as cosine or Lanczos windows, to address these limitations. Additionally, the observed negative backward gain suggests potential intrinsic isolation mechanisms, though the underlying coupling processes require further study. While our devices demonstrate excellent performance under ideal impedance matching, practical measurement environments with inevitable mismatches still constrain the maximum achievable gain, highlighting the importance of developing robust solutions for impedance matching in cryogenic microwave networks.

The noise characterization procedure could be improved in future implementations, as current systematic uncertainties arise primarily from package-dependent losses in the calibration translation between the qubit reference and CP-JTWPA, despite our effort in accounting for them. These uncertainties could be eliminated by using fully identical package designs or, more promisingly, by integrating a transmon qubit directly on the CP-JTWPA chip using compatible junction oxidation conditions in our fabrication process. Such integration would provide more reliable noise characterization while demonstrating the technology's readiness for integration with superconducting quantum circuits.

\section*{ACKNOWLEDGEMENT}
The authors acknowledge L. Szikszai for assistance on the development of the fabrication recipe. We also wish to thank K. O'Brien and his team at Massachuset Institute of Technology for sharing their invaluable harmonic balance Julia library. This work was supported in part by the Ministry of Education, Culture, Sports, Science and Technology (MEXT) Quantum Leap Flagship Program (Q-LEAP) (Grant No. JPMXS0118068682) and the JSPS Grant-in-Aid for Scientific Research (KAKENHI) (Grant No. JP22H04937).

\appendix
\numberwithin{equation}{section}  
\section{DEVICE PARAMETERS AND MODULATION CONFIGURATIONS}\label{sec:appendix_device_parameters}
\begin{table}[!htbp]
    \caption{\label{tab:device_params} Reference design parameters and measured characteristics of three CP-JTWPA devices discussed in this work. The measured characteristics correspond to the cases with ideal background impedance matching, except for the values in the parenthesis which represent the imperfect impedance-matching case used in device C's noise measurement where pump power was not estimated.}
    \begin{ruledtabular}
    \begin{tabular}{@{}lcccc@{}}
    \multicolumn{5}{c}{\textbf{Design parameters}} \\
    \colrule
    Device & & A & B & C \\ \hline
    Circuit parameters & & & & \\
    Number of cells (cells) & $N$ & 2400 & 2400 & 2397 \\
    Cell separation ($\mu$m) & $a$ & 20 & 20 & 20 \\
    Stub length\footnotemark[4] ($\mu$m) & $l_{\text{stub}}$ & 80 & 100 & 100 \\
    Shunt capacitance\footnotemark[4] (fF) & $C_\mathrm{g}$ & 33 & 40 & 41 \\
    Junc. inductance\footnotemark[4] (pH) & $L_\mathrm{J}$ & 67 & 94 & 78 \\
    Junc. capacitance\footnotemark[1]\footnotemark[4] (fF) & $C_\mathrm{J}$ & 48.5 & 54 & 54.5 \\
    Geom. inductance\footnotemark[4] (pH) & $L_\mathrm{s}$ & 9.8 & 9.6 & 9.8 \\ \hline
    Periodic modulation & & & & \\
    Type & & Sine & Sine & Sine \\
    Periodicity (cells) & $\Lambda$ & 40 & 30 & 30 \\ 
    Window function & & boxcar & Hann & Tukey\footnotemark[2] \\ \hline
    Modulation amplitudes & & & & \\
    Cell impedance (\%) & $m$ & 5 & 8 & 8 \\
    \colrule
    \noalign{\vskip 1pt}
    \colrule
    \multicolumn{5}{c}{\textbf{Measured characteristics}} \\
    \colrule
    Band gap (no pump) & & & & \\
    Center frequency (GHz) & $\omega_\mathrm{gap}/2\pi$ & 7.90 & 8.13 & 8.75 \\
    Bandwidth (GHz) & $\Delta \omega_\mathrm{gap}/2\pi$ & 0.39 & 0.63 & 0.69 \\ \hline
    Gain performance\footnotemark[3] & & & & \\
    Gain (dB) & & 10--13 & 17--20 & 20--23 \\
     & & & & (17--20) \\
    Bandwidth (GHz) & & 5.1 & 3.2 & 5 \\
     & & & & (4.8) \\
    Low band (GHz) & & 4.2--6.7 & 5.0--6.6 & 4.3--7.0 \\
     & & & & (4.5--7.0) \\
    High band (GHz) & & 7.9--10.5 & 8.3--9.9 & 9.1--11.4 \\
     & & & & (9.0--11.3) \\
    Pump frequency (GHz) & $\omega_\mathrm{p}/2\pi$
 & 7.29 & 7.43 & 8.01 \\
     & & & & (7.97) \\
    Pump power\footnotemark[5] (dBm) & $P_\mathrm{p}$ & $-68.8$ & $-71.4$ & $-70.3$ \\
     & & & & (n/a) \\
    \end{tabular}
    \end{ruledtabular}
\footnotetext[1]{This value represents the intrinsic junction capacitance $C_\mathrm{J}$ estimated from simulation parameters. It is calculated by subtracting the series capacitance contribution from adjacent open stubs (approximately $C_\mathrm{g}/2$) from the total parallel capacitance: $C_\mathrm{J} \approx C_\mathrm{parallel} - C_\mathrm{g}/2$.}
\footnotetext[2]{The alpha for the Tukey window in use is 0.5, which determines the fraction of the window that transitions using a cosine function. With $\alpha=0.5$, half of the window tapers smoothly at the edges (25\% at each end) while the central 50\% maintains constant amplitude identical to a uniform modulation.}
\footnotetext[3]{Refers to the gain performance achieved without impedance mismatch, measured by sandwiching the sample with cryogenic attenuators.}
\footnotetext[4]{Values shown refers roughly to the mean of the modulated parameter arrays.}
\footnotetext[5]{The pump power is estimated by the power applied in fitting the theoretical gain prediction to the gain curve together with other parameters of the CP-JTWPAs.}
\end{table}

The devices differ primarily in their modulation depth and window functions, while sharing similar basic architecture, unit-cell counts, and fabrication recipes. Due to the challenges in direct parameter measurements of JTWPAs, the device parameters presented in this work are determined through a combination of design considerations and experimental validation. The stub capacitances are initially estimated from finite-element electromagnetic simulations, the junction inductances from normal-state resistance measurements, and the junction capacitances from device geometry assuming 50~fF/$\mu$m$^2$. These parameters are then refined by fitting simulated transmission and gain characteristics to experimental measurements.

To achieve sinusoidal impedance modulation with amplitude $m$, we coordinate variations in both junction area and stub length. The characteristic impedance is given by $Z = \sqrt{(L_\mathrm{J} + L_\mathrm{s})/C_\mathrm{g}}$, where $L_\mathrm{s}$ is almost constant over a wide range of stub length. To achieve an 8\% impedance modulation through stub length alone would require roughly 16\% variation in $C_\mathrm{g}$, leading to large geometric changes. Instead, we apply a correlated modulation on both junction area (varying $L_\mathrm{J}$) and stub length (varying $C_\mathrm{g}$), such that their combined effect produces the target impedance profile $Z(z) = Z_0[1 + m \cdot W(z) \sin(2\pi z/\Lambda)]$. This reduces the required variation in the stub length to approximately 8\% while maintaining the same overall impedance-modulation amplitude.
\section{COUPLED-MODE ANALYSIS OF WINDOWED PERIODIC MODULATION}\label{sec:appendix_coupled_mode}

Consider a microwave transmission line of length $l$ composed of a chain of unit cells. Each unit cell contains a series inductor with constant inductance $L$ and a shunt capacitor with position-dependent capacitance $C(z)$, where $z$ is the position along the transmission line ($0 \le z \le l$). While our actual device implementation modulates both inductance and capacitance in a correlated manner, for analytical tractability we focus our theoretical analysis on capacitance modulation only. This simplification is sufficient to demonstrate the fundamental connection between window functions and transmission ripples, while avoiding the non-standard Helmholtz equation that would arise from simultaneous modulation of both parameters. 

The characteristic impedance of this transmission line in the unmodulated case is
\begin{equation}
    Z_0 = \sqrt{\frac{L}{C_0}}, \label{eq:Z_0}
\end{equation}
where $C_0$ is the mean capacitance.

To achieve the desired impedance modulation, we introduce a periodic modulation of the capacitance along the transmission line,
\begin{equation}
    C(z) = C_0 [1 + m W(z) \cos(2\beta_\mathrm{B} z)], \label{eq:modulated_capacitance}
\end{equation}
where $z$ is the position along the transmission line $(0 \leq z \leq l)$, $m$ is the modulation depth $(0 < m < 1)$, $\beta_\mathrm{B} = \pi/\Lambda$ is the Bragg wave vector with $\Lambda$ being the period of the modulation, and $W(z)$ is the window function. The resulting modulated impedance can be expressed as
\begin{equation}
    Z(z) = \sqrt{\frac{L}{C(z)}} \approx Z_0 [1 - \frac{1}{2}m W(z) \cos(2\beta_\mathrm{B} z)].
\end{equation}
This approximation holds for small modulation depths~$(m \ll 1)$.

The window function $W(z)$ is defined over the length $l$ of the waveguide,
\begin{equation}
    W(z) = \begin{cases}
        f(z), & 0 \leq z \leq l \\
        0, & \text{otherwise}
    \end{cases},
\end{equation}
where $f(z) = 1$ for boxcar window and $f(z) = \sin^2(\pi z/l)$ for Hann window.

Starting with the telegrapher's equations for a lossless transmission line with constant inductance and variable capacitance,
\begin{align}
    \frac{\partial v}{\partial z} &= -L\frac{\partial i}{\partial t}, \label{eq:telegrapher_1}\\
    \frac{\partial i}{\partial z} &= -C(z)\frac{\partial v}{\partial t}, \label{eq:telegrapher_2}
\end{align}
we look for solutions of the form
\begin{align}
v(z,t) &= V(z)e^{j\omega t}, \label{eq:voltage_time}\\
i(z,t) &= I(z)e^{j\omega t}, \label{eq:current_time}
\end{align}
where $V(z)$ and $I(z)$ are slowly-varying envelope functions. Substituting Eqs.~\eqref{eq:voltage_time} and \eqref{eq:current_time} into Eqs.~\eqref{eq:telegrapher_1} and \eqref{eq:telegrapher_2}, and eliminating $I(z)$, we obtain the wave equation
\begin{equation}
\frac{d^2V(z)}{dz^2} + \beta_0^2 [1 + m W(z) \cos(2\beta_\mathrm{B} z)] V(z) = 0, \label{eq:wave_equation}
\end{equation}
where $\beta_0^2 = \omega^2 LC_0$.
We express the voltage as a sum of forward and backward propagating waves,
\begin{equation}
    V(z) = V^+(z)e^{-j\beta_0 z} + V^-(z)e^{j\beta_0 z}, \label{eq:voltage_decomposition}
\end{equation}
where $V^+(z)$ and $V^-(z)$ represent the slowly-varying envelopes of the forward and backward waves. Substituting Eq.~\eqref{eq:voltage_decomposition} into the wave equation~\eqref{eq:wave_equation} and applying the slowly-varying envelope approximation, where we neglect the second derivatives of $V^+(z)$ and $V^-(z)$, we obtain the coupled-mode equations
\begin{align}
    \frac{dV^+(z)}{dz} &= j\kappa W(z) V^-(z) e^{j2\Delta_{\beta} z} \label{eq:coupled_mode_1},\\
    \frac{dV^-(z)}{dz} &= -j\kappa W(z) V^+(z) e^{-j2\Delta_{\beta} z}, \label{eq:coupled_mode_2}
\end{align}
where $\kappa = \beta_0 m/4$ is the coupling coefficient between the forward and backward propagating waves, and $\Delta_{\beta} = \beta_0 - \beta_\mathrm{B}$ is the detuning parameter. To obtain physically intuitive solutions that reveal the roles of window function and coupling strength, we employ a transfer-matrix formalism, which connects the voltage wave amplitudes at different points in a passive medium. Given the linear nature of Eqs.~\eqref{eq:coupled_mode_1} and \eqref{eq:coupled_mode_2}, we expect the field amplitudes at any position $z$ to be linearly related to those at position $z_0$ through a 2$\times$2 transfer matrix $T(z_0,z)$,
\begin{equation}
    \begin{pmatrix} V^+(z) \\ V^-(z) \end{pmatrix} = T(z_0,z)
    \begin{pmatrix} V^+(z_0) \\ V^-(z_0) \end{pmatrix}. \label{eq:transfer_matrix}
\end{equation}
Differentiating with respect to $z$ and using the coupled-mode equations \eqref{eq:coupled_mode_1} and \eqref{eq:coupled_mode_2}, we obtain
\begin{equation}
    \begin{split}
        &\begin{pmatrix} 
        0 & j\kappa W(z)e^{j2\Delta_{\beta} z} \\
        -j\kappa W(z)e^{-j2\Delta_{\beta} z} & 0
        \end{pmatrix}
        T(z_0,z)
        \begin{pmatrix} V^+(z_0) \\ V^-(z_0) \end{pmatrix} \\
        & = \frac{d}{dz}[T(z_0,z)]
        \begin{pmatrix} V^+(z_0) \\ V^-(z_0) \end{pmatrix}. \label{eq:substituted}
    \end{split}
\end{equation}
Since Eq.~\eqref{eq:substituted} must hold for any initial values $V^+(z_0)$ and $V^-(z_0)$, we obtain a differential equation for the transfer matrix,
\begin{equation}
    \frac{d}{dz}[T(z_0,z)] = A(z)T(z_0,z), \label{eq:diff_eq_transfer}
\end{equation}
where
\begin{equation}
    A(z) = \begin{pmatrix} 
    0 & j\kappa W(z)e^{j2\Delta_{\beta} z} \\
    -j\kappa W(z)e^{-j2\Delta_{\beta} z} & 0
    \end{pmatrix}. \label{eq:A_matrix}
\end{equation}
The solution to this matrix differential equation can be expressed using the Magnus expansion,
\begin{equation}
    T(z_0,z) = \exp(\Omega(z_0,z)), \label{eq:magnus_expansion}
\end{equation}
where $\Omega(z_0,z) = \Omega_1(z_0,z) + \Omega_2(z_0,z) + \Omega_3(z_0,z) + \cdots$, with
\begin{align}
    \Omega_1(z_0,z) &= \int_{z_0}^z A(z') dz', \label{eq:omega_1} \\
    \Omega_2(z_0,z) &= \frac{1}{2} \int_{z_0}^z \int_{z_0}^{z'} [A(z'), A(z'')] dz'' dz'. \label{eq:omega_2}
\end{align}
The commutator evaluates to
\begin{equation}
    [A(z'), A(z'')] = 2j\kappa^2 W(z')W(z'') \begin{pmatrix} 
    \sin\phi & 0 \\
    0 & -\sin\phi
    \end{pmatrix},
    \label{eq:commutator}
\end{equation}
where $\phi = 2\Delta_{\beta}(z'-z'')$.
This non-vanishing commutator, proportional to $\kappa^2$, indicates that higher-order terms in the Magnus expansion become significant for larger modulation amplitude. In order to obtain analytically tractable solutions that provide physical insight, we focus on the weak coupling regime ($\kappa \ll 1$) where the first-order approximation is sufficient:
\begin{equation}
    T(z_0,z) \approx \exp(\Omega_1(z_0,z)). \label{eq:first_order_approx}
\end{equation}
Evaluating $\Omega_1(z_0,z)$,
\begin{equation}
    \Omega_1(z_0,z) = 
    \begin{pmatrix} 
    0 & j\kappa F(z_0,z) \\
    -j\kappa F^*(z_0,z) & 0
    \end{pmatrix}, \label{eq:omega_1_eval}
\end{equation}
where
\begin{equation}
    F(z_0,z) = \int_{z_0}^z W(z') e^{j2\Delta_{\beta} z'} dz'. \label{eq:F_function}
\end{equation}
The matrix exponential of this form yields
\begin{equation}
    T(z_0,z) \approx \begin{pmatrix}
    \cosh(\kappa |F|) & j\frac{F}{|F|}\sinh(\kappa |F|) \\
    j\frac{F^*}{|F|}\sinh(\kappa |F|) & \cosh(\kappa |F|)
    \end{pmatrix}, \label{eq:T_first_order}
\end{equation}
where $F$ is shorthand for $F(z_0,z)$.

To analyze the transmission and reflection characteristics of the modulated line, we consider a wave incident from the left ($z_0=0$) with no incident wave from the right ($z=l$) with a matched load. This gives us the boundary conditions
\begin{equation}
    V^+(0) = 1, \quad V^-(l) = 0. \label{eq:boundary_conditions}
\end{equation}
The transmission coefficient $t$ is defined as the ratio of transmitted to incident wave amplitudes, $t = V^+(l)/V^+(0)$. Using the transfer matrix relation,
\begin{equation}
    \begin{pmatrix} V^+(l) \\ 0 \end{pmatrix} = T(0,l)
    \begin{pmatrix} 1 \\ V^-(0) \end{pmatrix}, \label{eq:boundary_matrix}
\end{equation}
we can solve for $V^-(0)$ from the second row:
\begin{equation}
    V^-(0) = -\frac{T_{21}(0,l)}{T_{22}(0,l)}. \label{eq:backward_wave}
\end{equation}
Substituting this back into the first row to find $V^+(l)$:
\begin{equation}
    V^+(l) = T_{11}(0,l) - \frac{T_{12}(0,l)T_{21}(0,l)}{T_{22}(0,l)}. \label{eq:forward_wave}
\end{equation}
Using Eq.~\eqref{eq:T_first_order}, we obtain an expression for the transmission coefficient:
\begin{equation}
    t \approx \frac{V^+(l)}{V^+(0)} = \frac{1}{\cosh(\kappa|F(0,l)|)}. \label{eq:transmission_coeff}
\end{equation}
Following a similar procedure, the reflection coefficient $r = V^-(0)/V^+(0)$ can be shown to be:
\begin{equation}
    r \approx j\frac{F^*(0,l)}{|F(0,l)|}\tanh(\kappa |F(0,l)|). \label{eq:reflection_coeff}
\end{equation}

These coefficients depend on the form of $F(0,l)$, which is determined by the choice of window function $W(z)$. For the boxcar window (uniform modulation),
\begin{equation}
    W_{\text{boxcar}}(z) = 1, \quad 0 \leq z \leq l, \label{eq:W_boxcar}
\end{equation}
which yields
\begin{equation}
    F_{\text{boxcar}}(0,l) = \frac{e^{j2\Delta_{\beta} l} - 1}{j2\Delta_{\beta}}. \label{eq:F_boxcar}
\end{equation}\\
For the Hann window (smooth tapering at the edges),
\begin{equation}
    W_\mathrm{Hann}(z) = \sin^2(\pi z/l) = \frac{1}{2}(1 - \cos(2\pi z/l)), \quad 0 \leq z \leq l, \label{eq:W_hann}
\end{equation}
which yields
\begin{align}
    F_\mathrm{Hann}(0,l) &= \int_0^l W_\mathrm{Hann}(z)e^{j2\Delta_{\beta} z}dz \nonumber \\
    &= \frac{1}{2}\int_0^l e^{j2\Delta_{\beta} z}dz - \frac{1}{2}\int_0^l \cos(2\pi z/l)e^{j2\Delta_{\beta} z}dz \nonumber \\
    &= \frac{1}{2}\frac{e^{j2\Delta_{\beta} l} - 1}{j2\Delta_{\beta}} - \nonumber \\
    &\quad \frac{1}{4}\left[\frac{e^{j(2\Delta_{\beta} + 2\pi/l)l} - 1}{j(2\Delta_{\beta} + 2\pi/l)} + \frac{e^{j(2\Delta_{\beta} - 2\pi/l)l} - 1}{j(2\Delta_{\beta} - 2\pi/l)}\right]. \label{eq:F_hann}
\end{align}

\section{NUMERICAL METHODS FOR LINEAR 1D ARTIFICIAL TRANSMISSION LINE}\label{sec:appendix_numerical_analysis_1dTL}
\subsection{ABCD-matrix analysis}
While the coupled-mode analysis in Appendix~\ref{sec:appendix_coupled_mode} provides valuable insights into the suppression of transmission ripples through windowing, it is limited to weak modulation ($m \ll 1$) and analytically tractable window functions. To analyze transmission lines with arbitrary modulation strength and window profiles, we employ the ABCD-matrix method (also known as transmission-matrix method). This approach can be utilized to study 1D linear transmission lines by dividing the line into small segments, each treated as a two-port network. For a lossless line composing of series inductance and shunt capacitance, the ABCD-matrix transforms the voltage and current over an infinitesimal length $\Delta z$ as~\cite{MicrowaveEngineering4th}
\begin{equation}
    \begin{pmatrix}
        V(z+\Delta z) \\
        I(z+\Delta z)
    \end{pmatrix} = 
    \begin{pmatrix}
        1 & Z(z) \\
        Y(z) & 1
    \end{pmatrix}
    \begin{pmatrix}
        V(z) \\
        I(z)
    \end{pmatrix}, \label{eq:ABCD_segment}
\end{equation}
where $Z(z) = j\omega L(z)\Delta z$ and $Y(z) = j\omega C(z)\Delta z$, with position-dependent series inductance $L$ and shunt capacitance $C$. For a lumped-element line with unit cell length $a$, we have $\Delta z = a$ and $Z(z), Y(z)$ are given by the discrete elements $Z_n = j\omega L_n, Y_n = j\omega C_n$ respectively.
\begin{equation}
    \begin{pmatrix}
        V_{n+1} \\
        I_{n+1}
    \end{pmatrix} = 
    \begin{pmatrix}
        1 & Z_n \\
        Y_n & 1
    \end{pmatrix}
    \begin{pmatrix}
        V_n \\
        I_n
    \end{pmatrix}. \label{eq:ABCD_segment_discrete}
\end{equation}
The discrete ABCD-matrix formalism extends naturally to lumped-element circuits, allowing us to model complex structures such as open-stub capacitances and Josephson-junction inductances in our JTWPA line. For a line with $N$ unit cells, the cascaded matrices relate input and output as
\begin{equation}
    \begin{pmatrix}
        V_N \\
        I_N
    \end{pmatrix} = 
    \begin{pmatrix}
        A & B \\
        C & D
    \end{pmatrix}_{\!\mathrm{total}}
    \begin{pmatrix}
        V_0 \\
        I_0
    \end{pmatrix} = 
    \prod_{n=N}^1
    \begin{pmatrix}
        1 & Z_n \\
        Y_n & 1
    \end{pmatrix}
    \begin{pmatrix}
        V_0 \\
        I_0
    \end{pmatrix}.
\end{equation}
The total ABCD matrix can be converted to scattering parameters:
\begin{align}
    S_{11} &= \frac{A + B/Z_0 - CZ_0 - D}{A + B/Z_0 + CZ_0 + D} \label{eq:S11_ABCD} \\
    S_{21} &= \frac{2(AD-BC)}{A + B/Z_0 + CZ_0 + D} \label{eq:S21_ABCD}
\end{align}
where $Z_0$ is the characteristic impedance of the input and output ports. This numerical method allows us to analyze transmission lines with arbitrary window functions and modulation depths beyond the weak-modulation regime.

In a more complex 1D microwave circuit such as the lumped-element model we used for our CP-JTWPA in Fig.~\ref{fig:devicePic}, $Z_n$ and $Y_n$ terms in Eq.~\eqref{eq:ABCD_segment_discrete} can be replaced by more complex expressions to include the effect of junction capacitance, and additional matrices can be inserted to account for stray geometric inductance in series with the junctions. 

\subsection{Example parameters for comparing analytical and numerical calculations}
To compare our coupled-mode analysis to the ABCD-matrix approach, we use the following test parameters which roughly resembles device B with a reduced modulation depth:
\begin{align*}
    l &= 48\text{ mm} && \text{(total length)} \nonumber \\
    L &= 5.3\text{ pH/}\mu\text{m} && \text{(series inductance per unit length)} \nonumber \\
    C &= 1.9\text{ fF/}\mu\text{m} && \text{(shunt capacitance per unit length)} \nonumber \\
    m &= 0.03 && \text{(modulation depth)} \nonumber \\
    \Lambda &= 600\text{ }\mu\text{m} && \text{(modulation period)}
\end{align*}

These parameters result in a characteristic impedance $Z_0 = \sqrt{L/C} \approx 53\text{ }\Omega$ and a Bragg frequency $\omega_\mathrm{B}/2\pi \approx 8.3\text{ GHz}$. The transmission line of length 48~mm is divided into $N = 50000$ segments for numerical accuracy. The corresponding results are shown in Fig.~\ref{fig:windowedModulation_theory}(c,d).
\section{GAIN SIMULATIONS AND QUANTUM EFFICIENCY ANALYSIS OF CP-JTWPAS}\label{sec:appendix_juliaPlots}

\begin{figure*}[t]
    \includegraphics[width=18.24cm]{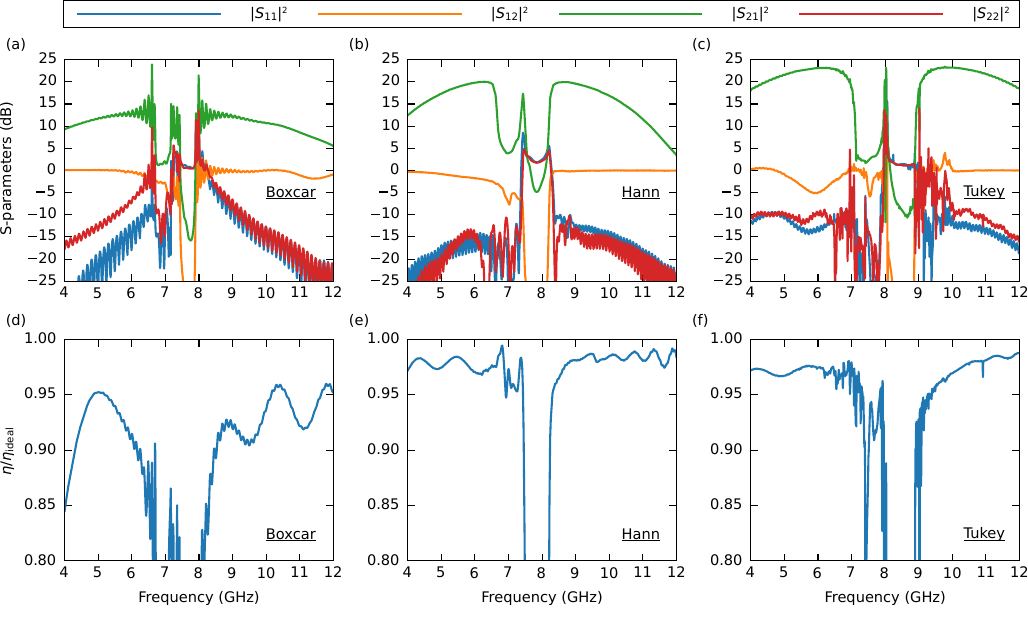}
    \caption{Harmonic balance simulations of the CP-JTWPA with different window functions. (a-c) Simulated scattering parameters for the three CP-JTWPA designs: (a) boxcar window~(device~A), (b) Hann window~(device~B), and (c) Tukey window~(device~C). All S-parameters ($|S_{11}|^2$, $|S_{12}|^2$, $|S_{21}|^2$, and $|S_{22}|^2$) are shown over the 4–12 GHz frequency range with configuration given in Appendix~\ref{sec:appendix_device_parameters}. The high-gain regions in $|S_{21}|^2$ (green) correspond to the phase-matched parametric amplification bands, while the band-gap regions appear at approximately 7.9 GHz, 8.1 GHz, and 8.8 GHz for the three devices, respectively. (d-f) Corresponding quantum efficiency ratios ($\eta/\eta_\mathrm{ideal}$) for each device, showing the deviation from ideal JTWPA behavior. The ratio approaches unity in most of the operating bandwidth but drops significantly within and near the band-gap frequencies.}
    \label{fig:juliaPlots}
\end{figure*}

\subsection{Gain simulations}

We simulate the gain of our CP-JTWPAs using harmonic-balance calculations with the JosephsonCircuits.jl library~\cite{obrienKpobrienJosephsonCircuitsjl2025} and the parameters given in Appendix~\ref{sec:appendix_device_parameters}, assuming perfect impedance matching at both ends. We swept with 10-MHz steps over the 4--12-GHz range to capture potential ripples in the S-parameters. To obtain sufficient accuracy in our nonlinear analysis, we used the harmonic-balance method with 20 harmonics of the pump frequency $\omega_\mathrm{p}$ to solve the nonlinear circuit response, and 10 harmonics in the signal-idler space for the linearized small-signal analysis. In our 4WM implementation, we included pump harmonics $n\omega_\mathrm{p}$ for $n = 1,3,5,..,19$ with odd harmonics dominating the mixing process, and signal/idler mixing products of the form $\omega_\mathrm{s,i}\pm2n\omega_\mathrm{p}$ for $n = 0,1,2,3,4$. This approach captured 4WM processes up to the 8th-order sideband, with the highest-frequency side bands showing gain as low as $-80$~dB compared to the input signal power, sufficiently negligible for accurate simulation results.

Shown in Fig.~\ref{fig:juliaPlots}, the $|S_{21}|^2$ parameters from these simulations were used for fitting experimental data in the main text. The simulations predict significant reflected gain~($|S_{11}|^2$ and $|S_{22}|^2$) near the band gap in the boxcar device, while reflections remain low~($<$10~dB) for the Hann and Tukey devices due to the suppression of reflections around the band gap by the windowed modulations. The backward gain $|S_{12}|^2$ generally remains below 0~dB in all simulations, with a small frequency range showing approximately 1~dB gain over the low-frequency region in the Tukey device with this pump configuration. The mechanism behind negative backward gain regions remains unclear. It is worth further investigations as it potentially allows a non-reciprocal amplification.

It should be noted that for the Tukey window case, we encountered convergence challenges in the simulation, particularly near the band-gap region. The sharp ripples visible in these regions in other S-parameters than $|S_{21}|^2$ may therefore be numerical artifacts rather than representing the actual device response. This convergence issue likely stems from two factors: first, the shorter taper length in the Tukey window compared to the Hann window may retain some abrupt transitions in the impedance profile; second, we positioned the pump frequency very close to the band gap where residual reflection ripples still exist. While residual ripples were also present in the boxcar case, we did not encounter similar convergence issues because the pump was necessarily placed further from the band-gap due to the slow decay of reflections in that design.

\subsection{Quantum efficiency analysis}
Regarding quantum efficiency $\eta$, our simulations predict that the amplifiers can theoretically reach over 90\% of the ideal quantum efficiency $\eta_\mathrm{ideal}$. This high performance can be attributed to the use of relatively weak pump power compared to the junction critical current, which reduced interactions between the signal and higher-frequency modes. The Hann and Tukey devices demonstrate higher overall $\eta$ than the boxcar device for two reasons. First, reflections around the band gap rapidly deteriorate $\eta$ in the boxcar device. Second, the additional parallel capacitance from larger $C_\mathrm{g}$ in the Hann and Tukey devices creates stronger curvature in their dispersion relations, further suppressing interactions between the signal mode and higher-frequency modes.

In experiments, with the measured gain $G$ and added noise $A$ of an amplifier, its $\eta$ can be calculated as~\cite{boutinEffectHigherOrderNonlinearities2017}
\begin{equation}
    \eta(A) = \frac{1}{1+2A},
\end{equation}
and it can be compared to that of an ideal quantum-noise-limited amplifier, given by
\begin{equation}
    \eta_\mathrm{ideal}(G) = \frac{1}{2-1/G}.
\end{equation}
The ratio $\eta/\eta_\mathrm{ideal}$ can serve as a figure of merit for near-quantum-noise-limited amplifiers.
\section{EFFECT OF IMPEDANCE MISMATCH ON JTWPA AND THE USE OF ATTENUATORS}\label{sec:appendix_impedance_mismatch_effect}

\begin{figure*}[t]
    \includegraphics[width=16.75cm]{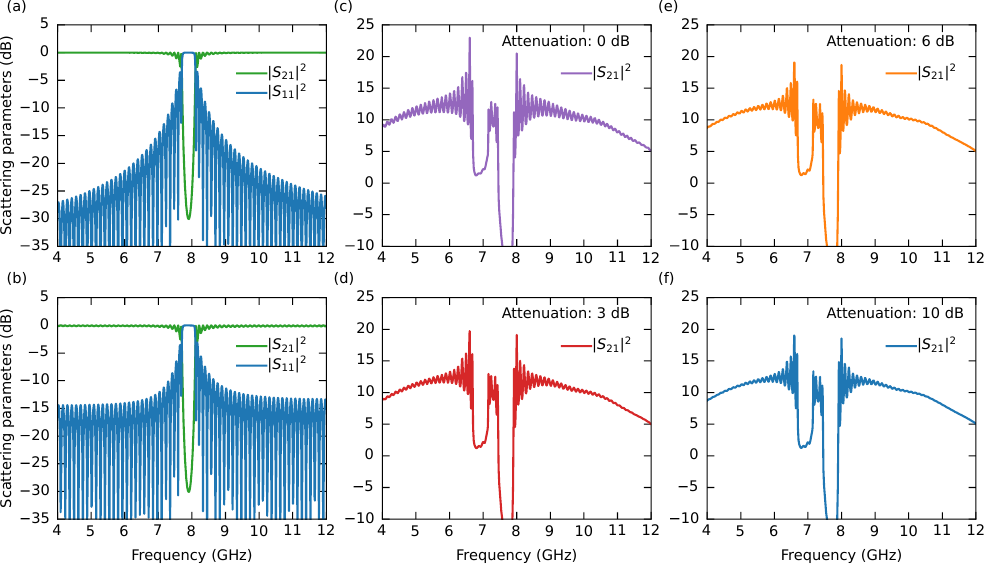}
    \caption{Harmonic-balance simulations of the boxcar CP-JTWPA under different impedance matching and attenuation conditions. (a,b)~Baseline-transmission characteristics with pump off (pump power below $-110$~dBm): (a)~ideal 50-$\Omega$ port impedance and (b)~mismatched 40-$\Omega$ port impedance yielding $|S_{11}|^2 = -15$~dB reflection. (c--f) Gain profiles with the CP-JTWPA sandwiched between attenuators each providing: (c)~0~dB, (d)~3~dB, (e)~6~dB, and (f)~10~dB attenuation.}
    \label{fig:mismatchCompare}
\end{figure*}

Throughout this work, when characterizing the intrinsic gain of the CP-JTWPAs, cryogenic attenuators were used to sandwich the JTWPA packages. These attenuators serve to isolate the JTWPAs from external impedance mismatches, with attenuation values chosen to minimize signal-to-noise ratio degradation while providing adequate isolation, allowing the measurements to accurately reflect the gain profiles of the individual designs. We progressively increased the installed attenuation from 3~dB (boxcar) to 6~dB (Hann) and finally 10~dB (Tukey), as we observed higher gains from successive samples and recognized that stronger attenuation was needed. While these different attenuation values might initially appear to compromise the fairness of device comparisons, the following analysis demonstrates that the observed ripple characteristics represent intrinsic device behavior rather than measurement artifacts.

\subsection{Origin of gain-profile ripples and attenuation required for their suppression}

The role of attenuators is to suppress external reflections that could create measurement artifacts, allowing us to observe the device's intrinsic gain behavior. An $N$-dB attenuator attached to one port of the CP-JTWPA provides a minimum return loss of $2N\!$~dB, since signals must traverse the attenuator twice when reflected by external components (assuming the attenuator itself has excellent return loss, typically $>$25~dB). This means that a device sandwiched between two $N$-dB attenuators experiences a minimum round-trip loss of $4N\!$~dB from the attenuators alone. When combined with the return losses $R$ from external components (cables and circulators typically providing 15\text{--}20~dB each), the total round-trip loss becomes $4N\!+2R\!$~dB.

Gain ripples arise when impedance mismatches at the JTWPA boundaries cause signals to undergo multiple round trips, with each forward pass through the device providing additional amplification. In our near-lossless devices, if the total round-trip loss approaches or falls below the single-pass gain, signals can bounce repeatedly between the input and output terminations, accumulating amplified power over multiple traversals. From the receiver's perspective, this results in enhanced power accumulation over the measurement integration window, manifesting as gain ripples when sweeping across frequency.

For our boxcar device sandwiched between 3-dB attenuators, the minimum round-trip loss from the attenuators alone is 12~dB. Adding the return losses from external components, the total round-trip loss typically exceeds 40~dB, significantly exceeding the 10\text{--}13-dB gain of our boxcar CP-JTWPA, which should adequately suppress ripples from external reflections. The observation of pronounced gain ripples despite this suppression provides strong evidence for significant internal reflections within the device itself, consistent with our theoretical predictions for periodic-modulation effects.

\subsection{Simulated gain profiles of boxcar CP-JTWPA under impedance mismatch}

To validate the attenuation levels needed for the boxcar device to observe its intrinsic gain profile, we simulated its baseline transmission and gain under different impedance-matching and pump configurations as shown in Fig.~\ref{fig:mismatchCompare}. Accurately modeling frequency-dependent impedance mismatches in components connecting to JTWPAs requires precise knowledge of their scattering parameters at cryogenic temperatures, which are difficult to obtain and demand extensive specialized measurements. Instead, we model impedance mismatch in our simulations by assuming a constant port impedance of 40~$\Omega$. This impedance value produces a return loss of 15~dB, representing a realistic worst-case scenario that exceeds the typical return loss specifications for our cryogenic circulators and SMA cables.

We then compare the boxcar CP-JTWPA gain profiles under different sandwiching attenuation values with this 40-$\Omega$ port impedance mismatch. Using the ideal impedance-matched transmission $|S_{21}|^2$ from Fig.~\ref{fig:juliaPlots} as reference, the 0-dB case (no added attenuators) shows some residual ripples from external reflections, despite the ~30-dB total round-trip loss from external components alone. With 3-dB attenuators, these residual external ripples are effectively suppressed, and the gain profiles converge to levels within $\pm$0.5~dB of the 6-dB- and 10-dB-attenuator cases. Importantly, stronger attenuators do not alter the intrinsic gain profiles, as demonstrated by the nearly identical curves between 6-dB and 10-dB sandwiching attenuators. This analysis confirms that 3-dB attenuators provide sufficient isolation for the boxcar CP-JTWPA to reveal its intrinsic characteristics without introducing detectable bias in our device comparisons.
\section{MONTE CARLO ANALYSIS OF JTWPA GAIN UNDER JUNCTION CRITICAL-CURRENT VARIATIONS}\label{sec:appendix_mc_dual_vs_mono}

\begin{figure}[t]
    \includegraphics[width=8.45cm]{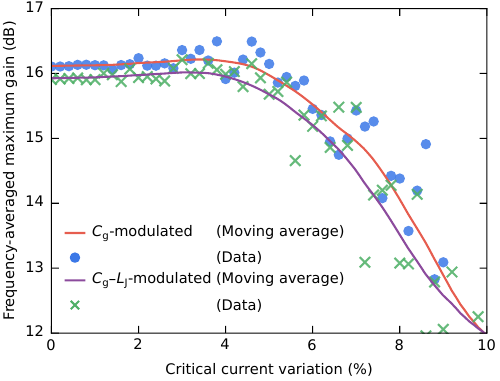}
    \caption{Performance comparison of capacitive-only ($C_\mathrm{G}$) and capacitive--inductive ($C_\mathrm{g}$--$L_\mathrm{J}$) modulated JTWPAs. The frequency-averaged maximum gain is plotted against the junction critical-current variation in standard-deviation percentage. Individual simulation points (dots) and their moving averages over twenty successive points (solid lines) show gain response over 0--10\% variation, showing a similar gain degradation in both modulation schemes.}
    \label{fig:mc_gain_vs_varIc}
\end{figure}

Our devices implement the dual-parameter modulation scheme described in Section~\ref{sec:device_design} and Appendix~\ref{sec:appendix_device_parameters}, where both junction area and stub length are varied in a correlated manner to achieve the required 8\% impedance modulation. While this approach reduces the required stub-length variation from 16\% to 8\% compared to capacitive-only modulation, it intentionally introduces variations in junction critical current along the transmission line. Since uniform junction parameters are typically desired in JTWPAs to minimize performance degradation from fabrication variations, we performed Monte Carlo simulations to verify that this intentional non-uniformity does not compromise the device's robustness against additional fabrication-induced junction variations.

We compared two approaches, both achieving 8\% impedance modulation based on a design similar to the Hann-window device given in Appendix~\ref{sec:appendix_device_parameters}. The first uses single-parameter modulation with 16\% stub-length variation (varying $C_\mathrm{G}$) alone, while the second implements our dual-parameter modulation scheme simultaneously varying both stub length and junction area (varying $C_\mathrm{g}$, $L_\mathrm{J}$) by 8\% in a correlated manner.

As shown in Fig.~\ref{fig:mc_gain_vs_varIc}, we analyzed gain degradation through harmonic-balance simulations by introducing normally distributed variations in the junction critical current. For each variation level (0\% to 10\% in 0.2\% steps), we performed 5--10 Monte Carlo iterations. In each iteration, we randomly varied the junction parameters according to the specified distribution and optimized the pump power ($-78$ to $-69$~dBm) through binary search to achieve the maximum gain. The gain was evaluated by averaging over the 5\text{--}6-GHz signal band with a 7.427-GHz pump frequency. The small gain difference between the two schemes arises from a slight shift in band-gap frequency due to the different modulation approaches when geometric inductance is considered.

Comparing the two impedance-modulation schemes, besides a slight offset in the average gain between the models, we found no significant difference in how JTWPA gain decreased with increasing junction critical-current variation. Both models showed similar robustness, with gain starting to decrease around 5\% variation for our designs, and reaching a 3-dB drop at roughly 8\% variation.
\section{SYSTEM-TRANSMISSION CALIBRATION AND NOISE CALCULATION}\label{sec:appendix_sysAttCal}

\subsection{Microwave network for the Tukey CP-JTWPA noise and backward gain characterization}

\begin{figure}[t]
   \includegraphics[width=8.54cm]{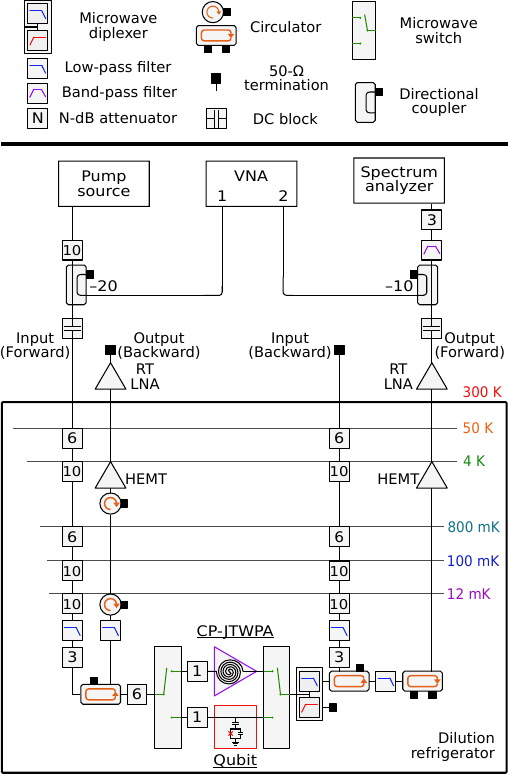}
   \caption{Measurement setup used in the noise and backward gain characterization of the CP-JTWPA. The device is cooled to 12 mK in a dilution refrigerator with microwave switches allowing alternating measurements between the JTWPA and the reference qubit-coupled transmission line. The room temperature equipment configuration shown here is arranged for forward transmission and noise measurements. For backward gain measurements, the VNA is reconfigured to connect to the backward input and output ports while keeping the pump turned on.}
   \label{fig:tukey_connection}
\end{figure}

The Tukey CP-JTWPA is characterized using a bi-directional microwave circuit that enables measurements in both forward- and backward-transmission directions. A vector network analyzer (VNA; Keysight N5232A) serves as the probe signal source for both transmission and noise measurements. The probe signal is combined with a pump signal generated by a microwave source (Rohde \& Schwarz SGS100A) at room temperature via a directional coupler before entering the dilution refrigerator. Along the forward path, signals propagate through a series of cryogenic attenuators, a low-pass filter (RLC Electronics F-30-12.4-R), and a cryogenic double-junction circulator~(QCY-G0401202AM) before reaching either the CP-JTWPA or the qubit-coupled transmission line, as determined by the states of two microwave switches (Radiall R572433000). The forward signals then pass through a microwave diplexer (Marki DPX-1114) installed to dissipate any higher-frequency signals generated by the JTWPA. The forward probe signal subsequently travels through an amplification chain comprising circulators, a filter, and a HEMT amplifier (LNF LNC-4-16B) back to room temperature, where it is further amplified and split by a directional coupler to both the VNA for transmission measurement and a spectrum analyzer (Keysight E4405B) for noise measurement. A band-pass filter was installed to remove the strong pump signal that risk overloading the ADC of the spectrum analyzer. In the backward-gain measurement, the VNA is reconfigured to connect to the backward input and output ports. Note that with a broadband noise source such as a JTWPA, image-noise rejection is necessary for an accurate noise-power characterization if superheterodyne detection is employed.

\subsection{System-attenuation calibration using qubit-coupled transmission line}
To calibrate the total system attenuation, we utilize a qubit-coupled transmission line, which exhibits power-dependent transmission characteristics. At low powers, the qubit behaves as an almost perfect mirror, completely reflecting weak continuous-wave signals. At high powers, the qubit saturates, allowing near-perfect transmission. The two cases are connected by a gradual transition. This power-dependent behavior enables accurate extraction of the system attenuation between the room-temperature microwave source and the qubit inside the cryostat.

We fit the measured transmission data to the theoretical model for a qubit-coupled transmission line~\cite{mirhosseiniCavityQuantumElectrodynamics2019b,kannanGeneratingSpatiallyEntangled2020}:
\begin{equation}
   t = 1 - e^{i\theta_\mathrm{F}}\frac{\xi\Gamma_1}{2\Gamma_2}\frac{1-\frac{i\Delta}{\Gamma_2}}{1+\left(\frac{\Delta}{\Gamma_2}\right)^2+\frac{\Omega^2}{\Gamma_1\Gamma_2}}, \label{eq:qubitTrans}
\end{equation}
where $\xi$ represents the ratio of emission into the transmission line compared to all loss channels, which we assumed to be unity given a large designed coupling to the line. $\Gamma_1$ is the spontaneous emission rate into the coupled line, and $\Gamma_2$ is the transverse decoherence rate related to the pure dephasing rate $\Gamma_\phi$ through $\Gamma_2 = \Gamma_1/2+\Gamma_\phi$. The detuning between qubit frequency $\omega_\mathrm{q}$ and probe frequency $\omega_\mathrm{p}$ is denoted by $\Delta=\omega_\mathrm{q}-\omega_\mathrm{p}$, while $\Omega$ represents the drive amplitude in rad/s. The probe signal power $P$ is given by $P=\hbar\omega_\mathrm{q}\Omega^2/2\Gamma_1$, which is reduced from the room-temperature source power $P_\mathrm{RT}$ by the calibrated system attenuation $A_\mathrm{sys}^\mathrm{cal}$ such that $P_\mathrm{RT}=A_\mathrm{sys}^\mathrm{cal}P$. We include a Fano phase parameter $\theta_\mathrm{F}$ to account for transmission asymmetry around the resonance due to Fano interference~\cite{probstEfficientRobustAnalysis2015,joshiResonanceFluorescenceChiral2023}.

\begin{figure}[t]
   \includegraphics[width=8.88cm]{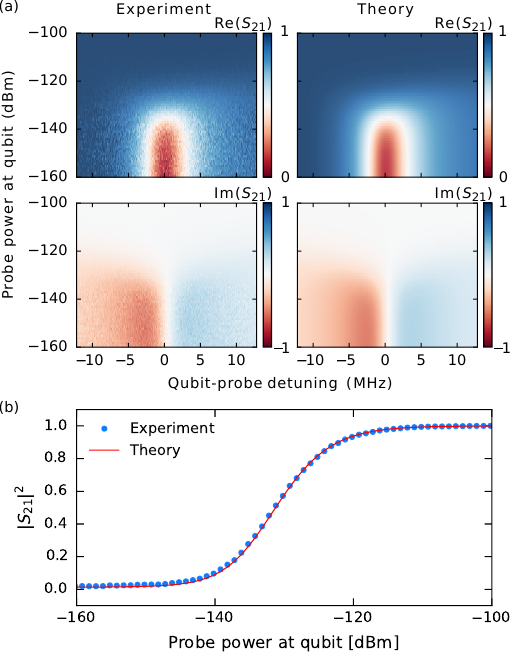}
   \caption{System calibration using qubit-coupled transmission line. (a)~Complex transmission characteristics through the cryostat shown in a 2D scan over room-temperature probe power and frequency. Experimental (left) and theoretical (right) data show excellent agreement in both real (top) and imaginary (bottom) components of the transmission coefficient $S_{21}$. (b)~High-averaging power sweep showing transmission magnitude $|S_{21}|^2$ versus probe power at the lowest-transmission frequency (5.5633~GHz) around the transmission dip due to the qubit.}
   \label{fig:qubitTrans}
\end{figure}

Fig.~\ref{fig:qubitTrans}(a) shows transmission data from a 2D scan over probe power and frequency using a VNA, with data normalized to the near-unity transmission obtained by driving the qubit at high probe power past saturation ($\Omega\gg\Gamma_1, \Gamma_2$). The data was fitted using Eq.~\eqref{eq:qubitTrans}, showing good agreement between experiment and theory. To further check the fit accuracy, we performed an additional power sweep with more averaging at $\omega_\mathrm{p}/2\pi=5.5633$~GHz where the transmission minimum occurs. Fig.~\ref{fig:qubitTrans}(b) shows this data overlaid with a theoretical curve generated using parameters extracted from the initial 2D fit, demonstrating good agreement without further fitting. The obtained parameters are summarized in Table~\ref{tab:qubitFitParams}.

\begin{table}[!htb]
   \caption{\label{tab:qubitFitParams}Fitted parameters from the qubit-coupled transmission-line calibration.}
   \begin{ruledtabular}
   \begin{tabular}{lll}
   System attenuation & $A_\mathrm{sys}^\mathrm{cal}$ & $88.20\pm0.02$~dB \\
   Qubit relaxation rate & $\Gamma_1/2\pi$ & $5.508\pm0.004$~MHz \\
   Qubit pure-dephasing rate & $\Gamma_\phi/2\pi$ & $259\pm2$~kHz \\
   Qubit frequency & $\omega_\mathrm{q}/2\pi$ & $5.5626\pm0$~GHz \\
   Fano phase parameter & $\theta_\mathrm{F}$ & $0.260\pm0.001$~rad \\
   \end{tabular}
   \end{ruledtabular}
\end{table}

\subsection{Translation of system calibration}
The noise measurement requires precise knowledge of the system gain from the Tukey CP-JTWPA output to the room-temperature digitizer. Since the qubit-based calibration used a separate sample package, the calibration must be translated between the two configurations. 

\begin{figure}[!ht]
   \includegraphics[width=8.53cm]{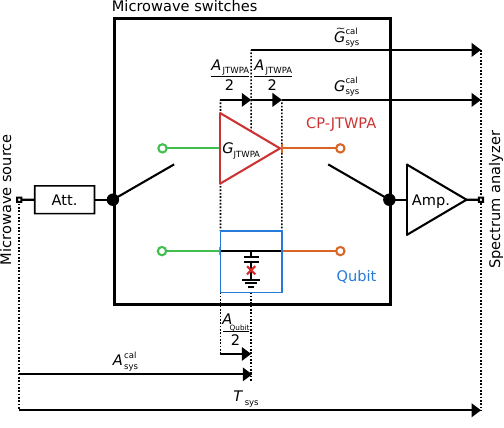}
   \caption{Measurement setup for system calibration. Two microwave switches enable alternate measurements between the Tukey CP-JTWPA path and qubit-coupled-line path, where paths of the same color share near-identical transmissions. System parameters labeled include: calibrated system attenuation $A_\mathrm{sys}^\mathrm{cal}$, calibrated system gain $G_\mathrm{sys}^\mathrm{cal}$, qubit package loss $A_\mathrm{Qubit}$, and CP-JTWPA insertion loss $A_\mathrm{JTWPA}$. The device-referenced system gain $\widetilde{G}_\mathrm{sys}^\mathrm{cal}$ is obtained by adjusting $G_\mathrm{sys}^\mathrm{cal}$ for half of the CP-JTWPA package loss to reference the gain to the center of the JTWPA device.}
   \label{fig:calCircuit}
\end{figure}

The measurement setup uses two microwave switches to alternately connect either the CP-JTWPA or the qubit-coupled line to the measurement path (see Fig.~\ref{fig:calCircuit}). Each measurement path employs matched cable pairs, with near-identical cables connecting each sample to the input and output switches, respectively, to the minimize transmission difference between the two measurement configurations. We characterized the package losses for both devices in separate cool downs using a 50-$\Omega$ union as reference, then incorporated these losses by assuming symmetric loss distribution.

The system gain $G_\mathrm{sys}^\mathrm{cal}$ from the CP-JTWPA package output port to the room-temperature spectrum analyzer is calculated from the measured system transmission $|S_\mathrm{21,sys}|^2$ between the VNA source and spectrum analyzer:
\begin{equation}
G_\mathrm{sys}^\mathrm{cal} = \left[|S_\mathrm{21,sys}|^2 + (A_\mathrm{sys}^\mathrm{cal} - A_\mathrm{Qubit}/2 + A_\mathrm{JTWPA})\right]\,\,\mathrm{dB},
\end{equation}
where $A_\mathrm{sys}^\mathrm{cal} = 88.20$~dB is the calibrated system attenuation obtained from the qubit-coupled transmission line, $A_\mathrm{Qubit} = 0.57$~dB is the qubit package loss, and $A_\mathrm{JTWPA} = 0.29$~dB is the total CP-JTWPA insertion loss including its package. This calculation yields $G_\mathrm{sys}^\mathrm{cal} = 56.9$~dB.

For calculating the JTWPA added noise, we model package losses as lumped attenuators at the input and output of an otherwise lossless JTWPA. While this approach can overestimate added noise compared to distributed noise models used in previous works~\cite{macklinQuantumLimitedJosephson2015, planatPhotonicCrystalJosephsonTravelingWave2020}, the difference is negligible given our low JTWPA and package insertion losses. To properly account for the package loss in our noise calculation, we shift the reference plane of system gain from the JTWPA package output port to the center of the JTWPA package. This yields the device-referenced system gain $\widetilde{G}_\mathrm{sys}^\mathrm{cal} = G_\mathrm{sys}^\mathrm{cal}/(A_\mathrm{JTWPA}/2) = 56.76$~dB.

\subsection{Noise calculation} 
\begin{figure}[!b]
   \includegraphics[width=7.33cm]{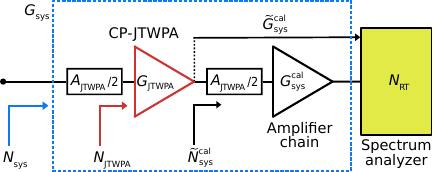}
   \caption{Noise-propagation schematic in our measurement setup. The input signal experiences half of the CP-JTWPA package loss ($A_\mathrm{JTWPA}/2$), then undergoes parametric amplification ($G_\mathrm{JTWPA}$) where the JTWPA added noise ($N_\mathrm{JTWPA}$) is introduced, followed by the remaining half of the package loss after. The signal then propagates through the amplification chain ($G_\mathrm{sys}^\mathrm{cal}$) before being measured by the spectrum analyzer to yield $N_\mathrm{RT}$. The blue dashed box indicates the system boundary for calculating the input-referred system noise $N_\mathrm{sys}$. The black dashed line indicates the reference plane for the CP-JTWPA added noise calculation, located at the center of the CP-JTWPA package.}
   \label{fig:noiseProp}
\end{figure}

Using $\widetilde{G}_\mathrm{sys}^\mathrm{cal}$, we calculate the added noise of the CP-JTWPA as illustrated in Fig.~\ref{fig:noiseProp}. We determine the noise power at room temperature in units of photon flux per hertz, $N_\mathrm{RT} = P_\mathrm{SA}/\mathrm{B}/\hbar\omega_\mathrm{p}$, where $P_\mathrm{SA}$ is the average noise power measured by the spectrum analyzer, $\mathrm{B} = 100$~Hz is the measurement bandwidth, and $\hbar\omega_\mathrm{p}$ is the photon energy at the measurement frequency of $5.5633$~GHz. We assume the input noise remains at the vacuum fluctuation level due to cryogenic attenuation and thermal anchoring of the input line.

To ensure measurement accuracy and account for system noise fluctuations, we performed 100 repeated measurements with interleaved CP-JTWPA pump-on and pump-off states, each lasting approximately one second. This interleaved measurement scheme minimizes the impact of amplifier chain noise drift on the added noise measurement. When the CP-JTWPA is off, the measured noise at room temperature $N_\mathrm{RT}^\mathrm{off}$ consists of amplified vacuum noise $N_\mathrm{vac}$ and system added noise $\widetilde{N}_\mathrm{sys}^\mathrm{cal}$ (at JTWPA reference plane):
\begin{equation}
N_\mathrm{RT}^\mathrm{off} = \widetilde{G}_\mathrm{sys}^\mathrm{cal} (N_\mathrm{vac} + \widetilde{N}_\mathrm{sys}^\mathrm{cal})
\end{equation}
When the CP-JTWPA is turned on, it contributes both parametric gain $G_\mathrm{JTWPA}$ and its own added noise $N_\mathrm{JTWPA}$ to the signal. The measured noise becomes:
\begin{equation}
N_\mathrm{RT}^\mathrm{on} = \widetilde{G}_\mathrm{sys}^\mathrm{cal} \left(G_\mathrm{JTWPA} (N_\mathrm{vac} + N_\mathrm{JTWPA}) + \widetilde{N}_\mathrm{sys}^\mathrm{cal} \right)
\end{equation}
The difference between these two states isolates the CP-JTWPA's contribution:
\begin{equation}
N_\mathrm{RT}^\mathrm{on} - N_\mathrm{RT}^\mathrm{off} = \widetilde{G}_\mathrm{sys}^\mathrm{cal} \left( G_\mathrm{JTWPA} (N_\mathrm{vac} + N_\mathrm{JTWPA}) - N_\mathrm{vac} \right)
\end{equation}
Solving for the CP-JTWPA added noise yields:
\begin{equation}
N_\mathrm{JTWPA} = \frac{(N_\mathrm{RT}^\mathrm{on} - N_\mathrm{RT}^\mathrm{off}) / \widetilde{G}_\mathrm{sys}^\mathrm{cal} + N_\mathrm{vac}}{G_\mathrm{JTWPA}} - N_\mathrm{vac} \label{eq:N_JTWPA}
\end{equation}

The input-referred system noise in photon flux per hertz $N_\mathrm{sys} = N_\mathrm{RT}/G_\mathrm{sys}$ is calculated for both on and off states by treating the CP-JTWPA and subsequent amplification chain (indicated by the dashed box in Fig.~\ref{fig:noiseProp}) as a single amplification system, with a system gain $G_\mathrm{sys}=G_\mathrm{sys}^\mathrm{cal}/A_\mathrm{JTWPA}$. The noise temperature is converted from the system noise quanta as $T_\mathrm{N} = N_\mathrm{sys}\hbar\omega_\mathrm{p}/k_\mathrm{B}$.

Our uncertainty analysis accounts for both statistical and systematic contributions. Statistical uncertainties are calculated as standard deviations from repeated measurements. For systematic uncertainties, we estimate a $\pm$0.5-dB uncertainty in $\widetilde{G}_\mathrm{sys}^\mathrm{cal}$, based on typical variations observed in our cryogenic measurements, including cable differences and package-loss variations. To incorporate this systematic uncertainty, we calculate the added noise and system noise at both the upper and lower bounds of this $\pm$0.5-dB range, creating an uncertainty envelope. We then report the total uncertainty as half the width of this envelope, representing the maximum deviation from our central estimate.
\section{FABRICATION RECIPE}\label{sec:appendix_fabrication}

This section outlines the recipe used for fabricating CP-JTWPAs. The process employs techniques common to transmon-qubit fabrication, requiring similar cleanroom equipment. Due to the high pattern density of a CP-JTWPA, minimizing visible defects (dust, particles, scratches) is critical for producing operational samples.

\subsection{Part 1: Base circuit fabrication by ultraviolet lithography}

\subsubsection{Substrate preparation}
We begin with a high-resistivity silicon substrate~($\approx$20~k$\Omega\cdot$cm). For new wafers, cleaning consists of sequential 5-minute ultrasonic baths in acetone and IPA at room temperature, followed by a 30-second DI water rinse and 5-minute dehydration bake at 240°C on a hot plate (all baking steps in the recipe are done on a hot plate). Alternatively, RCA/HF cleaning can be used for better surface preparation, eliminating the need for dehydration baking.

A 200-nm aluminum film is deposited using electron-beam evaporation (Plassys), preceded by argon-ion milling~(400~V, 2~minutes) for adhesion. Post-deposition cleaning consists of a 5-minute ultrasonic bath in IPA followed by a DI water rinse to prepare the surface for spin coating.

\subsubsection{Photolithography}
To define the open-stub geometry, AZ1500 photoresist~(4.4-cp viscosity) is spin-coated to roughly 600-nm thickness using: 2-second ramp-up, 5~seconds at 500~rpm, 10-second ramp-up, 1~minute at 2000~rpm, and 5-second ramp-down. After a 5-minute soft bake at 100°C and cooling, the resist is exposed using a Heidelberg MLA150 UV writer with 375-nm light at 120 mJ/cm$^2$. Development follows in 1:1 AZ-developer:DI water for 2 minutes, with a 1-minute DI rinse and nitrogen blow-dry.

\subsubsection{Pattern transfer}
After lithography, a 5-minute post-development bake at 120°C hardens the resist, followed by cooling. A 20-second oxygen plasma descum (50-sccm O$_2$, 100~W) removes residual resist in exposed areas. Aluminum etching begins with a brief DI water dip followed by a 6-minute immersion in aluminum etchant (Fujifilm aluminum mixed acid solution). After observing complete pattern development~($\approx$4~minutes~45~seconds), an additional minute of over-etching ensures complete aluminum removal. A thorough 5-minute DI water rinse follows using multiple beakers of fresh DI water. The wafer can then be directly transferred to a remover bath for resist removal without nitrogen blow-dry.

\subsubsection{Resist removal and final cleaning}
The resist mask is removed through two-stage cleaning in NMP-based remover (Tok Remover 104, 60\% DMSO, 40\% NMP) at 80°C (1 hour in each bath), followed by a 5-minute ultrasonic cleaning in acetone and IPA and a final DI water rinse. After transferring to a new bath, using a pipette to direct solvent flow across the wafer surface helps dislodge and remove residues.

\subsection{Part 2: Junction fabrication by electron-beam lithography}

\subsubsection{Bilayer resist application}
A bilayer resist stack provides the undercut profile for Manhattan-type junction fabrication. The bottom layer~(Copolymer EL6, 200-nm thick) is applied via spin coating: 2-second ramp-up, 5~seconds at 500~rpm, 5-second ramp-up, 1~minute at 2000~rpm, and 5-second ramp-down, followed by a 5-minute pre-bake at 180°C and cooling to room temperature.

The top imaging layer (950 PMMA A9, 2-\textmu m thick) is applied via: 5-second ramp-up, 10~seconds at 1000~rpm, 40-second ramp-up, 2~minutes at 2500~rpm, and 10-second ramp-down, with a 15-minute pre-bake at 180°C and cooling.

\subsubsection{Electron-beam exposure and development}
The resist stack is exposed using a Raith EBPG5150 100-kV system with: main dose of 600--700~\textmu C/cm$^2$ (700 for junctions, 600 for larger features) and an undercut dose of 100~\textmu C/cm$^2$, using 10 nA beam current at 100 kV. To minimize stitching errors in critical junction regions, implement short stage movements between successive junction pattern exposures and ensure both junction electrodes of each junction site are contained within a single write field.

Development uses roughly 3:7 DI water:IPA prepared by adding DI water to 350~mL IPA to reach 500~mL total volume. After a 10-minute mixing, the wafer is immersed face-down for 4~minutes, followed by a 30-second IPA rinse and nitrogen blow-dry. A 10-second oxygen plasma descum (50-sccm O$_2$, 50~W) removes residual resist.

\subsubsection{Junction evaporation and liftoff}
Josephson junctions are formed by double-angle shadow evaporation, with $>$3-hour pump-down to below 10$^{-7}$ Torr. We avoid argon-ion milling to prevent contamination of the silicon substrate by resist sputtering~\cite{dunsworthCharacterizationReductionCapacitive2017}. The first layer deposition consists of 2-minute titanium gettering at 0.1 nm/s, 30-nm aluminum electrode at 0.2 nm/s, and 8-minute oxidation at 0.12 Torr (yielding $\approx$75 $\Omega$/\textmu m$^2$). The second layer includes 2-minute titanium gettering, 90-nm aluminum at 0.2 nm/s, and 5-minute oxidation at 15 Torr for passivation.

Liftoff uses two 2-hour remover baths at 120°C, removing floating aluminum before transferring to the second bath, followed by a 5-minute acetone rinse with 10-second ultrasonic agitation and a 5-minute IPA rinse. At this stage, junctions lack galvanic contact with the open-stub structure.

\subsection{Part 3: Airbridge fabrication and final processing}

\subsubsection{Support layer formation}
Next, we create airbridges to suppress slot-line modes and provide ground continuity, while forming bandages for galvanic contacts between junction and the open stubs. SPR220-3.0 photoresist ($\approx$3 \textmu m thick) is applied after a 3-minute dry bake at 115°C and cooling. Spin coating is then performed with 2-second ramp-up, 10~seconds at 300~rpm, 2-second ramp-up, 1~minute at 2500~rpm, and 5-second ramp-down. After a 90-second pre-bake at 115°C and cooling, exposure uses a Heidelberg UV writer with 405-nm light at 170 mJ/cm$^2$.

Post-exposure processing includes a 90-second bake at 115°C, cooling, development in 1:1 AZ-developer:DI water for 3~minutes, a 30-second DI rinse, and nitrogen blow-dry. A critical 3-minute reflow bake at 140°C creates the rounded profile for airbridges. A 20-second oxygen plasma descum (50-sccm O$_2$, 100~W) removes residual resist.

\subsubsection{Airbridge metallization}
Metal deposition begins with a 6-minute argon-ion milling (5-sccm flow, 400~V, 65~mA) at 3-rpm rotation for clean surfaces and galvanic contact. A 300-nm aluminum layer is deposited at normal angle with 3-rpm rotation, followed by a 5-minute oxidation at 15~Torr for surface passivation.

\subsubsection{Pattern definition and etching}
Airbridge patterns are defined using AZ1500 photoresist (38~cp, $\approx$2 \textmu m thickness) without pre-bake. Spin coating uses: 2-second ramp-up, 5 seconds at 500~rpm, 2-second ramp-up, 1~minute at 2500~rpm, and 5-second ramp-down, followed by a 10-minute bake at 80°C and cooling.

Exposure uses the Heidelberg MLA150 with 405-nm light at 100 mJ/cm$^2$, with pattern inversion for positive resist. After a 2-minute development in 1:1 AZ-developer:DI water, a 30-second DI rinse, and nitrogen blow-dry, a 20-second descum~(50-sccm O$_2$, 100~W) removes residual resist.

Aluminum etching begins with a brief DI water dip before an 8.5-minute immersion in aluminum etchant, with 1-minute over-etching after visual clearing. A 5-minute DI water rinse with multiple beakers follows. A 3-minute oxygen plasma ashing (50-sccm O$_2$, 100 W) removes resist damaged by argon-ion milling.~\cite{chenFabricationCharacterizationAluminum2014}.
\vspace{0.5cm}

\subsubsection{Final dicing and cleaning}
Before dicing, a protective layer of AZ1500 (38~cp) is applied with identical spin profile to the last step, but with a 10-minute bake at 70°C for easier removal. After dicing, cleaning consists of two 3-hour remover baths at 80°C, followed by 5-minute acetone and IPA rinses. A pipette is used for flowing solvent onto the chips after each transfers to remove trapped residue. 

Note that the airbridge process can increase the junction resistance on an average of 8-10\%, while typically maintaining the same resistance variation.

\bibliographystyle{apsrev4-2}
\bibliography{main}

\begin{thebibliography}{50}%
\makeatletter
\providecommand \@ifxundefined [1]{%
 \@ifx{#1\undefined}
}%
\providecommand \@ifnum [1]{%
 \ifnum #1\expandafter \@firstoftwo
 \else \expandafter \@secondoftwo
 \fi
}%
\providecommand \@ifx [1]{%
 \ifx #1\expandafter \@firstoftwo
 \else \expandafter \@secondoftwo
 \fi
}%
\providecommand \natexlab [1]{#1}%
\providecommand \enquote  [1]{``#1''}%
\providecommand \bibnamefont  [1]{#1}%
\providecommand \bibfnamefont [1]{#1}%
\providecommand \citenamefont [1]{#1}%
\providecommand \href@noop [0]{\@secondoftwo}%
\providecommand \href [0]{\begingroup \@sanitize@url \@href}%
\providecommand \@href[1]{\@@startlink{#1}\@@href}%
\providecommand \@@href[1]{\endgroup#1\@@endlink}%
\providecommand \@sanitize@url [0]{\catcode `\\12\catcode `\$12\catcode `\&12\catcode `\#12\catcode `\^12\catcode `\_12\catcode `\%12\relax}%
\providecommand \@@startlink[1]{}%
\providecommand \@@endlink[0]{}%
\providecommand \url  [0]{\begingroup\@sanitize@url \@url }%
\providecommand \@url [1]{\endgroup\@href {#1}{\urlprefix }}%
\providecommand \urlprefix  [0]{URL }%
\providecommand \Eprint [0]{\href }%
\providecommand \doibase [0]{https://doi.org/}%
\providecommand \selectlanguage [0]{\@gobble}%
\providecommand \bibinfo  [0]{\@secondoftwo}%
\providecommand \bibfield  [0]{\@secondoftwo}%
\providecommand \translation [1]{[#1]}%
\providecommand \BibitemOpen [0]{}%
\providecommand \bibitemStop [0]{}%
\providecommand \bibitemNoStop [0]{.\EOS\space}%
\providecommand \EOS [0]{\spacefactor3000\relax}%
\providecommand \BibitemShut  [1]{\csname bibitem#1\endcsname}%
\let\auto@bib@innerbib\@empty
\bibitem [{\citenamefont {Sank}\ \emph {et~al.}(2016)\citenamefont {Sank}, \citenamefont {Chen}, \citenamefont {Khezri}, \citenamefont {Kelly}, \citenamefont {Barends}, \citenamefont {Campbell}, \citenamefont {Chen}, \citenamefont {Chiaro}, \citenamefont {Dunsworth}, \citenamefont {Fowler}, \citenamefont {Jeffrey}, \citenamefont {Lucero}, \citenamefont {Megrant}, \citenamefont {Mutus}, \citenamefont {Neeley}, \citenamefont {Neill}, \citenamefont {O'Malley}, \citenamefont {Quintana}, \citenamefont {Roushan}, \citenamefont {Vainsencher}, \citenamefont {White}, \citenamefont {Wenner}, \citenamefont {Korotkov},\ and\ \citenamefont {Martinis}}]{sankMeasurementInducedStateTransitions2016}%
  \BibitemOpen
  \bibfield  {author} {\bibinfo {author} {\bibfnamefont {D.}~\bibnamefont {Sank}}, \bibinfo {author} {\bibfnamefont {Z.}~\bibnamefont {Chen}}, \bibinfo {author} {\bibfnamefont {M.}~\bibnamefont {Khezri}}, \bibinfo {author} {\bibfnamefont {J.}~\bibnamefont {Kelly}}, \bibinfo {author} {\bibfnamefont {R.}~\bibnamefont {Barends}}, \bibinfo {author} {\bibfnamefont {B.}~\bibnamefont {Campbell}}, \bibinfo {author} {\bibfnamefont {Y.}~\bibnamefont {Chen}}, \bibinfo {author} {\bibfnamefont {B.}~\bibnamefont {Chiaro}}, \bibinfo {author} {\bibfnamefont {A.}~\bibnamefont {Dunsworth}}, \bibinfo {author} {\bibfnamefont {A.}~\bibnamefont {Fowler}}, \bibinfo {author} {\bibfnamefont {E.}~\bibnamefont {Jeffrey}}, \bibinfo {author} {\bibfnamefont {E.}~\bibnamefont {Lucero}}, \bibinfo {author} {\bibfnamefont {A.}~\bibnamefont {Megrant}}, \bibinfo {author} {\bibfnamefont {J.}~\bibnamefont {Mutus}}, \bibinfo {author} {\bibfnamefont {M.}~\bibnamefont {Neeley}}, \bibinfo {author} {\bibfnamefont {C.}~\bibnamefont {Neill}}, \bibinfo {author} {\bibfnamefont {P.~J.~J.}\ \bibnamefont {O'Malley}}, \bibinfo {author} {\bibfnamefont {C.}~\bibnamefont {Quintana}}, \bibinfo {author} {\bibfnamefont {P.}~\bibnamefont {Roushan}}, \bibinfo {author} {\bibfnamefont {A.}~\bibnamefont {Vainsencher}}, \bibinfo {author} {\bibfnamefont {T.}~\bibnamefont {White}}, \bibinfo {author} {\bibfnamefont {J.}~\bibnamefont {Wenner}}, \bibinfo {author} {\bibfnamefont {A.~N.}\ \bibnamefont {Korotkov}},\ and\ \bibinfo {author} {\bibfnamefont {J.~M.}\ \bibnamefont {Martinis}},\ }\href {https://doi.org/10.1103/PhysRevLett.117.190503} {\bibfield  {journal} {\bibinfo  {journal} {Physical Review Letters}\ }\textbf {\bibinfo {volume} {117}},\ \bibinfo {pages} {190503} (\bibinfo {year} {2016})}\BibitemShut {NoStop}%
\bibitem [{\citenamefont {Dumas}\ \emph {et~al.}(2024)\citenamefont {Dumas}, \citenamefont {{Groleau-Par{\'e}}}, \citenamefont {McDonald}, \citenamefont {{Mu{\~n}oz-Arias}}, \citenamefont {Lled{\'o}}, \citenamefont {D'Anjou},\ and\ \citenamefont {Blais}}]{dumasMeasurementInducedTransmonIonization2024}%
  \BibitemOpen
  \bibfield  {author} {\bibinfo {author} {\bibfnamefont {M.~F.}\ \bibnamefont {Dumas}}, \bibinfo {author} {\bibfnamefont {B.}~\bibnamefont {{Groleau-Par{\'e}}}}, \bibinfo {author} {\bibfnamefont {A.}~\bibnamefont {McDonald}}, \bibinfo {author} {\bibfnamefont {M.~H.}\ \bibnamefont {{Mu{\~n}oz-Arias}}}, \bibinfo {author} {\bibfnamefont {C.}~\bibnamefont {Lled{\'o}}}, \bibinfo {author} {\bibfnamefont {B.}~\bibnamefont {D'Anjou}},\ and\ \bibinfo {author} {\bibfnamefont {A.}~\bibnamefont {Blais}},\ }\href {https://doi.org/10.1103/PhysRevX.14.041023} {\bibfield  {journal} {\bibinfo  {journal} {Physical Review X}\ }\textbf {\bibinfo {volume} {14}},\ \bibinfo {pages} {041023} (\bibinfo {year} {2024})}\BibitemShut {NoStop}%
\bibitem [{\citenamefont {{Castellanos-Beltran}}\ \emph {et~al.}(2008)\citenamefont {{Castellanos-Beltran}}, \citenamefont {Irwin}, \citenamefont {Hilton}, \citenamefont {Vale},\ and\ \citenamefont {Lehnert}}]{castellanos-beltranAmplificationSqueezingQuantum2008}%
  \BibitemOpen
  \bibfield  {author} {\bibinfo {author} {\bibfnamefont {M.~A.}\ \bibnamefont {{Castellanos-Beltran}}}, \bibinfo {author} {\bibfnamefont {K.~D.}\ \bibnamefont {Irwin}}, \bibinfo {author} {\bibfnamefont {G.~C.}\ \bibnamefont {Hilton}}, \bibinfo {author} {\bibfnamefont {L.~R.}\ \bibnamefont {Vale}},\ and\ \bibinfo {author} {\bibfnamefont {K.~W.}\ \bibnamefont {Lehnert}},\ }\href {https://doi.org/10.1038/nphys1090} {\bibfield  {journal} {\bibinfo  {journal} {Nature Physics}\ }\textbf {\bibinfo {volume} {4}},\ \bibinfo {pages} {929} (\bibinfo {year} {2008})}\BibitemShut {NoStop}%
\bibitem [{\citenamefont {Yamamoto}\ \emph {et~al.}(2008)\citenamefont {Yamamoto}, \citenamefont {Inomata}, \citenamefont {Watanabe}, \citenamefont {Matsuba}, \citenamefont {Miyazaki}, \citenamefont {Oliver}, \citenamefont {Nakamura},\ and\ \citenamefont {Tsai}}]{yamamotoFluxdrivenJosephsonParametric2008}%
  \BibitemOpen
  \bibfield  {author} {\bibinfo {author} {\bibfnamefont {T.}~\bibnamefont {Yamamoto}}, \bibinfo {author} {\bibfnamefont {K.}~\bibnamefont {Inomata}}, \bibinfo {author} {\bibfnamefont {M.}~\bibnamefont {Watanabe}}, \bibinfo {author} {\bibfnamefont {K.}~\bibnamefont {Matsuba}}, \bibinfo {author} {\bibfnamefont {T.}~\bibnamefont {Miyazaki}}, \bibinfo {author} {\bibfnamefont {W.~D.}\ \bibnamefont {Oliver}}, \bibinfo {author} {\bibfnamefont {Y.}~\bibnamefont {Nakamura}},\ and\ \bibinfo {author} {\bibfnamefont {J.~S.}\ \bibnamefont {Tsai}},\ }\href {https://doi.org/10.1063/1.2964182} {\bibfield  {journal} {\bibinfo  {journal} {Applied Physics Letters}\ }\textbf {\bibinfo {volume} {93}},\ \bibinfo {pages} {042510} (\bibinfo {year} {2008})}\BibitemShut {NoStop}%
\bibitem [{\citenamefont {Bergeal}\ \emph {et~al.}(2010)\citenamefont {Bergeal}, \citenamefont {Schackert}, \citenamefont {Metcalfe}, \citenamefont {Vijay}, \citenamefont {Manucharyan}, \citenamefont {Frunzio}, \citenamefont {Prober}, \citenamefont {Schoelkopf}, \citenamefont {Girvin},\ and\ \citenamefont {Devoret}}]{bergealPhasepreservingAmplificationQuantum2010}%
  \BibitemOpen
  \bibfield  {author} {\bibinfo {author} {\bibfnamefont {N.}~\bibnamefont {Bergeal}}, \bibinfo {author} {\bibfnamefont {F.}~\bibnamefont {Schackert}}, \bibinfo {author} {\bibfnamefont {M.}~\bibnamefont {Metcalfe}}, \bibinfo {author} {\bibfnamefont {R.}~\bibnamefont {Vijay}}, \bibinfo {author} {\bibfnamefont {V.~E.}\ \bibnamefont {Manucharyan}}, \bibinfo {author} {\bibfnamefont {L.}~\bibnamefont {Frunzio}}, \bibinfo {author} {\bibfnamefont {D.~E.}\ \bibnamefont {Prober}}, \bibinfo {author} {\bibfnamefont {R.~J.}\ \bibnamefont {Schoelkopf}}, \bibinfo {author} {\bibfnamefont {S.~M.}\ \bibnamefont {Girvin}},\ and\ \bibinfo {author} {\bibfnamefont {M.~H.}\ \bibnamefont {Devoret}},\ }\href {https://doi.org/10.1038/nature09035} {\bibfield  {journal} {\bibinfo  {journal} {Nature}\ }\textbf {\bibinfo {volume} {465}},\ \bibinfo {pages} {64} (\bibinfo {year} {2010})}\BibitemShut {NoStop}%
\bibitem [{\citenamefont {Roy}\ \emph {et~al.}(2015)\citenamefont {Roy}, \citenamefont {Kundu}, \citenamefont {Chand}, \citenamefont {Vadiraj}, \citenamefont {Ranadive}, \citenamefont {Nehra}, \citenamefont {Patankar}, \citenamefont {Aumentado}, \citenamefont {Clerk},\ and\ \citenamefont {Vijay}}]{royBroadbandParametricAmplification2015}%
  \BibitemOpen
  \bibfield  {author} {\bibinfo {author} {\bibfnamefont {T.}~\bibnamefont {Roy}}, \bibinfo {author} {\bibfnamefont {S.}~\bibnamefont {Kundu}}, \bibinfo {author} {\bibfnamefont {M.}~\bibnamefont {Chand}}, \bibinfo {author} {\bibfnamefont {A.~M.}\ \bibnamefont {Vadiraj}}, \bibinfo {author} {\bibfnamefont {A.}~\bibnamefont {Ranadive}}, \bibinfo {author} {\bibfnamefont {N.}~\bibnamefont {Nehra}}, \bibinfo {author} {\bibfnamefont {M.~P.}\ \bibnamefont {Patankar}}, \bibinfo {author} {\bibfnamefont {J.}~\bibnamefont {Aumentado}}, \bibinfo {author} {\bibfnamefont {A.~A.}\ \bibnamefont {Clerk}},\ and\ \bibinfo {author} {\bibfnamefont {R.}~\bibnamefont {Vijay}},\ }\href {https://doi.org/10.1063/1.4939148} {\bibfield  {journal} {\bibinfo  {journal} {Applied Physics Letters}\ }\textbf {\bibinfo {volume} {107}},\ \bibinfo {pages} {262601} (\bibinfo {year} {2015})}\BibitemShut {NoStop}%
\bibitem [{\citenamefont {Grebel}\ \emph {et~al.}(2021)\citenamefont {Grebel}, \citenamefont {Bienfait}, \citenamefont {Dumur}, \citenamefont {Chang}, \citenamefont {Chou}, \citenamefont {Conner}, \citenamefont {Peairs}, \citenamefont {Povey}, \citenamefont {Zhong},\ and\ \citenamefont {Cleland}}]{grebelFluxpumpedImpedanceengineeredBroadband2021}%
  \BibitemOpen
  \bibfield  {author} {\bibinfo {author} {\bibfnamefont {J.}~\bibnamefont {Grebel}}, \bibinfo {author} {\bibfnamefont {A.}~\bibnamefont {Bienfait}}, \bibinfo {author} {\bibfnamefont {{\'E}.}~\bibnamefont {Dumur}}, \bibinfo {author} {\bibfnamefont {H.-S.}\ \bibnamefont {Chang}}, \bibinfo {author} {\bibfnamefont {M.-H.}\ \bibnamefont {Chou}}, \bibinfo {author} {\bibfnamefont {C.~R.}\ \bibnamefont {Conner}}, \bibinfo {author} {\bibfnamefont {G.~A.}\ \bibnamefont {Peairs}}, \bibinfo {author} {\bibfnamefont {R.~G.}\ \bibnamefont {Povey}}, \bibinfo {author} {\bibfnamefont {Y.~P.}\ \bibnamefont {Zhong}},\ and\ \bibinfo {author} {\bibfnamefont {A.~N.}\ \bibnamefont {Cleland}},\ }\href {https://doi.org/10.1063/5.0035945} {\bibfield  {journal} {\bibinfo  {journal} {Applied Physics Letters}\ }\textbf {\bibinfo {volume} {118}},\ \bibinfo {pages} {142601} (\bibinfo {year} {2021})}\BibitemShut {NoStop}%
\bibitem [{\citenamefont {White}\ \emph {et~al.}(2023)\citenamefont {White}, \citenamefont {Opremcak}, \citenamefont {Sterling}, \citenamefont {Korotkov}, \citenamefont {Sank}, \citenamefont {Acharya}, \citenamefont {Ansmann}, \citenamefont {Arute}, \citenamefont {Arya}, \citenamefont {Bardin}, \citenamefont {Bengtsson}, \citenamefont {Bourassa}, \citenamefont {Bovaird}, \citenamefont {Brill}, \citenamefont {Buckley}, \citenamefont {Buell}, \citenamefont {Burger}, \citenamefont {Burkett}, \citenamefont {Bushnell}, \citenamefont {Chen}, \citenamefont {Chiaro}, \citenamefont {Cogan}, \citenamefont {Collins}, \citenamefont {Crook}, \citenamefont {Curtin}, \citenamefont {Demura}, \citenamefont {Dunsworth}, \citenamefont {Erickson}, \citenamefont {Fatemi}, \citenamefont {Burgos}, \citenamefont {Forati}, \citenamefont {Foxen}, \citenamefont {Giang}, \citenamefont {Giustina}, \citenamefont {Grajales~Dau}, \citenamefont {Hamilton}, \citenamefont {Harrington}, \citenamefont {Hilton}, \citenamefont {Hoffmann}, \citenamefont {Hong}, \citenamefont {Huang}, \citenamefont {Huff}, \citenamefont {Iveland}, \citenamefont {Jeffrey}, \citenamefont {Kieferov{\'a}}, \citenamefont {Kim}, \citenamefont {Klimov}, \citenamefont {Kostritsa}, \citenamefont {Kreikebaum}, \citenamefont {Landhuis}, \citenamefont {Laptev}, \citenamefont {Laws}, \citenamefont {Lee}, \citenamefont {Lester}, \citenamefont {Lill}, \citenamefont {Liu}, \citenamefont {Locharla}, \citenamefont {Lucero}, \citenamefont {McCourt}, \citenamefont {McEwen}, \citenamefont {Mi}, \citenamefont {Miao}, \citenamefont {Montazeri}, \citenamefont {Morvan}, \citenamefont {Neeley}, \citenamefont {Neill}, \citenamefont {Nersisyan}, \citenamefont {Ng}, \citenamefont {Nguyen}, \citenamefont {Nguyen}, \citenamefont {Potter}, \citenamefont {Quintana}, \citenamefont {Roushan}, \citenamefont {Sankaragomathi}, \citenamefont {Satzinger}, \citenamefont {Schuster}, \citenamefont {Shearn}, \citenamefont {Shorter}, \citenamefont {Shvarts}, \citenamefont {Skruzny}, \citenamefont {Smith}, \citenamefont {Szalay}, \citenamefont {Torres}, \citenamefont {Woo}, \citenamefont {Yao}, \citenamefont {Yeh}, \citenamefont {Yoo}, \citenamefont {Young}, \citenamefont {Zhu}, \citenamefont {Zobrist}, \citenamefont {Chen}, \citenamefont {Megrant}, \citenamefont {Kelly},\ and\ \citenamefont {Naaman}}]{whiteReadoutQuantumProcessor2023}%
  \BibitemOpen
  \bibfield  {author} {\bibinfo {author} {\bibfnamefont {T.}~\bibnamefont {White}}, \bibinfo {author} {\bibfnamefont {A.}~\bibnamefont {Opremcak}}, \bibinfo {author} {\bibfnamefont {G.}~\bibnamefont {Sterling}}, \bibinfo {author} {\bibfnamefont {A.}~\bibnamefont {Korotkov}}, \bibinfo {author} {\bibfnamefont {D.}~\bibnamefont {Sank}}, \bibinfo {author} {\bibfnamefont {R.}~\bibnamefont {Acharya}}, \bibinfo {author} {\bibfnamefont {M.}~\bibnamefont {Ansmann}}, \bibinfo {author} {\bibfnamefont {F.}~\bibnamefont {Arute}}, \bibinfo {author} {\bibfnamefont {K.}~\bibnamefont {Arya}}, \bibinfo {author} {\bibfnamefont {J.~C.}\ \bibnamefont {Bardin}}, \bibinfo {author} {\bibfnamefont {A.}~\bibnamefont {Bengtsson}}, \bibinfo {author} {\bibfnamefont {A.}~\bibnamefont {Bourassa}}, \bibinfo {author} {\bibfnamefont {J.}~\bibnamefont {Bovaird}}, \bibinfo {author} {\bibfnamefont {L.}~\bibnamefont {Brill}}, \bibinfo {author} {\bibfnamefont {B.~B.}\ \bibnamefont {Buckley}}, \bibinfo {author} {\bibfnamefont {D.~A.}\ \bibnamefont {Buell}}, \bibinfo {author} {\bibfnamefont {T.}~\bibnamefont {Burger}}, \bibinfo {author} {\bibfnamefont {B.}~\bibnamefont {Burkett}}, \bibinfo {author} {\bibfnamefont {N.}~\bibnamefont {Bushnell}}, \bibinfo {author} {\bibfnamefont {Z.}~\bibnamefont {Chen}}, \bibinfo {author} {\bibfnamefont {B.}~\bibnamefont {Chiaro}}, \bibinfo {author} {\bibfnamefont {J.}~\bibnamefont {Cogan}}, \bibinfo {author} {\bibfnamefont {R.}~\bibnamefont {Collins}}, \bibinfo {author} {\bibfnamefont {A.~L.}\ \bibnamefont {Crook}}, \bibinfo {author} {\bibfnamefont {B.}~\bibnamefont {Curtin}}, \bibinfo {author} {\bibfnamefont {S.}~\bibnamefont {Demura}}, \bibinfo {author} {\bibfnamefont {A.}~\bibnamefont {Dunsworth}}, \bibinfo {author} {\bibfnamefont {C.}~\bibnamefont {Erickson}}, \bibinfo {author} {\bibfnamefont {R.}~\bibnamefont {Fatemi}}, \bibinfo {author} {\bibfnamefont {L.~F.}\ \bibnamefont {Burgos}}, \bibinfo {author} {\bibfnamefont {E.}~\bibnamefont {Forati}}, \bibinfo {author} {\bibfnamefont {B.}~\bibnamefont {Foxen}}, \bibinfo {author} {\bibfnamefont {W.}~\bibnamefont {Giang}}, \bibinfo {author} {\bibfnamefont {M.}~\bibnamefont {Giustina}}, \bibinfo {author} {\bibfnamefont {A.}~\bibnamefont {Grajales~Dau}}, \bibinfo {author} {\bibfnamefont {M.~C.}\ \bibnamefont {Hamilton}}, \bibinfo {author} {\bibfnamefont {S.~D.}\ \bibnamefont {Harrington}}, \bibinfo {author} {\bibfnamefont {J.}~\bibnamefont {Hilton}}, \bibinfo {author} {\bibfnamefont {M.}~\bibnamefont {Hoffmann}}, \bibinfo {author} {\bibfnamefont {S.}~\bibnamefont {Hong}}, \bibinfo {author} {\bibfnamefont {T.}~\bibnamefont {Huang}}, \bibinfo {author} {\bibfnamefont {A.}~\bibnamefont {Huff}}, \bibinfo {author} {\bibfnamefont {J.}~\bibnamefont {Iveland}}, \bibinfo {author} {\bibfnamefont {E.}~\bibnamefont {Jeffrey}}, \bibinfo {author} {\bibfnamefont {M.}~\bibnamefont {Kieferov{\'a}}}, \bibinfo {author} {\bibfnamefont {S.}~\bibnamefont {Kim}}, \bibinfo {author} {\bibfnamefont {P.~V.}\ \bibnamefont {Klimov}}, \bibinfo {author} {\bibfnamefont {F.}~\bibnamefont {Kostritsa}}, \bibinfo {author} {\bibfnamefont {J.~M.}\ \bibnamefont {Kreikebaum}}, \bibinfo {author} {\bibfnamefont {D.}~\bibnamefont {Landhuis}}, \bibinfo {author} {\bibfnamefont {P.}~\bibnamefont {Laptev}}, \bibinfo {author} {\bibfnamefont {L.}~\bibnamefont {Laws}}, \bibinfo {author} {\bibfnamefont {K.}~\bibnamefont {Lee}}, \bibinfo {author} {\bibfnamefont {B.~J.}\ \bibnamefont {Lester}}, \bibinfo {author} {\bibfnamefont {A.}~\bibnamefont {Lill}}, \bibinfo {author} {\bibfnamefont {W.}~\bibnamefont {Liu}}, \bibinfo {author} {\bibfnamefont {A.}~\bibnamefont {Locharla}}, \bibinfo {author} {\bibfnamefont {E.}~\bibnamefont {Lucero}}, \bibinfo {author} {\bibfnamefont {T.}~\bibnamefont {McCourt}}, \bibinfo {author} {\bibfnamefont {M.}~\bibnamefont {McEwen}}, \bibinfo {author} {\bibfnamefont {X.}~\bibnamefont {Mi}}, \bibinfo {author} {\bibfnamefont {K.~C.}\ \bibnamefont {Miao}}, \bibinfo {author} {\bibfnamefont {S.}~\bibnamefont {Montazeri}}, \bibinfo {author} {\bibfnamefont {A.}~\bibnamefont {Morvan}}, \bibinfo {author} {\bibfnamefont {M.}~\bibnamefont {Neeley}}, \bibinfo {author} {\bibfnamefont {C.}~\bibnamefont {Neill}}, \bibinfo {author} {\bibfnamefont {A.}~\bibnamefont {Nersisyan}}, \bibinfo {author} {\bibfnamefont {J.~H.}\ \bibnamefont {Ng}}, \bibinfo {author} {\bibfnamefont {A.}~\bibnamefont {Nguyen}}, \bibinfo {author} {\bibfnamefont {M.}~\bibnamefont {Nguyen}}, \bibinfo {author} {\bibfnamefont {R.}~\bibnamefont {Potter}}, \bibinfo {author} {\bibfnamefont {C.}~\bibnamefont {Quintana}}, \bibinfo {author} {\bibfnamefont {P.}~\bibnamefont {Roushan}}, \bibinfo {author} {\bibfnamefont {K.}~\bibnamefont {Sankaragomathi}}, \bibinfo {author} {\bibfnamefont {K.~J.}\ \bibnamefont {Satzinger}}, \bibinfo {author} {\bibfnamefont {C.}~\bibnamefont {Schuster}}, \bibinfo {author} {\bibfnamefont {M.~J.}\ \bibnamefont {Shearn}}, \bibinfo {author} {\bibfnamefont {A.}~\bibnamefont {Shorter}}, \bibinfo {author} {\bibfnamefont {V.}~\bibnamefont {Shvarts}}, \bibinfo {author} {\bibfnamefont {J.}~\bibnamefont {Skruzny}}, \bibinfo {author} {\bibfnamefont {W.~C.}\ \bibnamefont {Smith}}, \bibinfo {author} {\bibfnamefont {M.}~\bibnamefont {Szalay}}, \bibinfo {author} {\bibfnamefont {A.}~\bibnamefont {Torres}}, \bibinfo {author} {\bibfnamefont {B.~W.~K.}\ \bibnamefont {Woo}}, \bibinfo {author} {\bibfnamefont {Z.~J.}\ \bibnamefont {Yao}}, \bibinfo {author} {\bibfnamefont {P.}~\bibnamefont {Yeh}}, \bibinfo {author} {\bibfnamefont {J.}~\bibnamefont {Yoo}}, \bibinfo {author} {\bibfnamefont {G.}~\bibnamefont {Young}}, \bibinfo {author} {\bibfnamefont {N.}~\bibnamefont {Zhu}}, \bibinfo {author} {\bibfnamefont {N.}~\bibnamefont {Zobrist}}, \bibinfo {author} {\bibfnamefont {Y.}~\bibnamefont {Chen}}, \bibinfo {author} {\bibfnamefont {A.}~\bibnamefont {Megrant}}, \bibinfo {author} {\bibfnamefont {J.}~\bibnamefont {Kelly}},\ and\ \bibinfo {author} {\bibfnamefont {O.}~\bibnamefont {Naaman}},\ }\href {https://doi.org/10.1063/5.0127375} {\bibfield  {journal} {\bibinfo  {journal} {Applied Physics Letters}\ }\textbf {\bibinfo {volume} {122}},\ \bibinfo {pages} {014001} (\bibinfo {year} {2023})}\BibitemShut {NoStop}%
\bibitem [{\citenamefont {Macklin}\ \emph {et~al.}(2015)\citenamefont {Macklin}, \citenamefont {O'Brien}, \citenamefont {Hover}, \citenamefont {Schwartz}, \citenamefont {Bolkhovsky}, \citenamefont {Zhang}, \citenamefont {Oliver},\ and\ \citenamefont {Siddiqi}}]{macklinQuantumLimitedJosephson2015}%
  \BibitemOpen
  \bibfield  {author} {\bibinfo {author} {\bibfnamefont {C.}~\bibnamefont {Macklin}}, \bibinfo {author} {\bibfnamefont {K.}~\bibnamefont {O'Brien}}, \bibinfo {author} {\bibfnamefont {D.}~\bibnamefont {Hover}}, \bibinfo {author} {\bibfnamefont {M.~E.}\ \bibnamefont {Schwartz}}, \bibinfo {author} {\bibfnamefont {V.}~\bibnamefont {Bolkhovsky}}, \bibinfo {author} {\bibfnamefont {X.}~\bibnamefont {Zhang}}, \bibinfo {author} {\bibfnamefont {W.~D.}\ \bibnamefont {Oliver}},\ and\ \bibinfo {author} {\bibfnamefont {I.}~\bibnamefont {Siddiqi}},\ }\href {https://doi.org/10.1126/science.aaa8525} {\bibfield  {journal} {\bibinfo  {journal} {Science}\ }\textbf {\bibinfo {volume} {350}},\ \bibinfo {pages} {307} (\bibinfo {year} {2015})}\BibitemShut {NoStop}%
\bibitem [{\citenamefont {White}\ \emph {et~al.}(2015)\citenamefont {White}, \citenamefont {Mutus}, \citenamefont {Hoi}, \citenamefont {Barends}, \citenamefont {Campbell}, \citenamefont {Chen}, \citenamefont {Chen}, \citenamefont {Chiaro}, \citenamefont {Dunsworth}, \citenamefont {Jeffrey}, \citenamefont {Kelly}, \citenamefont {Megrant}, \citenamefont {Neill}, \citenamefont {O'Malley}, \citenamefont {Roushan}, \citenamefont {Sank}, \citenamefont {Vainsencher}, \citenamefont {Wenner}, \citenamefont {Chaudhuri}, \citenamefont {Gao},\ and\ \citenamefont {Martinis}}]{whiteTravelingWaveParametric2015}%
  \BibitemOpen
  \bibfield  {author} {\bibinfo {author} {\bibfnamefont {T.~C.}\ \bibnamefont {White}}, \bibinfo {author} {\bibfnamefont {J.~Y.}\ \bibnamefont {Mutus}}, \bibinfo {author} {\bibfnamefont {I.-C.}\ \bibnamefont {Hoi}}, \bibinfo {author} {\bibfnamefont {R.}~\bibnamefont {Barends}}, \bibinfo {author} {\bibfnamefont {B.}~\bibnamefont {Campbell}}, \bibinfo {author} {\bibfnamefont {Y.}~\bibnamefont {Chen}}, \bibinfo {author} {\bibfnamefont {Z.}~\bibnamefont {Chen}}, \bibinfo {author} {\bibfnamefont {B.}~\bibnamefont {Chiaro}}, \bibinfo {author} {\bibfnamefont {A.}~\bibnamefont {Dunsworth}}, \bibinfo {author} {\bibfnamefont {E.}~\bibnamefont {Jeffrey}}, \bibinfo {author} {\bibfnamefont {J.}~\bibnamefont {Kelly}}, \bibinfo {author} {\bibfnamefont {A.}~\bibnamefont {Megrant}}, \bibinfo {author} {\bibfnamefont {C.}~\bibnamefont {Neill}}, \bibinfo {author} {\bibfnamefont {P.~J.~J.}\ \bibnamefont {O'Malley}}, \bibinfo {author} {\bibfnamefont {P.}~\bibnamefont {Roushan}}, \bibinfo {author} {\bibfnamefont {D.}~\bibnamefont {Sank}}, \bibinfo {author} {\bibfnamefont {A.}~\bibnamefont {Vainsencher}}, \bibinfo {author} {\bibfnamefont {J.}~\bibnamefont {Wenner}}, \bibinfo {author} {\bibfnamefont {S.}~\bibnamefont {Chaudhuri}}, \bibinfo {author} {\bibfnamefont {J.}~\bibnamefont {Gao}},\ and\ \bibinfo {author} {\bibfnamefont {J.~M.}\ \bibnamefont {Martinis}},\ }\href {https://doi.org/10.1063/1.4922348} {\bibfield  {journal} {\bibinfo  {journal} {Applied Physics Letters}\ }\textbf {\bibinfo {volume} {106}},\ \bibinfo {pages} {242601} (\bibinfo {year} {2015})}\BibitemShut {NoStop}%
\bibitem [{\citenamefont {Planat}\ \emph {et~al.}(2020)\citenamefont {Planat}, \citenamefont {Ranadive}, \citenamefont {Dassonneville}, \citenamefont {Puertas~Mart{\'i}nez}, \citenamefont {L{\'e}ger}, \citenamefont {Naud}, \citenamefont {Buisson}, \citenamefont {{Hasch-Guichard}}, \citenamefont {Basko},\ and\ \citenamefont {Roch}}]{planatPhotonicCrystalJosephsonTravelingWave2020}%
  \BibitemOpen
  \bibfield  {author} {\bibinfo {author} {\bibfnamefont {L.}~\bibnamefont {Planat}}, \bibinfo {author} {\bibfnamefont {A.}~\bibnamefont {Ranadive}}, \bibinfo {author} {\bibfnamefont {R.}~\bibnamefont {Dassonneville}}, \bibinfo {author} {\bibfnamefont {J.}~\bibnamefont {Puertas~Mart{\'i}nez}}, \bibinfo {author} {\bibfnamefont {S.}~\bibnamefont {L{\'e}ger}}, \bibinfo {author} {\bibfnamefont {C.}~\bibnamefont {Naud}}, \bibinfo {author} {\bibfnamefont {O.}~\bibnamefont {Buisson}}, \bibinfo {author} {\bibfnamefont {W.}~\bibnamefont {{Hasch-Guichard}}}, \bibinfo {author} {\bibfnamefont {D.~M.}\ \bibnamefont {Basko}},\ and\ \bibinfo {author} {\bibfnamefont {N.}~\bibnamefont {Roch}},\ }\href {https://doi.org/10.1103/PhysRevX.10.021021} {\bibfield  {journal} {\bibinfo  {journal} {Physical Review X}\ }\textbf {\bibinfo {volume} {10}},\ \bibinfo {pages} {021021} (\bibinfo {year} {2020})}\BibitemShut {NoStop}%
\bibitem [{\citenamefont {Feng}\ \emph {et~al.}(2020)\citenamefont {Feng}, \citenamefont {Vahidpour}, \citenamefont {Mohan}, \citenamefont {Sharac}, \citenamefont {Whyland}, \citenamefont {Stanwyck}, \citenamefont {Ramachandran},\ and\ \citenamefont {Selvanayagam}}]{fengDesignMeasurementJosephson2020}%
  \BibitemOpen
  \bibfield  {author} {\bibinfo {author} {\bibfnamefont {D.~C.}\ \bibnamefont {Feng}}, \bibinfo {author} {\bibfnamefont {M.}~\bibnamefont {Vahidpour}}, \bibinfo {author} {\bibfnamefont {Y.}~\bibnamefont {Mohan}}, \bibinfo {author} {\bibfnamefont {N.}~\bibnamefont {Sharac}}, \bibinfo {author} {\bibfnamefont {T.}~\bibnamefont {Whyland}}, \bibinfo {author} {\bibfnamefont {S.}~\bibnamefont {Stanwyck}}, \bibinfo {author} {\bibfnamefont {G.}~\bibnamefont {Ramachandran}},\ and\ \bibinfo {author} {\bibfnamefont {M.}~\bibnamefont {Selvanayagam}},\ }in\ \href {https://doi.org/10.1109/IMS30576.2020.9223912} {\emph {\bibinfo {booktitle} {2020 {{IEEE}}/{{MTT-S International Microwave Symposium}} ({{IMS}})}}}\ (\bibinfo  {publisher} {IEEE},\ \bibinfo {address} {Los Angeles, CA, USA},\ \bibinfo {year} {2020})\ pp.\ \bibinfo {pages} {940--943}\BibitemShut {NoStop}%
\bibitem [{\citenamefont {Perelshtein}\ \emph {et~al.}(2022)\citenamefont {Perelshtein}, \citenamefont {Petrovnin}, \citenamefont {Vesterinen}, \citenamefont {Hamedani~Raja}, \citenamefont {Lilja}, \citenamefont {Will}, \citenamefont {Savin}, \citenamefont {Simbierowicz}, \citenamefont {Jabdaraghi}, \citenamefont {Lehtinen}, \citenamefont {Gr{\"o}nberg}, \citenamefont {Hassel}, \citenamefont {Prunnila}, \citenamefont {Govenius}, \citenamefont {Paraoanu},\ and\ \citenamefont {Hakonen}}]{perelshteinBroadbandContinuousVariableEntanglement2022}%
  \BibitemOpen
  \bibfield  {author} {\bibinfo {author} {\bibfnamefont {M.}~\bibnamefont {Perelshtein}}, \bibinfo {author} {\bibfnamefont {K.}~\bibnamefont {Petrovnin}}, \bibinfo {author} {\bibfnamefont {V.}~\bibnamefont {Vesterinen}}, \bibinfo {author} {\bibfnamefont {S.}~\bibnamefont {Hamedani~Raja}}, \bibinfo {author} {\bibfnamefont {I.}~\bibnamefont {Lilja}}, \bibinfo {author} {\bibfnamefont {M.}~\bibnamefont {Will}}, \bibinfo {author} {\bibfnamefont {A.}~\bibnamefont {Savin}}, \bibinfo {author} {\bibfnamefont {S.}~\bibnamefont {Simbierowicz}}, \bibinfo {author} {\bibfnamefont {R.}~\bibnamefont {Jabdaraghi}}, \bibinfo {author} {\bibfnamefont {J.}~\bibnamefont {Lehtinen}}, \bibinfo {author} {\bibfnamefont {L.}~\bibnamefont {Gr{\"o}nberg}}, \bibinfo {author} {\bibfnamefont {J.}~\bibnamefont {Hassel}}, \bibinfo {author} {\bibfnamefont {M.}~\bibnamefont {Prunnila}}, \bibinfo {author} {\bibfnamefont {J.}~\bibnamefont {Govenius}}, \bibinfo {author} {\bibfnamefont {G.}~\bibnamefont {Paraoanu}},\ and\ \bibinfo {author} {\bibfnamefont {P.}~\bibnamefont {Hakonen}},\ }\href {https://doi.org/10.1103/PhysRevApplied.18.024063} {\bibfield  {journal} {\bibinfo  {journal} {Physical Review Applied}\ }\textbf {\bibinfo {volume} {18}},\ \bibinfo {pages} {024063} (\bibinfo {year} {2022})}\BibitemShut {NoStop}%
\bibitem [{\citenamefont {Ranadive}\ \emph {et~al.}(2022)\citenamefont {Ranadive}, \citenamefont {Esposito}, \citenamefont {Planat}, \citenamefont {Bonet}, \citenamefont {Naud}, \citenamefont {Buisson}, \citenamefont {Guichard},\ and\ \citenamefont {Roch}}]{ranadiveKerrReversalJosephson2022}%
  \BibitemOpen
  \bibfield  {author} {\bibinfo {author} {\bibfnamefont {A.}~\bibnamefont {Ranadive}}, \bibinfo {author} {\bibfnamefont {M.}~\bibnamefont {Esposito}}, \bibinfo {author} {\bibfnamefont {L.}~\bibnamefont {Planat}}, \bibinfo {author} {\bibfnamefont {E.}~\bibnamefont {Bonet}}, \bibinfo {author} {\bibfnamefont {C.}~\bibnamefont {Naud}}, \bibinfo {author} {\bibfnamefont {O.}~\bibnamefont {Buisson}}, \bibinfo {author} {\bibfnamefont {W.}~\bibnamefont {Guichard}},\ and\ \bibinfo {author} {\bibfnamefont {N.}~\bibnamefont {Roch}},\ }\href {https://doi.org/10.1038/s41467-022-29375-5} {\bibfield  {journal} {\bibinfo  {journal} {Nature Communications}\ }\textbf {\bibinfo {volume} {13}},\ \bibinfo {pages} {1737} (\bibinfo {year} {2022})}\BibitemShut {NoStop}%
\bibitem [{\citenamefont {Kono}\ \emph {et~al.}(2024)\citenamefont {Kono}, \citenamefont {Pan}, \citenamefont {Chegnizadeh}, \citenamefont {Wang}, \citenamefont {Youssefi}, \citenamefont {Scigliuzzo},\ and\ \citenamefont {Kippenberg}}]{konoMechanicallyInducedCorrelated2024}%
  \BibitemOpen
  \bibfield  {author} {\bibinfo {author} {\bibfnamefont {S.}~\bibnamefont {Kono}}, \bibinfo {author} {\bibfnamefont {J.}~\bibnamefont {Pan}}, \bibinfo {author} {\bibfnamefont {M.}~\bibnamefont {Chegnizadeh}}, \bibinfo {author} {\bibfnamefont {X.}~\bibnamefont {Wang}}, \bibinfo {author} {\bibfnamefont {A.}~\bibnamefont {Youssefi}}, \bibinfo {author} {\bibfnamefont {M.}~\bibnamefont {Scigliuzzo}},\ and\ \bibinfo {author} {\bibfnamefont {T.~J.}\ \bibnamefont {Kippenberg}},\ }\href {https://doi.org/10.1038/s41467-024-48230-3} {\bibfield  {journal} {\bibinfo  {journal} {Nature Communications}\ }\textbf {\bibinfo {volume} {15}},\ \bibinfo {pages} {3950} (\bibinfo {year} {2024})}\BibitemShut {NoStop}%
\bibitem [{\citenamefont {Swiadek}\ \emph {et~al.}(2024)\citenamefont {Swiadek}, \citenamefont {Shillito}, \citenamefont {Magnard}, \citenamefont {Remm}, \citenamefont {Hellings}, \citenamefont {Lacroix}, \citenamefont {Ficheux}, \citenamefont {Zanuz}, \citenamefont {Norris}, \citenamefont {Blais}, \citenamefont {Krinner},\ and\ \citenamefont {Wallraff}}]{swiadekEnhancingDispersiveReadout2024}%
  \BibitemOpen
  \bibfield  {author} {\bibinfo {author} {\bibfnamefont {F.}~\bibnamefont {Swiadek}}, \bibinfo {author} {\bibfnamefont {R.}~\bibnamefont {Shillito}}, \bibinfo {author} {\bibfnamefont {P.}~\bibnamefont {Magnard}}, \bibinfo {author} {\bibfnamefont {A.}~\bibnamefont {Remm}}, \bibinfo {author} {\bibfnamefont {C.}~\bibnamefont {Hellings}}, \bibinfo {author} {\bibfnamefont {N.}~\bibnamefont {Lacroix}}, \bibinfo {author} {\bibfnamefont {Q.}~\bibnamefont {Ficheux}}, \bibinfo {author} {\bibfnamefont {D.~C.}\ \bibnamefont {Zanuz}}, \bibinfo {author} {\bibfnamefont {G.~J.}\ \bibnamefont {Norris}}, \bibinfo {author} {\bibfnamefont {A.}~\bibnamefont {Blais}}, \bibinfo {author} {\bibfnamefont {S.}~\bibnamefont {Krinner}},\ and\ \bibinfo {author} {\bibfnamefont {A.}~\bibnamefont {Wallraff}},\ }\href {https://doi.org/10.1103/PRXQuantum.5.040326} {\bibfield  {journal} {\bibinfo  {journal} {PRX Quantum}\ }\textbf {\bibinfo {volume} {5}},\ \bibinfo {pages} {040326} (\bibinfo {year} {2024})}\BibitemShut {NoStop}%
\bibitem [{\citenamefont {Nie}\ \emph {et~al.}(2024)\citenamefont {Nie}, \citenamefont {Bista}, \citenamefont {Chow}, \citenamefont {Pfaff},\ and\ \citenamefont {Kou}}]{nieParametricallyControlledMicrowavephotonic2024}%
  \BibitemOpen
  \bibfield  {author} {\bibinfo {author} {\bibfnamefont {K.}~\bibnamefont {Nie}}, \bibinfo {author} {\bibfnamefont {A.}~\bibnamefont {Bista}}, \bibinfo {author} {\bibfnamefont {K.}~\bibnamefont {Chow}}, \bibinfo {author} {\bibfnamefont {W.}~\bibnamefont {Pfaff}},\ and\ \bibinfo {author} {\bibfnamefont {A.}~\bibnamefont {Kou}},\ }\href {https://doi.org/10.1103/PhysRevApplied.22.054021} {\bibfield  {journal} {\bibinfo  {journal} {Physical Review Applied}\ }\textbf {\bibinfo {volume} {22}},\ \bibinfo {pages} {054021} (\bibinfo {year} {2024})}\BibitemShut {NoStop}%
\bibitem [{\citenamefont {Meesala}\ \emph {et~al.}(2024)\citenamefont {Meesala}, \citenamefont {Lake}, \citenamefont {Wood}, \citenamefont {Chiappina}, \citenamefont {Zhong}, \citenamefont {Beyer}, \citenamefont {Shaw}, \citenamefont {Jiang},\ and\ \citenamefont {Painter}}]{meesalaQuantumEntanglementOptical2024}%
  \BibitemOpen
  \bibfield  {author} {\bibinfo {author} {\bibfnamefont {S.}~\bibnamefont {Meesala}}, \bibinfo {author} {\bibfnamefont {D.}~\bibnamefont {Lake}}, \bibinfo {author} {\bibfnamefont {S.}~\bibnamefont {Wood}}, \bibinfo {author} {\bibfnamefont {P.}~\bibnamefont {Chiappina}}, \bibinfo {author} {\bibfnamefont {C.}~\bibnamefont {Zhong}}, \bibinfo {author} {\bibfnamefont {A.~D.}\ \bibnamefont {Beyer}}, \bibinfo {author} {\bibfnamefont {M.~D.}\ \bibnamefont {Shaw}}, \bibinfo {author} {\bibfnamefont {L.}~\bibnamefont {Jiang}},\ and\ \bibinfo {author} {\bibfnamefont {O.}~\bibnamefont {Painter}},\ }\href {https://doi.org/10.1103/PhysRevX.14.031055} {\bibfield  {journal} {\bibinfo  {journal} {Physical Review X}\ }\textbf {\bibinfo {volume} {14}},\ \bibinfo {pages} {031055} (\bibinfo {year} {2024})}\BibitemShut {NoStop}%
\bibitem [{\citenamefont {Ahn}\ \emph {et~al.}(2024)\citenamefont {Ahn}, \citenamefont {Kim}, \citenamefont {Ivanov}, \citenamefont {Kwon}, \citenamefont {Byun}, \citenamefont {{van Loo}}, \citenamefont {Park}, \citenamefont {Jeong}, \citenamefont {Lee}, \citenamefont {Kim}, \citenamefont {Kutlu}, \citenamefont {Yi}, \citenamefont {Nakamura}, \citenamefont {Oh}, \citenamefont {Ahn}, \citenamefont {Bae}, \citenamefont {Choi}, \citenamefont {Choi}, \citenamefont {Chong}, \citenamefont {Chung}, \citenamefont {Gkika}, \citenamefont {Kim}, \citenamefont {Kim}, \citenamefont {Ko}, \citenamefont {Miceli}, \citenamefont {Lee}, \citenamefont {Lee}, \citenamefont {Lee}, \citenamefont {Lee}, \citenamefont {Matlashov}, \citenamefont {Parashar}, \citenamefont {Seong}, \citenamefont {Shin}, \citenamefont {Uchaikin}, \citenamefont {Youn},\ and\ \citenamefont {Semertzidis}}]{ahnExtensiveSearchAxion2024}%
  \BibitemOpen
  \bibfield  {author} {\bibinfo {author} {\bibfnamefont {S.}~\bibnamefont {Ahn}}, \bibinfo {author} {\bibfnamefont {J.}~\bibnamefont {Kim}}, \bibinfo {author} {\bibfnamefont {B.~I.}\ \bibnamefont {Ivanov}}, \bibinfo {author} {\bibfnamefont {O.}~\bibnamefont {Kwon}}, \bibinfo {author} {\bibfnamefont {H.}~\bibnamefont {Byun}}, \bibinfo {author} {\bibfnamefont {A.~F.}\ \bibnamefont {{van Loo}}}, \bibinfo {author} {\bibfnamefont {S.}~\bibnamefont {Park}}, \bibinfo {author} {\bibfnamefont {J.}~\bibnamefont {Jeong}}, \bibinfo {author} {\bibfnamefont {S.}~\bibnamefont {Lee}}, \bibinfo {author} {\bibfnamefont {J.}~\bibnamefont {Kim}}, \bibinfo {author} {\bibfnamefont {{\c C}.}~\bibnamefont {Kutlu}}, \bibinfo {author} {\bibfnamefont {A.~K.}\ \bibnamefont {Yi}}, \bibinfo {author} {\bibfnamefont {Y.}~\bibnamefont {Nakamura}}, \bibinfo {author} {\bibfnamefont {S.}~\bibnamefont {Oh}}, \bibinfo {author} {\bibfnamefont {D.}~\bibnamefont {Ahn}}, \bibinfo {author} {\bibfnamefont {S.}~\bibnamefont {Bae}}, \bibinfo {author} {\bibfnamefont {H.}~\bibnamefont {Choi}}, \bibinfo {author} {\bibfnamefont {J.}~\bibnamefont {Choi}}, \bibinfo {author} {\bibfnamefont {Y.}~\bibnamefont {Chong}}, \bibinfo {author} {\bibfnamefont {W.}~\bibnamefont {Chung}}, \bibinfo {author} {\bibfnamefont {V.}~\bibnamefont {Gkika}}, \bibinfo {author} {\bibfnamefont {J.~E.}\ \bibnamefont {Kim}}, \bibinfo {author} {\bibfnamefont {Y.}~\bibnamefont {Kim}}, \bibinfo {author} {\bibfnamefont {B.~R.}\ \bibnamefont {Ko}}, \bibinfo {author} {\bibfnamefont {L.}~\bibnamefont {Miceli}}, \bibinfo {author} {\bibfnamefont {D.}~\bibnamefont {Lee}}, \bibinfo {author} {\bibfnamefont {J.}~\bibnamefont {Lee}}, \bibinfo {author} {\bibfnamefont {K.~W.}\ \bibnamefont {Lee}}, \bibinfo {author} {\bibfnamefont {M.}~\bibnamefont {Lee}}, \bibinfo {author} {\bibfnamefont {A.}~\bibnamefont {Matlashov}}, \bibinfo {author} {\bibfnamefont {P.}~\bibnamefont {Parashar}}, \bibinfo {author} {\bibfnamefont {T.}~\bibnamefont {Seong}}, \bibinfo {author} {\bibfnamefont {Y.~C.}\ \bibnamefont {Shin}}, \bibinfo {author} {\bibfnamefont {S.~V.}\ \bibnamefont {Uchaikin}}, \bibinfo {author} {\bibfnamefont {S.}~\bibnamefont {Youn}},\ and\ \bibinfo {author} {\bibfnamefont {Y.~K.}\ \bibnamefont {Semertzidis}},\ }\href {https://doi.org/10.1103/PhysRevX.14.031023} {\bibfield  {journal} {\bibinfo  {journal} {Physical Review X}\ }\textbf {\bibinfo {volume} {14}},\ \bibinfo {pages} {031023} (\bibinfo {year} {2024})}\BibitemShut {NoStop}%
\bibitem [{\citenamefont {Fraudet}\ \emph {et~al.}(2025)\citenamefont {Fraudet}, \citenamefont {Snyman}, \citenamefont {Basko}, \citenamefont {L{\'e}ger}, \citenamefont {S{\'e}pulcre}, \citenamefont {Ranadive}, \citenamefont {Le~Gal}, \citenamefont {{Torras-Coloma}}, \citenamefont {Guichard}, \citenamefont {Florens},\ and\ \citenamefont {Roch}}]{fraudetDirectDetectionDownConverted2025}%
  \BibitemOpen
  \bibfield  {author} {\bibinfo {author} {\bibfnamefont {D.}~\bibnamefont {Fraudet}}, \bibinfo {author} {\bibfnamefont {I.}~\bibnamefont {Snyman}}, \bibinfo {author} {\bibfnamefont {D.~M.}\ \bibnamefont {Basko}}, \bibinfo {author} {\bibfnamefont {S.}~\bibnamefont {L{\'e}ger}}, \bibinfo {author} {\bibfnamefont {T.}~\bibnamefont {S{\'e}pulcre}}, \bibinfo {author} {\bibfnamefont {A.}~\bibnamefont {Ranadive}}, \bibinfo {author} {\bibfnamefont {G.}~\bibnamefont {Le~Gal}}, \bibinfo {author} {\bibfnamefont {A.}~\bibnamefont {{Torras-Coloma}}}, \bibinfo {author} {\bibfnamefont {W.}~\bibnamefont {Guichard}}, \bibinfo {author} {\bibfnamefont {S.}~\bibnamefont {Florens}},\ and\ \bibinfo {author} {\bibfnamefont {N.}~\bibnamefont {Roch}},\ }\href {https://doi.org/10.1103/PhysRevLett.134.013804} {\bibfield  {journal} {\bibinfo  {journal} {Physical Review Letters}\ }\textbf {\bibinfo {volume} {134}},\ \bibinfo {pages} {013804} (\bibinfo {year} {2025})}\BibitemShut {NoStop}%
\bibitem [{\citenamefont {Heinsoo}\ \emph {et~al.}(2018)\citenamefont {Heinsoo}, \citenamefont {Andersen}, \citenamefont {Remm}, \citenamefont {Krinner}, \citenamefont {Walter}, \citenamefont {Salath{\'e}}, \citenamefont {Gasparinetti}, \citenamefont {Besse}, \citenamefont {Poto{\v c}nik}, \citenamefont {Wallraff},\ and\ \citenamefont {Eichler}}]{heinsooRapidHighfidelityMultiplexed2018b}%
  \BibitemOpen
  \bibfield  {author} {\bibinfo {author} {\bibfnamefont {J.}~\bibnamefont {Heinsoo}}, \bibinfo {author} {\bibfnamefont {C.~K.}\ \bibnamefont {Andersen}}, \bibinfo {author} {\bibfnamefont {A.}~\bibnamefont {Remm}}, \bibinfo {author} {\bibfnamefont {S.}~\bibnamefont {Krinner}}, \bibinfo {author} {\bibfnamefont {T.}~\bibnamefont {Walter}}, \bibinfo {author} {\bibfnamefont {Y.}~\bibnamefont {Salath{\'e}}}, \bibinfo {author} {\bibfnamefont {S.}~\bibnamefont {Gasparinetti}}, \bibinfo {author} {\bibfnamefont {J.-C.}\ \bibnamefont {Besse}}, \bibinfo {author} {\bibfnamefont {A.}~\bibnamefont {Poto{\v c}nik}}, \bibinfo {author} {\bibfnamefont {A.}~\bibnamefont {Wallraff}},\ and\ \bibinfo {author} {\bibfnamefont {C.}~\bibnamefont {Eichler}},\ }\href {https://doi.org/10.1103/PhysRevApplied.10.034040} {\bibfield  {journal} {\bibinfo  {journal} {Physical Review Applied}\ }\textbf {\bibinfo {volume} {10}},\ \bibinfo {pages} {034040} (\bibinfo {year} {2018})}\BibitemShut {NoStop}%
\bibitem [{\citenamefont {Remm}\ \emph {et~al.}(2022)\citenamefont {Remm}, \citenamefont {Krinner}, \citenamefont {Lacroix}, \citenamefont {Hellings}, \citenamefont {Swiadek}, \citenamefont {Norris}, \citenamefont {Eichler},\ and\ \citenamefont {Wallraff}}]{remmIntermodulationDistortionJosephson2022}%
  \BibitemOpen
  \bibfield  {author} {\bibinfo {author} {\bibfnamefont {A.}~\bibnamefont {Remm}}, \bibinfo {author} {\bibfnamefont {S.}~\bibnamefont {Krinner}}, \bibinfo {author} {\bibfnamefont {N.}~\bibnamefont {Lacroix}}, \bibinfo {author} {\bibfnamefont {C.}~\bibnamefont {Hellings}}, \bibinfo {author} {\bibfnamefont {F.}~\bibnamefont {Swiadek}}, \bibinfo {author} {\bibfnamefont {G.}~\bibnamefont {Norris}}, \bibinfo {author} {\bibfnamefont {C.}~\bibnamefont {Eichler}},\ and\ \bibinfo {author} {\bibfnamefont {A.}~\bibnamefont {Wallraff}},\ }\href {https://doi.org/10.48550/arXiv.2210.04799} {\bibinfo {title} {Intermodulation {{Distortion}} in a {{Josephson Traveling Wave Parametric Amplifier}}}} (\bibinfo {year} {2022}),\ \Eprint {https://arxiv.org/abs/2210.04799} {arXiv:2210.04799 [cond-mat, physics:quant-ph]} \BibitemShut {NoStop}%
\bibitem [{\citenamefont {Kissling}\ \emph {et~al.}(2023)\citenamefont {Kissling}, \citenamefont {Gaydamachenko}, \citenamefont {Kaap}, \citenamefont {Khabipov}, \citenamefont {Dolata}, \citenamefont {Zorin},\ and\ \citenamefont {Grunhaupt}}]{kisslingVulnerabilityParameterSpread2023}%
  \BibitemOpen
  \bibfield  {author} {\bibinfo {author} {\bibfnamefont {C.}~\bibnamefont {Kissling}}, \bibinfo {author} {\bibfnamefont {V.}~\bibnamefont {Gaydamachenko}}, \bibinfo {author} {\bibfnamefont {F.}~\bibnamefont {Kaap}}, \bibinfo {author} {\bibfnamefont {M.}~\bibnamefont {Khabipov}}, \bibinfo {author} {\bibfnamefont {R.}~\bibnamefont {Dolata}}, \bibinfo {author} {\bibfnamefont {A.~B.}\ \bibnamefont {Zorin}},\ and\ \bibinfo {author} {\bibfnamefont {L.}~\bibnamefont {Grunhaupt}},\ }\href {https://doi.org/10.1109/TASC.2023.3242927} {\bibfield  {journal} {\bibinfo  {journal} {IEEE Transactions on Applied Superconductivity}\ }\textbf {\bibinfo {volume} {33}},\ \bibinfo {pages} {1} (\bibinfo {year} {2023})}\BibitemShut {NoStop}%
\bibitem [{\citenamefont {Peat{\'a}in}\ \emph {et~al.}(2023)\citenamefont {Peat{\'a}in}, \citenamefont {Dixon}, \citenamefont {Meeson}, \citenamefont {Williams}, \citenamefont {Kafanov},\ and\ \citenamefont {Pashkin}}]{peatainSimulatingEffectsFabrication2023}%
  \BibitemOpen
  \bibfield  {author} {\bibinfo {author} {\bibfnamefont {S.~{\'O}.}\ \bibnamefont {Peat{\'a}in}}, \bibinfo {author} {\bibfnamefont {T.}~\bibnamefont {Dixon}}, \bibinfo {author} {\bibfnamefont {P.~J.}\ \bibnamefont {Meeson}}, \bibinfo {author} {\bibfnamefont {J.~M.}\ \bibnamefont {Williams}}, \bibinfo {author} {\bibfnamefont {S.}~\bibnamefont {Kafanov}},\ and\ \bibinfo {author} {\bibfnamefont {Y.~A.}\ \bibnamefont {Pashkin}},\ }\href {https://doi.org/10.1088/1361-6668/acba4e} {\bibfield  {journal} {\bibinfo  {journal} {Superconductor Science and Technology}\ }\textbf {\bibinfo {volume} {36}},\ \bibinfo {pages} {045017} (\bibinfo {year} {2023})}\BibitemShut {NoStop}%
\bibitem [{\citenamefont {Sweeny}\ and\ \citenamefont {Mahler}(1985)}]{sweenyTravellingwaveParametricAmplifier1985a}%
  \BibitemOpen
  \bibfield  {author} {\bibinfo {author} {\bibfnamefont {M.}~\bibnamefont {Sweeny}}\ and\ \bibinfo {author} {\bibfnamefont {R.}~\bibnamefont {Mahler}},\ }\href {https://doi.org/10.1109/TMAG.1985.1063777} {\bibfield  {journal} {\bibinfo  {journal} {IEEE Transactions on Magnetics}\ }\textbf {\bibinfo {volume} {21}},\ \bibinfo {pages} {654} (\bibinfo {year} {1985})}\BibitemShut {NoStop}%
\bibitem [{\citenamefont {Mohebbi}\ and\ \citenamefont {Majedi}(2009)}]{mohebbiAnalysisSeriesConnectedDiscrete2009}%
  \BibitemOpen
  \bibfield  {author} {\bibinfo {author} {\bibfnamefont {H.}~\bibnamefont {Mohebbi}}\ and\ \bibinfo {author} {\bibfnamefont {A.}~\bibnamefont {Majedi}},\ }\href {https://doi.org/10.1109/TMTT.2009.2025413} {\bibfield  {journal} {\bibinfo  {journal} {IEEE Transactions on Microwave Theory and Techniques}\ }\textbf {\bibinfo {volume} {57}},\ \bibinfo {pages} {1865} (\bibinfo {year} {2009})}\BibitemShut {NoStop}%
\bibitem [{\citenamefont {Yaakobi}\ \emph {et~al.}(2013)\citenamefont {Yaakobi}, \citenamefont {Friedland}, \citenamefont {Macklin},\ and\ \citenamefont {Siddiqi}}]{yaakobiParametricAmplificationJosephson2013}%
  \BibitemOpen
  \bibfield  {author} {\bibinfo {author} {\bibfnamefont {O.}~\bibnamefont {Yaakobi}}, \bibinfo {author} {\bibfnamefont {L.}~\bibnamefont {Friedland}}, \bibinfo {author} {\bibfnamefont {C.}~\bibnamefont {Macklin}},\ and\ \bibinfo {author} {\bibfnamefont {I.}~\bibnamefont {Siddiqi}},\ }\href {https://doi.org/10.1103/PhysRevB.87.144301} {\bibfield  {journal} {\bibinfo  {journal} {Physical Review B}\ }\textbf {\bibinfo {volume} {87}},\ \bibinfo {pages} {144301} (\bibinfo {year} {2013})}\BibitemShut {NoStop}%
\bibitem [{\citenamefont {O'Brien}\ \emph {et~al.}(2014)\citenamefont {O'Brien}, \citenamefont {Macklin}, \citenamefont {Siddiqi},\ and\ \citenamefont {Zhang}}]{obrienResonantPhaseMatching2014}%
  \BibitemOpen
  \bibfield  {author} {\bibinfo {author} {\bibfnamefont {K.}~\bibnamefont {O'Brien}}, \bibinfo {author} {\bibfnamefont {C.}~\bibnamefont {Macklin}}, \bibinfo {author} {\bibfnamefont {I.}~\bibnamefont {Siddiqi}},\ and\ \bibinfo {author} {\bibfnamefont {X.}~\bibnamefont {Zhang}},\ }\href {https://doi.org/10.1103/PhysRevLett.113.157001} {\bibfield  {journal} {\bibinfo  {journal} {Physical Review Letters}\ }\textbf {\bibinfo {volume} {113}},\ \bibinfo {pages} {157001} (\bibinfo {year} {2014})}\BibitemShut {NoStop}%
\bibitem [{\citenamefont {Frattini}\ \emph {et~al.}(2017)\citenamefont {Frattini}, \citenamefont {Vool}, \citenamefont {Shankar}, \citenamefont {Narla}, \citenamefont {Sliwa},\ and\ \citenamefont {Devoret}}]{frattini3waveMixingJosephson2017a}%
  \BibitemOpen
  \bibfield  {author} {\bibinfo {author} {\bibfnamefont {N.~E.}\ \bibnamefont {Frattini}}, \bibinfo {author} {\bibfnamefont {U.}~\bibnamefont {Vool}}, \bibinfo {author} {\bibfnamefont {S.}~\bibnamefont {Shankar}}, \bibinfo {author} {\bibfnamefont {A.}~\bibnamefont {Narla}}, \bibinfo {author} {\bibfnamefont {K.~M.}\ \bibnamefont {Sliwa}},\ and\ \bibinfo {author} {\bibfnamefont {M.~H.}\ \bibnamefont {Devoret}},\ }\href {https://doi.org/10.1063/1.4984142} {\bibfield  {journal} {\bibinfo  {journal} {Applied Physics Letters}\ }\textbf {\bibinfo {volume} {110}},\ \bibinfo {pages} {222603} (\bibinfo {year} {2017})}\BibitemShut {NoStop}%
\bibitem [{\citenamefont {Fadavi~Roudsari}\ \emph {et~al.}(2023)\citenamefont {Fadavi~Roudsari}, \citenamefont {Shiri}, \citenamefont {Renberg~Nilsson}, \citenamefont {Tancredi}, \citenamefont {Osman}, \citenamefont {Svensson}, \citenamefont {Kudra}, \citenamefont {Rommel}, \citenamefont {Bylander}, \citenamefont {Shumeiko},\ and\ \citenamefont {Delsing}}]{fadaviroudsariThreewaveMixingTravelingwave2023}%
  \BibitemOpen
  \bibfield  {author} {\bibinfo {author} {\bibfnamefont {A.}~\bibnamefont {Fadavi~Roudsari}}, \bibinfo {author} {\bibfnamefont {D.}~\bibnamefont {Shiri}}, \bibinfo {author} {\bibfnamefont {H.}~\bibnamefont {Renberg~Nilsson}}, \bibinfo {author} {\bibfnamefont {G.}~\bibnamefont {Tancredi}}, \bibinfo {author} {\bibfnamefont {A.}~\bibnamefont {Osman}}, \bibinfo {author} {\bibfnamefont {I.-M.}\ \bibnamefont {Svensson}}, \bibinfo {author} {\bibfnamefont {M.}~\bibnamefont {Kudra}}, \bibinfo {author} {\bibfnamefont {M.}~\bibnamefont {Rommel}}, \bibinfo {author} {\bibfnamefont {J.}~\bibnamefont {Bylander}}, \bibinfo {author} {\bibfnamefont {V.}~\bibnamefont {Shumeiko}},\ and\ \bibinfo {author} {\bibfnamefont {P.}~\bibnamefont {Delsing}},\ }\href {https://doi.org/10.1063/5.0127690} {\bibfield  {journal} {\bibinfo  {journal} {Applied Physics Letters}\ }\textbf {\bibinfo {volume} {122}},\ \bibinfo {pages} {052601} (\bibinfo {year} {2023})}\BibitemShut {NoStop}%
\bibitem [{\citenamefont {Chaudhuri}\ \emph {et~al.}(2017)\citenamefont {Chaudhuri}, \citenamefont {Li}, \citenamefont {Irwin}, \citenamefont {Bockstiegel}, \citenamefont {Hubmayr}, \citenamefont {Ullom}, \citenamefont {Vissers},\ and\ \citenamefont {Gao}}]{chaudhuriBroadbandParametricAmplifiers2017}%
  \BibitemOpen
  \bibfield  {author} {\bibinfo {author} {\bibfnamefont {S.}~\bibnamefont {Chaudhuri}}, \bibinfo {author} {\bibfnamefont {D.}~\bibnamefont {Li}}, \bibinfo {author} {\bibfnamefont {K.~D.}\ \bibnamefont {Irwin}}, \bibinfo {author} {\bibfnamefont {C.}~\bibnamefont {Bockstiegel}}, \bibinfo {author} {\bibfnamefont {J.}~\bibnamefont {Hubmayr}}, \bibinfo {author} {\bibfnamefont {J.~N.}\ \bibnamefont {Ullom}}, \bibinfo {author} {\bibfnamefont {M.~R.}\ \bibnamefont {Vissers}},\ and\ \bibinfo {author} {\bibfnamefont {J.}~\bibnamefont {Gao}},\ }\href {https://doi.org/10.1063/1.4980102} {\bibfield  {journal} {\bibinfo  {journal} {Applied Physics Letters}\ }\textbf {\bibinfo {volume} {110}},\ \bibinfo {pages} {152601} (\bibinfo {year} {2017})}\BibitemShut {NoStop}%
\bibitem [{\citenamefont {Malnou}\ \emph {et~al.}(2021)\citenamefont {Malnou}, \citenamefont {Vissers}, \citenamefont {Wheeler}, \citenamefont {Aumentado}, \citenamefont {Hubmayr}, \citenamefont {Ullom},\ and\ \citenamefont {Gao}}]{malnouThreeWaveMixingKinetic2021}%
  \BibitemOpen
  \bibfield  {author} {\bibinfo {author} {\bibfnamefont {M.}~\bibnamefont {Malnou}}, \bibinfo {author} {\bibfnamefont {M.}~\bibnamefont {Vissers}}, \bibinfo {author} {\bibfnamefont {J.}~\bibnamefont {Wheeler}}, \bibinfo {author} {\bibfnamefont {J.}~\bibnamefont {Aumentado}}, \bibinfo {author} {\bibfnamefont {J.}~\bibnamefont {Hubmayr}}, \bibinfo {author} {\bibfnamefont {J.}~\bibnamefont {Ullom}},\ and\ \bibinfo {author} {\bibfnamefont {J.}~\bibnamefont {Gao}},\ }\href {https://doi.org/10.1103/PRXQuantum.2.010302} {\bibfield  {journal} {\bibinfo  {journal} {PRX Quantum}\ }\textbf {\bibinfo {volume} {2}},\ \bibinfo {pages} {010302} (\bibinfo {year} {2021})}\BibitemShut {NoStop}%
\bibitem [{\citenamefont {Faramarzi}\ \emph {et~al.}(2024)\citenamefont {Faramarzi}, \citenamefont {Stephenson}, \citenamefont {Sypkens}, \citenamefont {Eom}, \citenamefont {LeDuc},\ and\ \citenamefont {Day}}]{faramarzi48GHzKinetic2024}%
  \BibitemOpen
  \bibfield  {author} {\bibinfo {author} {\bibfnamefont {F.}~\bibnamefont {Faramarzi}}, \bibinfo {author} {\bibfnamefont {R.}~\bibnamefont {Stephenson}}, \bibinfo {author} {\bibfnamefont {S.}~\bibnamefont {Sypkens}}, \bibinfo {author} {\bibfnamefont {B.~H.}\ \bibnamefont {Eom}}, \bibinfo {author} {\bibfnamefont {H.}~\bibnamefont {LeDuc}},\ and\ \bibinfo {author} {\bibfnamefont {P.}~\bibnamefont {Day}},\ }\href {https://doi.org/10.1063/5.0208110} {\bibfield  {journal} {\bibinfo  {journal} {APL Quantum}\ }\textbf {\bibinfo {volume} {1}},\ \bibinfo {pages} {036107} (\bibinfo {year} {2024})}\BibitemShut {NoStop}%
\bibitem [{\citenamefont {Sage}\ \emph {et~al.}(2011)\citenamefont {Sage}, \citenamefont {Bolkhovsky}, \citenamefont {Oliver}, \citenamefont {Turek},\ and\ \citenamefont {Welander}}]{sageStudyLossSuperconducting2011a}%
  \BibitemOpen
  \bibfield  {author} {\bibinfo {author} {\bibfnamefont {J.~M.}\ \bibnamefont {Sage}}, \bibinfo {author} {\bibfnamefont {V.}~\bibnamefont {Bolkhovsky}}, \bibinfo {author} {\bibfnamefont {W.~D.}\ \bibnamefont {Oliver}}, \bibinfo {author} {\bibfnamefont {B.}~\bibnamefont {Turek}},\ and\ \bibinfo {author} {\bibfnamefont {P.~B.}\ \bibnamefont {Welander}},\ }\href {https://doi.org/10.1063/1.3552890} {\bibfield  {journal} {\bibinfo  {journal} {Journal of Applied Physics}\ }\textbf {\bibinfo {volume} {109}},\ \bibinfo {pages} {063915} (\bibinfo {year} {2011})}\BibitemShut {NoStop}%
\bibitem [{\citenamefont {Wang}\ \emph {et~al.}(2025)\citenamefont {Wang}, \citenamefont {Peng}, \citenamefont {Knecht}, \citenamefont {Cunningham}, \citenamefont {Lombo}, \citenamefont {Yen}, \citenamefont {Zaidenberg}, \citenamefont {Gingras}, \citenamefont {Niedzielski}, \citenamefont {Stickler}, \citenamefont {Sliwa}, \citenamefont {Serniak}, \citenamefont {Schwartz}, \citenamefont {Oliver},\ and\ \citenamefont {O'Brien}}]{wangHighEfficiencyLowLossFloquetmode2025}%
  \BibitemOpen
  \bibfield  {author} {\bibinfo {author} {\bibfnamefont {J.}~\bibnamefont {Wang}}, \bibinfo {author} {\bibfnamefont {K.}~\bibnamefont {Peng}}, \bibinfo {author} {\bibfnamefont {J.~M.}\ \bibnamefont {Knecht}}, \bibinfo {author} {\bibfnamefont {G.~D.}\ \bibnamefont {Cunningham}}, \bibinfo {author} {\bibfnamefont {A.~E.}\ \bibnamefont {Lombo}}, \bibinfo {author} {\bibfnamefont {A.}~\bibnamefont {Yen}}, \bibinfo {author} {\bibfnamefont {D.~A.}\ \bibnamefont {Zaidenberg}}, \bibinfo {author} {\bibfnamefont {M.}~\bibnamefont {Gingras}}, \bibinfo {author} {\bibfnamefont {B.~M.}\ \bibnamefont {Niedzielski}}, \bibinfo {author} {\bibfnamefont {H.}~\bibnamefont {Stickler}}, \bibinfo {author} {\bibfnamefont {K.}~\bibnamefont {Sliwa}}, \bibinfo {author} {\bibfnamefont {K.}~\bibnamefont {Serniak}}, \bibinfo {author} {\bibfnamefont {M.~E.}\ \bibnamefont {Schwartz}}, \bibinfo {author} {\bibfnamefont {W.~D.}\ \bibnamefont {Oliver}},\ and\ \bibinfo {author} {\bibfnamefont {K.~P.}\ \bibnamefont {O'Brien}},\ }\href {https://doi.org/10.48550/arXiv.2503.11812} {\bibinfo {title} {High-{{Efficiency}}, {{Low-Loss Floquet-mode Traveling Wave Parametric Amplifier}}}} (\bibinfo {year} {2025}),\ \Eprint {https://arxiv.org/abs/2503.11812} {arXiv:2503.11812 [quant-ph]} \BibitemShut {NoStop}%
\bibitem [{\citenamefont {Chen}\ \emph {et~al.}(2014)\citenamefont {Chen}, \citenamefont {Megrant}, \citenamefont {Kelly}, \citenamefont {Barends}, \citenamefont {Bochmann}, \citenamefont {Chen}, \citenamefont {Chiaro}, \citenamefont {Dunsworth}, \citenamefont {Jeffrey}, \citenamefont {Mutus}, \citenamefont {O'Malley}, \citenamefont {Neill}, \citenamefont {Roushan}, \citenamefont {Sank}, \citenamefont {Vainsencher}, \citenamefont {Wenner}, \citenamefont {White}, \citenamefont {Cleland},\ and\ \citenamefont {Martinis}}]{chenFabricationCharacterizationAluminum2014}%
  \BibitemOpen
  \bibfield  {author} {\bibinfo {author} {\bibfnamefont {Z.}~\bibnamefont {Chen}}, \bibinfo {author} {\bibfnamefont {A.}~\bibnamefont {Megrant}}, \bibinfo {author} {\bibfnamefont {J.}~\bibnamefont {Kelly}}, \bibinfo {author} {\bibfnamefont {R.}~\bibnamefont {Barends}}, \bibinfo {author} {\bibfnamefont {J.}~\bibnamefont {Bochmann}}, \bibinfo {author} {\bibfnamefont {Y.}~\bibnamefont {Chen}}, \bibinfo {author} {\bibfnamefont {B.}~\bibnamefont {Chiaro}}, \bibinfo {author} {\bibfnamefont {A.}~\bibnamefont {Dunsworth}}, \bibinfo {author} {\bibfnamefont {E.}~\bibnamefont {Jeffrey}}, \bibinfo {author} {\bibfnamefont {J.~Y.}\ \bibnamefont {Mutus}}, \bibinfo {author} {\bibfnamefont {P.~J.~J.}\ \bibnamefont {O'Malley}}, \bibinfo {author} {\bibfnamefont {C.}~\bibnamefont {Neill}}, \bibinfo {author} {\bibfnamefont {P.}~\bibnamefont {Roushan}}, \bibinfo {author} {\bibfnamefont {D.}~\bibnamefont {Sank}}, \bibinfo {author} {\bibfnamefont {A.}~\bibnamefont {Vainsencher}}, \bibinfo {author} {\bibfnamefont {J.}~\bibnamefont {Wenner}}, \bibinfo {author} {\bibfnamefont {T.~C.}\ \bibnamefont {White}}, \bibinfo {author} {\bibfnamefont {A.~N.}\ \bibnamefont {Cleland}},\ and\ \bibinfo {author} {\bibfnamefont {J.~M.}\ \bibnamefont {Martinis}},\ }\href {https://doi.org/10.1063/1.4863745} {\bibfield  {journal} {\bibinfo  {journal} {Applied Physics Letters}\ }\textbf {\bibinfo {volume} {104}},\ \bibinfo {pages} {052602} (\bibinfo {year} {2014})}\BibitemShut {NoStop}%
\bibitem [{\citenamefont {Dolan}(1977)}]{dolanOffsetMasksLiftoff1977}%
  \BibitemOpen
  \bibfield  {author} {\bibinfo {author} {\bibfnamefont {G.~J.}\ \bibnamefont {Dolan}},\ }\href {https://doi.org/10.1063/1.89690} {\bibfield  {journal} {\bibinfo  {journal} {Applied Physics Letters}\ }\textbf {\bibinfo {volume} {31}},\ \bibinfo {pages} {337} (\bibinfo {year} {1977})}\BibitemShut {NoStop}%
\bibitem [{\citenamefont {Lecocq}\ \emph {et~al.}(2011)\citenamefont {Lecocq}, \citenamefont {Pop}, \citenamefont {Peng}, \citenamefont {Matei}, \citenamefont {Crozes}, \citenamefont {Fournier}, \citenamefont {Naud}, \citenamefont {Guichard},\ and\ \citenamefont {Buisson}}]{lecocqJunctionFabricationShadow2011a}%
  \BibitemOpen
  \bibfield  {author} {\bibinfo {author} {\bibfnamefont {F.}~\bibnamefont {Lecocq}}, \bibinfo {author} {\bibfnamefont {I.~M.}\ \bibnamefont {Pop}}, \bibinfo {author} {\bibfnamefont {Z.}~\bibnamefont {Peng}}, \bibinfo {author} {\bibfnamefont {I.}~\bibnamefont {Matei}}, \bibinfo {author} {\bibfnamefont {T.}~\bibnamefont {Crozes}}, \bibinfo {author} {\bibfnamefont {T.}~\bibnamefont {Fournier}}, \bibinfo {author} {\bibfnamefont {C.}~\bibnamefont {Naud}}, \bibinfo {author} {\bibfnamefont {W.}~\bibnamefont {Guichard}},\ and\ \bibinfo {author} {\bibfnamefont {O.}~\bibnamefont {Buisson}},\ }\href {https://doi.org/10.1088/0957-4484/22/31/315302} {\bibfield  {journal} {\bibinfo  {journal} {Nanotechnology}\ }\textbf {\bibinfo {volume} {22}},\ \bibinfo {pages} {315302} (\bibinfo {year} {2011})}\BibitemShut {NoStop}%
\bibitem [{\citenamefont {Osman}\ \emph {et~al.}(2021)\citenamefont {Osman}, \citenamefont {Simon}, \citenamefont {Bengtsson}, \citenamefont {Kosen}, \citenamefont {Krantz}, \citenamefont {P.~Lozano}, \citenamefont {Scigliuzzo}, \citenamefont {Delsing}, \citenamefont {Bylander},\ and\ \citenamefont {Fadavi~Roudsari}}]{osmanSimplifiedJosephsonjunctionFabrication2021}%
  \BibitemOpen
  \bibfield  {author} {\bibinfo {author} {\bibfnamefont {A.}~\bibnamefont {Osman}}, \bibinfo {author} {\bibfnamefont {J.}~\bibnamefont {Simon}}, \bibinfo {author} {\bibfnamefont {A.}~\bibnamefont {Bengtsson}}, \bibinfo {author} {\bibfnamefont {S.}~\bibnamefont {Kosen}}, \bibinfo {author} {\bibfnamefont {P.}~\bibnamefont {Krantz}}, \bibinfo {author} {\bibfnamefont {D.}~\bibnamefont {P.~Lozano}}, \bibinfo {author} {\bibfnamefont {M.}~\bibnamefont {Scigliuzzo}}, \bibinfo {author} {\bibfnamefont {P.}~\bibnamefont {Delsing}}, \bibinfo {author} {\bibfnamefont {J.}~\bibnamefont {Bylander}},\ and\ \bibinfo {author} {\bibfnamefont {A.}~\bibnamefont {Fadavi~Roudsari}},\ }\href {https://doi.org/10.1063/5.0037093} {\bibfield  {journal} {\bibinfo  {journal} {Applied Physics Letters}\ }\textbf {\bibinfo {volume} {118}},\ \bibinfo {pages} {064002} (\bibinfo {year} {2021})}\BibitemShut {NoStop}%
\bibitem [{\citenamefont {Ho~Eom}\ \emph {et~al.}(2012)\citenamefont {Ho~Eom}, \citenamefont {Day}, \citenamefont {LeDuc},\ and\ \citenamefont {Zmuidzinas}}]{hoeomWidebandLownoiseSuperconducting2012b}%
  \BibitemOpen
  \bibfield  {author} {\bibinfo {author} {\bibfnamefont {B.}~\bibnamefont {Ho~Eom}}, \bibinfo {author} {\bibfnamefont {P.~K.}\ \bibnamefont {Day}}, \bibinfo {author} {\bibfnamefont {H.~G.}\ \bibnamefont {LeDuc}},\ and\ \bibinfo {author} {\bibfnamefont {J.}~\bibnamefont {Zmuidzinas}},\ }\href {https://doi.org/10.1038/nphys2356} {\bibfield  {journal} {\bibinfo  {journal} {Nature Physics}\ }\textbf {\bibinfo {volume} {8}},\ \bibinfo {pages} {623} (\bibinfo {year} {2012})}\BibitemShut {NoStop}%
\bibitem [{\citenamefont {Takagi}(2001)}]{takagiFrequencyDependenceBloch2001a}%
  \BibitemOpen
  \bibfield  {author} {\bibinfo {author} {\bibfnamefont {E.}~\bibnamefont {Takagi}},\ }in\ \href {https://doi.org/10.1109/MWSYM.2001.967008} {\emph {\bibinfo {booktitle} {2001 {{IEEE MTT-S International Microwave Sympsoium Digest}} ({{Cat}}. {{No}}.{{01CH37157}})}}},\ Vol.~\bibinfo {volume} {2}\ (\bibinfo  {publisher} {IEEE},\ \bibinfo {address} {Phoenix, AZ, USA},\ \bibinfo {year} {2001})\ pp.\ \bibinfo {pages} {779--782}\BibitemShut {NoStop}%
\bibitem [{\citenamefont {Gaydamachenko}\ \emph {et~al.}(2022)\citenamefont {Gaydamachenko}, \citenamefont {Kissling}, \citenamefont {Dolata},\ and\ \citenamefont {Zorin}}]{gaydamachenkoNumericalAnalysisThreewavemixing2022}%
  \BibitemOpen
  \bibfield  {author} {\bibinfo {author} {\bibfnamefont {V.}~\bibnamefont {Gaydamachenko}}, \bibinfo {author} {\bibfnamefont {C.}~\bibnamefont {Kissling}}, \bibinfo {author} {\bibfnamefont {R.}~\bibnamefont {Dolata}},\ and\ \bibinfo {author} {\bibfnamefont {A.~B.}\ \bibnamefont {Zorin}},\ }\href {https://doi.org/10.1063/5.0111111} {\bibfield  {journal} {\bibinfo  {journal} {Journal of Applied Physics}\ }\textbf {\bibinfo {volume} {132}},\ \bibinfo {pages} {154401} (\bibinfo {year} {2022})}\BibitemShut {NoStop}%
\bibitem [{\citenamefont {O'Brien}(2025)}]{obrienKpobrienJosephsonCircuitsjl2025}%
  \BibitemOpen
  \bibfield  {author} {\bibinfo {author} {\bibfnamefont {K.~P.}\ \bibnamefont {O'Brien}},\ }\href@noop {} {\bibinfo {title} {Kpobrien/{{JosephsonCircuits}}.jl}} (\bibinfo {year} {2025})\BibitemShut {NoStop}%
\bibitem [{\citenamefont {Kannan}\ \emph {et~al.}(2020)\citenamefont {Kannan}, \citenamefont {Campbell}, \citenamefont {Vasconcelos}, \citenamefont {Winik}, \citenamefont {Kim}, \citenamefont {Kjaergaard}, \citenamefont {Krantz}, \citenamefont {Melville}, \citenamefont {Niedzielski}, \citenamefont {Yoder}, \citenamefont {Orlando}, \citenamefont {Gustavsson},\ and\ \citenamefont {Oliver}}]{kannanGeneratingSpatiallyEntangled2020}%
  \BibitemOpen
  \bibfield  {author} {\bibinfo {author} {\bibfnamefont {B.}~\bibnamefont {Kannan}}, \bibinfo {author} {\bibfnamefont {D.~L.}\ \bibnamefont {Campbell}}, \bibinfo {author} {\bibfnamefont {F.}~\bibnamefont {Vasconcelos}}, \bibinfo {author} {\bibfnamefont {R.}~\bibnamefont {Winik}}, \bibinfo {author} {\bibfnamefont {D.~K.}\ \bibnamefont {Kim}}, \bibinfo {author} {\bibfnamefont {M.}~\bibnamefont {Kjaergaard}}, \bibinfo {author} {\bibfnamefont {P.}~\bibnamefont {Krantz}}, \bibinfo {author} {\bibfnamefont {A.}~\bibnamefont {Melville}}, \bibinfo {author} {\bibfnamefont {B.~M.}\ \bibnamefont {Niedzielski}}, \bibinfo {author} {\bibfnamefont {J.~L.}\ \bibnamefont {Yoder}}, \bibinfo {author} {\bibfnamefont {T.~P.}\ \bibnamefont {Orlando}}, \bibinfo {author} {\bibfnamefont {S.}~\bibnamefont {Gustavsson}},\ and\ \bibinfo {author} {\bibfnamefont {W.~D.}\ \bibnamefont {Oliver}},\ }\href {https://doi.org/10.1126/sciadv.abb8780} {\bibfield  {journal} {\bibinfo  {journal} {Science Advances}\ }\textbf {\bibinfo {volume} {6}},\ \bibinfo {pages} {eabb8780} (\bibinfo {year} {2020})}\BibitemShut {NoStop}%
\bibitem [{Mic()}]{MicrowaveEngineering4th}%
  \BibitemOpen
  \href@noop {} {\bibinfo {title} {Microwave {{Engineering}}, 4th {{Edition}} {\textbar} {{Wiley}}}}\BibitemShut {NoStop}%
\bibitem [{\citenamefont {Boutin}\ \emph {et~al.}(2017)\citenamefont {Boutin}, \citenamefont {Toyli}, \citenamefont {Venkatramani}, \citenamefont {Eddins}, \citenamefont {Siddiqi},\ and\ \citenamefont {Blais}}]{boutinEffectHigherOrderNonlinearities2017}%
  \BibitemOpen
  \bibfield  {author} {\bibinfo {author} {\bibfnamefont {S.}~\bibnamefont {Boutin}}, \bibinfo {author} {\bibfnamefont {D.~M.}\ \bibnamefont {Toyli}}, \bibinfo {author} {\bibfnamefont {A.~V.}\ \bibnamefont {Venkatramani}}, \bibinfo {author} {\bibfnamefont {A.~W.}\ \bibnamefont {Eddins}}, \bibinfo {author} {\bibfnamefont {I.}~\bibnamefont {Siddiqi}},\ and\ \bibinfo {author} {\bibfnamefont {A.}~\bibnamefont {Blais}},\ }\href {https://doi.org/10.1103/PhysRevApplied.8.054030} {\bibfield  {journal} {\bibinfo  {journal} {Physical Review Applied}\ }\textbf {\bibinfo {volume} {8}},\ \bibinfo {pages} {054030} (\bibinfo {year} {2017})}\BibitemShut {NoStop}%
\bibitem [{\citenamefont {Mirhosseini}\ \emph {et~al.}(2019)\citenamefont {Mirhosseini}, \citenamefont {Kim}, \citenamefont {Zhang}, \citenamefont {Sipahigil}, \citenamefont {Dieterle}, \citenamefont {Keller}, \citenamefont {{Asenjo-Garcia}}, \citenamefont {Chang},\ and\ \citenamefont {Painter}}]{mirhosseiniCavityQuantumElectrodynamics2019b}%
  \BibitemOpen
  \bibfield  {author} {\bibinfo {author} {\bibfnamefont {M.}~\bibnamefont {Mirhosseini}}, \bibinfo {author} {\bibfnamefont {E.}~\bibnamefont {Kim}}, \bibinfo {author} {\bibfnamefont {X.}~\bibnamefont {Zhang}}, \bibinfo {author} {\bibfnamefont {A.}~\bibnamefont {Sipahigil}}, \bibinfo {author} {\bibfnamefont {P.~B.}\ \bibnamefont {Dieterle}}, \bibinfo {author} {\bibfnamefont {A.~J.}\ \bibnamefont {Keller}}, \bibinfo {author} {\bibfnamefont {A.}~\bibnamefont {{Asenjo-Garcia}}}, \bibinfo {author} {\bibfnamefont {D.~E.}\ \bibnamefont {Chang}},\ and\ \bibinfo {author} {\bibfnamefont {O.}~\bibnamefont {Painter}},\ }\href {https://doi.org/10.1038/s41586-019-1196-1} {\bibfield  {journal} {\bibinfo  {journal} {Nature}\ }\textbf {\bibinfo {volume} {569}},\ \bibinfo {pages} {692} (\bibinfo {year} {2019})}\BibitemShut {NoStop}%
\bibitem [{\citenamefont {Probst}\ \emph {et~al.}(2015)\citenamefont {Probst}, \citenamefont {Song}, \citenamefont {Bushev}, \citenamefont {Ustinov},\ and\ \citenamefont {Weides}}]{probstEfficientRobustAnalysis2015}%
  \BibitemOpen
  \bibfield  {author} {\bibinfo {author} {\bibfnamefont {S.}~\bibnamefont {Probst}}, \bibinfo {author} {\bibfnamefont {F.~B.}\ \bibnamefont {Song}}, \bibinfo {author} {\bibfnamefont {P.~A.}\ \bibnamefont {Bushev}}, \bibinfo {author} {\bibfnamefont {A.~V.}\ \bibnamefont {Ustinov}},\ and\ \bibinfo {author} {\bibfnamefont {M.}~\bibnamefont {Weides}},\ }\href {https://doi.org/10.1063/1.4907935} {\bibfield  {journal} {\bibinfo  {journal} {Review of Scientific Instruments}\ }\textbf {\bibinfo {volume} {86}},\ \bibinfo {pages} {024706} (\bibinfo {year} {2015})}\BibitemShut {NoStop}%
\bibitem [{\citenamefont {Joshi}\ \emph {et~al.}(2023)\citenamefont {Joshi}, \citenamefont {Yang},\ and\ \citenamefont {Mirhosseini}}]{joshiResonanceFluorescenceChiral2023}%
  \BibitemOpen
  \bibfield  {author} {\bibinfo {author} {\bibfnamefont {C.}~\bibnamefont {Joshi}}, \bibinfo {author} {\bibfnamefont {F.}~\bibnamefont {Yang}},\ and\ \bibinfo {author} {\bibfnamefont {M.}~\bibnamefont {Mirhosseini}},\ }\href {https://doi.org/10.1103/PhysRevX.13.021039} {\bibfield  {journal} {\bibinfo  {journal} {Physical Review X}\ }\textbf {\bibinfo {volume} {13}},\ \bibinfo {pages} {021039} (\bibinfo {year} {2023})}\BibitemShut {NoStop}%
\bibitem [{\citenamefont {Dunsworth}\ \emph {et~al.}(2017)\citenamefont {Dunsworth}, \citenamefont {Megrant}, \citenamefont {Quintana}, \citenamefont {Chen}, \citenamefont {Barends}, \citenamefont {Burkett}, \citenamefont {Foxen}, \citenamefont {Chen}, \citenamefont {Chiaro}, \citenamefont {Fowler}, \citenamefont {Graff}, \citenamefont {Jeffrey}, \citenamefont {Kelly}, \citenamefont {Lucero}, \citenamefont {Mutus}, \citenamefont {Neeley}, \citenamefont {Neill}, \citenamefont {Roushan}, \citenamefont {Sank}, \citenamefont {Vainsencher}, \citenamefont {Wenner}, \citenamefont {White},\ and\ \citenamefont {Martinis}}]{dunsworthCharacterizationReductionCapacitive2017}%
  \BibitemOpen
  \bibfield  {author} {\bibinfo {author} {\bibfnamefont {A.}~\bibnamefont {Dunsworth}}, \bibinfo {author} {\bibfnamefont {A.}~\bibnamefont {Megrant}}, \bibinfo {author} {\bibfnamefont {C.}~\bibnamefont {Quintana}}, \bibinfo {author} {\bibfnamefont {Z.}~\bibnamefont {Chen}}, \bibinfo {author} {\bibfnamefont {R.}~\bibnamefont {Barends}}, \bibinfo {author} {\bibfnamefont {B.}~\bibnamefont {Burkett}}, \bibinfo {author} {\bibfnamefont {B.}~\bibnamefont {Foxen}}, \bibinfo {author} {\bibfnamefont {Y.}~\bibnamefont {Chen}}, \bibinfo {author} {\bibfnamefont {B.}~\bibnamefont {Chiaro}}, \bibinfo {author} {\bibfnamefont {A.}~\bibnamefont {Fowler}}, \bibinfo {author} {\bibfnamefont {R.}~\bibnamefont {Graff}}, \bibinfo {author} {\bibfnamefont {E.}~\bibnamefont {Jeffrey}}, \bibinfo {author} {\bibfnamefont {J.}~\bibnamefont {Kelly}}, \bibinfo {author} {\bibfnamefont {E.}~\bibnamefont {Lucero}}, \bibinfo {author} {\bibfnamefont {J.~Y.}\ \bibnamefont {Mutus}}, \bibinfo {author} {\bibfnamefont {M.}~\bibnamefont {Neeley}}, \bibinfo {author} {\bibfnamefont {C.}~\bibnamefont {Neill}}, \bibinfo {author} {\bibfnamefont {P.}~\bibnamefont {Roushan}}, \bibinfo {author} {\bibfnamefont {D.}~\bibnamefont {Sank}}, \bibinfo {author} {\bibfnamefont {A.}~\bibnamefont {Vainsencher}}, \bibinfo {author} {\bibfnamefont {J.}~\bibnamefont {Wenner}}, \bibinfo {author} {\bibfnamefont {T.~C.}\ \bibnamefont {White}},\ and\ \bibinfo {author} {\bibfnamefont {J.~M.}\ \bibnamefont {Martinis}},\ }\href {https://doi.org/10.1063/1.4993577} {\bibfield  {journal} {\bibinfo  {journal} {Applied Physics Letters}\ }\textbf {\bibinfo {volume} {111}},\ \bibinfo {pages} {022601} (\bibinfo {year} {2017})}\BibitemShut {NoStop}%
\end{thebibliography}%

\end{document}